\documentclass[11pt,a4paper]{article}

\usepackage{jheppub}
\usepackage[usenames,dvipsnames]{xcolor}
\usepackage{amssymb,bm,cancel} 
\usepackage{amsmath}
\usepackage{slashed}
\usepackage{mathtools}
\usepackage{amsfonts}    
\usepackage{dsfont}
\usepackage{pdfpages}
\usepackage{verbatim}
\usepackage{stmaryrd}
\usepackage{graphicx}
\usepackage{hyperref}
\usepackage{bbm}
\usepackage{tikz}
\usepackage{pgfplots}
\usetikzlibrary{intersections, pgfplots.fillbetween}
\usetikzlibrary{arrows}
\usetikzlibrary{decorations.pathmorphing}
\usetikzlibrary{ decorations.markings}
\usetikzlibrary{shapes.misc}
\tikzset{cross/.style={cross out, draw=black, minimum size=2*(#1-\pgflinewidth), inner sep=0pt, outer sep=0pt},
cross/.default={1pt}}
\usetikzlibrary{shapes.geometric}
\tikzset{
    partial ellipse/.style args={#1:#2:#3}{
        insert path={+ (#1:#3) arc (#1:#2:#3)}
    }
}

\renewcommand{\figurename}{Fig.}

\newcommand{\nn}{\nonumber}

\newcommand{\bi}{\begin{itemize}}
\newcommand{\ei}{\end{itemize}}
\newcommand{\bea}{\begin{eqnarray}}
\newcommand{\eea}{\end{eqnarray}}
\newcommand{\be}{\begin{equation}}
\newcommand{\ee}{\end{equation}}

\definecolor{pink}{rgb}{1.0, 0.13, 0.32}

\numberwithin{equation}{section}
\begin{document}

\vspace*{2.5cm}
\begin{center}
{ \LARGE \text{Fermionic fields of higher spin in de Sitter space}\\}
\vspace*{1.7cm}
Dionysios Anninos,$^{1,2}$ Chiara Baracco,$^1$ Vasileios A. Letsios,$^3$ and Guillermo A. Silva$^{4}$
\vspace*{0.6cm}
\vskip4mm
\begin{center}
{
\footnotesize
{
$^1$Department of Mathematics, King's College London, Strand, London WC2R 2LS, UK \\
$^2$ Instituut voor Theoretische Fysica, KU Leuven, Celestijnenlaan 200D, B-3001 Leuven, Belgium  \\
$^3$ Physique de l’Univers, Champs et Gravitation, UMONS,
Place du Parc 20, 7000 Mons, Belgium \\
$^4$ Instituto de Física de La Plata - CONICET, Diagonal 113 e/ 63 y 64, 1900, La Plata, Argentina\\
  \& Departamento de F\'\i sica, Universidad Nacional de La Plata,
C.C. 67, 1900, La Plata, Argentina\\
}}
\end{center}

\begin{center}
{\textsf{\footnotesize{
dionysios.anninos@kcl.ac.uk, chiara.baracco@kcl.ac.uk, vasileios.letsios@umons.ac.be, silva@fisica.unlp.edu.ar}} } 
\end{center}

\vspace*{0.6cm}

\vspace*{0.6cm}


\end{center}
\vspace*{1.5cm}
\begin{abstract}
\noindent
We consider fermionic fields of higher spin on a four-dimensional de Sitter background. A particular emphasis is placed on the Rarita-Schwinger spin-$\tfrac{3}{2}$ case. Both massive fields and gauge fields are considered, and their relation to the representation theory of $SO(4,1)$ is discussed. In Lorentzian signature, we study properties of the Bunch-Davies mode functions, and the late time structure of their two-point functions. For the Rarita-Schwinger gauge field, we consider a quantisation procedure based on the Minkowskian limit of the field operator. In Euclidean signature, the fields are placed on a four-sphere and the Euclidean path integral is computed at one-loop. The resulting Euclidean partition function is expressed in terms of unitary Lorentzian group characters with edge corrections. The unitary nature of the characters contrasts the lack of a conventional real action for the Rarita-Schwinger gauge field in de Sitter space. We speculate on the microscopic properties of a theory comprised of an infinite tower of interacting integer and half-integer gauge fields in de Sitter space. Along the way, we discuss a potentially interesting expression for the higher-spin path integral on the four-sphere.

\end{abstract}

\newpage

\tableofcontents

\section{Introduction}

The purpose of this article is to explore quantum fields of half-integer intrinsic spin on a four-dimensional de Sitter background. We will primarily focus on the case of a spin-$\tfrac{3}{2}$ field, which we refer to as the massive Rarita-Schwinger field when massive, and the Rarita-Schwinger gauge field when it displays a gauge-redundancy.  The study of such fields on a Minkowski background is an old problem \cite{fierz1939relativistic,PhysRev.60.61,Fang:1979hq}. Moreover, early considerations of the Rarita-Schwinger fields on arbitrary curved backgrounds \cite{Buchdahl:1958xv} (see also \cite{Hack:2011yv}) indicated that it is not straightforward to maintain the constraints of the theory unless the background metric satisfies an Einstein condition. Such findings are accompanied by results  \cite{Johnson:1960vt,Velo:1969bt} on the acausal propagation of the Rarita-Schwinger field placed in an external electromagnetic field. At least when massive, however, the free Rarita-Schwinger field can be formulated on a de Sitter background, in addition to Minkowski and anti-de Sitter space \cite{Schenkel:2011nv}. 
 \newline\newline
In what follows, we will elaborate on some of the properties of the Rarita-Schwinger field on a fixed de Sitter background. Some relevant literature considering this problem includes \cite{vasiliev1988free,metsaev1998fermionic,Deser:2001us,Deser:2001xr,Deser:2003gw,Metsaev:2006zy,Metsaev:2006ui,Zinoviev:2007ig,Anguelova:2003kf,Redkov:2011nes,Pejhan:2019ech,Nolland:2000fx,Letsios:2022tsq,Letsios:2023qzq, Higuchi-Letsios, Boulanger:2023lgd, Letsios:2023awz, Letsios:2023tuc}. We will mostly consider the problem at the free level, both in the classical and quantum case, and in both Lorentzian and Euclidean signature. We will, however, discuss a potential avenue towards the interacting problem in the concluding remarks. Part of our philosophy is that a more comprehensive understanding of general particle species on a de Sitter background may aid in the construction of broader classes of tractable quantum models \cite{Anninos:2020hfj}. Moreover, one might hope for the unlikely possibility that the presence of such particles in the early Universe may yield observable consequences. 
\newline\newline
When considering the Rarita-Schwinger gauge field on a de Sitter background certain issues arise. This appears to have been first discussed in the work of \cite{Pilch:1984aw}, who showed that there can be no four-dimensional de Sitter supergravity action satisfying standard reality or positivity properties. What the authors of \cite{Pilch:1984aw} showed more concretely, is that attempting to write down a supergravity action endowed with a positive cosmological constant comes at the price of either a gravitino with imaginary `mass' or a graviphoton whose kinetic term has the wrong sign. Such a statement naturally fits with the more general idea that de Sitter space is not compatible with supersymmetry \cite{lukierski1985all}. At the level of quantum field theory, this is most clearly expressed as the fact that none of the de Sitter isometry generators  are globally positive definite. This is in conflict with the presence of a supercharge that would automatically imply a positive bosonic charge. For supergravity quantised on closed spatial slices, one expects physical states to be annihilated by the supercharge so as to hopefully circumvent the issue, but  \cite{Pilch:1984aw} appears to tell us otherwise. Sensible supergravity theories with de Sitter solutions permitting a non-linearly realised supersymmetry were constructed in \cite{Bergshoeff:2015tra}, where a `massive' gravitino was shown to be present.
\newline\newline
In contrast to the above situation, there have been a few instances where the aforementioned troubles might be alleviated. In \cite{Anous:2014lia,Cassani:2012ri} it was shown, through fairly clear reasoning, that superconformal field theories make sense on a de Sitter background due to the presence of enhanced conformal symmetry generators whose charges can be indeed be positive (see also \cite{Higuchi-Letsios}). In \cite{Anninos:2023exn,Anninos:2022ujl,Muhlmann:2025ngz}, noting that the graviton multiplet in two spacetime dimensions does not carry propagating local degrees of freedom, a two-dimensional supergravity theory with $\Lambda>0$ coupled to a superconformal field theory was examined. In \cite{Sezgin:2012ag,Hertog:2017ymy}, it was hypothesised that a supersymmetric extension of higher-spin theories with four-dimensional de Sitter vacua, containing an infinite tower of massless integer and half-integer gauge fields, exists. Similar to their purely bosonic counterparts, there is no known spacetime action formalism in terms of Fronsdal fields for these supersymmetric higher-spin theories, so it is hard to assess the quantum completion of the putative theories in \cite{Sezgin:2012ag,Hertog:2017ymy}. Nonetheless, a proposal for their Hartle-Hawking wavefunctional was considered in \cite{Hertog:2017ymy}. Finally, in \cite{Hull:1998vg}, a type of superstring theory was envisioned that contains supersymmetric de Sitter solutions. 
\newline\newline
We will also be interested in the complementary Euclidean setting, whereby  the fermionic fields are placed on a four-sphere background. The Euclidean action for a fermionic field generally does not have any reality properties \cite{Stone:2020vva}. Rather, one should consider complexifying all fermionic fields in Euclidean signature. Although ultimately the constraints of Lorentzian unitarity will impose the choice of contour for the complexified Euclidean fermions, it may be of interest to consider the Euclidean theories  in and of their own right, dispensed of such constraints. For instance, reflection positivity or standard reality conditions might be foregone when implementing methods of supersymmetric localisation to mass deformed theories on the Euclidean sphere \cite{Jafferis:2010un}, and considering twisted versions of supersymmetric theories that compute topological invariants \cite{Witten:1988ze}. (This was almost certainly not the mindset in \cite{Pilch:1984aw}, who in addition to assuming the existence of an action, were strictly considering the problem in Lorentzian signature.) 
\newline\newline
Bearing all this in mind, one might  imagine implementing some of the methods of supersymmetry and supergravity toward a `twisted' Euclidean $\mathcal{N}=2$ de Sitter theory. This, in the general hope of generating a sharper theoretical dataset for Euclidean quantities of interest in $\Lambda>0$ theories of quantum gravity. 

\subsection*{Outline of the paper}

In section \ref{UIRsec}, we recall certain facts about half-integer spin unitary irreducible representations (UIRs) of the four-dimensional de Sitter isometry group $SO(4,1)$, or better said, its double cover Spin$(4,1)$. Rather than `mass', which is not a particularly sensible notion on a curved space, we discuss the properties of the UIRs. We note that there exist discrete series UIRs for each half-integer spin, much like they do for integer spin. These discrete series are typically associated with the single particle Hilbert space of free gauge fields on a de Sitter background \cite{Higuchi:1991tn,RiosFukelman:2023mgq, Letsios:2023qzq, Higuchi-Letsios}. 
\newline\newline
In sections \ref{sec:mrs} and \ref{section 4}, we study the massive Rarita-Schwinger field and compute its explicit field modes in the planar coordinate system of a four-dimensional de Sitter background. We discuss the quantisation of the massive Rarita-Schwinger field, emphasising both late and early time limits. We compute some properties of the equal time two-point function. 
\newline\newline
In section \ref{section 5}, we study the solutions of the equation of motion governing the Rarita-Schwinger gauge field. We note their late time behaviour which has a growing mode, in parallel to bosonic higher-spin gauge fields in de Sitter space. We consider a possible quantisation for the Rarita-Schwinger gauge field. The quantisation procedure we consider is not based on an action formalism, but the requirement that the quantum field operator reduces to its Minkowskian counterpart in the flat space limit. 
\newline\newline
In section \ref{EPI}, we study the problem in Euclidean signature, and compute the one-loop sphere path integral, $\mathcal{Z}$, for both the massive Rarita-Schwinger field, and the Rarita-Schwinger gauge field. The path integral admits an expression in terms of the Harish-Chandra characters of unitary irreducible representations for both cases, despite the imaginary `mass' of the Rarita-Schwinger gauge field \cite{Anninos:2023exn}. Along the way we compute the coefficients of the logarithmic divergences of $\log \mathcal{Z}$, for both cases. We also consider the one-loop partition function for theories with infinite towers of bosonic and fermionic higher-spin gauge fields. We uncover a potentially interesting formula (\ref{chsS4}).
\newline\newline
In section \ref{outlook}, we conclude with an outlook on the ingredients of a putative microscopic theory describing an interacting tower of fermionic and bosonic higher-spin gauge fields explored in \cite{Sezgin:2012ag,Hertog:2017ymy}. We follow the procedure of \cite{Anninos:2017eib} where the analogous problem was considered for the better understood case of a spectrum with purely bosonic fields. 
\newline\newline
Our choice of conventions, along with many details of the path integral computations, can be found in the various appendices.

\section{Fermionic representations of $SO(4,1)$}\label{UIRsec}

In this section, we review fermionic unitary irreducible representations of the four-dimensional de Sitter group $SO(4,1)$, or better stated its simply connected double cover Spin$(4,1)$. Its representation theory has been analysed in several mathematical works \cite{dixmier1961representations,ottoson1968classification,schwarz1971unitary,hirai,Dobrev:1977qv,Schaub:2023scu}.
It should be noted that, although the four-dimensional de Sitter group, $SO(4,1)$, and anti-de Sitter group, $SO(3,2)$, are naively similar, their unitary representations are strikingly different. This is a reflection of the spacetimes themselves exhibiting different physics, as seen from a group theoretic point of view. We denote the ten generators of $SO(4,1)$ by $M_{AB} = -M_{BA}$, with $A,B=0,1,2,3,4$. They satisfy the algebra 
\begin{equation}
[M_{AB},M_{CD}]=\eta_{AD} M_{BC}  -\eta_{BD} M_{AC} +  \eta_{BC}M_{AD} - \eta_{AC} M_{BD}~,
\end{equation}
with $\eta_{AB} =\text{diag}(-1,1,1,1,1)$. $SO(4)$, the maximally compact subgroup of $SO(4,1)$, is generated by the six $M_{ab}$ with $a,b=1,2,3,4$ and $a>b$.

\subsection{Fermionic representation content of $SO(4,1)$}
We follow the notation of \cite{schwarz1971unitary}, and label the representations by their $SO(4)$ content. It was shown in \cite{dixmier1961representations} that a given unitary irreducible representation (UIR) of $SO(4)$ can be contained at most once in the corresponding $SO(4,1)$ irreducible representation. UIRs of the $so(4)$ Lie algebra can be labeled by two numbers, $q$ and $p$ with $|p|\le q$, either both integer or half-integer. For each UIR of $SO(4)$ one finds a corresponding UIR for its double cover. The states within the UIR can be further characterized by two indices $l$ and $m$ satisfying $|p| \le l \le q$ and $-l \le m \le l$, which correspond to the decomposition $so(4) \supset so(3) \supset so(2)$. The dimension of the $SO(4)$ UIR $\mathbb{Y}_{q,p}$ is found using the Weyl dimension formula
\begin{equation}
\dim \mathbb{Y}_{q,p} = (q+1)^2 - p^2 ~. 
\end{equation}
Given the UIRs of $SO(4)$, one can proceed to obtain those of $SO(4,1)$ by ensuring that the generators obey the appropriate Hermiticity conditions. The complete set of UIRs for the double cover of $SO(4,1)$ can be found in Table V of \cite{schwarz1971unitary}. Here we list the fermionic subset which is the case of most interest in what follows. 
\newline\newline
\textbf{Fermionic Principal series $\mathcal{D}_{s,\nu}$.} These are the continuous representations, labeled by a half-integer parameter $s =\tfrac{1}{2}+n$, with $n \in \mathbb{N}$, and a continuous parameter $\nu \in \mathbb{R}^+$. Their $SO(4)$ content is labeled by the set of half-integers $(p,q)$ satisfying $|p| \le s\le q$. The representations are unbounded because $q$ can increase indefinitely. Physically, we can view these UIRs as being furnished by single-particle states of a massive fermionic field with intrinsic spin $s$. We note that there are no complementary series for half-integer spin. Rather than `mass', a more invariant quantity is the eigenvalue of the quadratic Casimir, which is given by $\mathcal{C}_2=-\nu^2 - \tfrac{9}{4}+s(s+1)$ and it is negative provided $\nu^2>s(s+1)-\tfrac{9}{4}$.
\newline\newline
\textbf{Fermionic Discrete series $\mathcal{D}^\pm_{s,t}$.} The remaining fermionic UIRs are the two discrete series representations of $SO(4,1)$, labeled by $s=\tfrac{1}{2}+n$ and $t$ such that  $\tfrac{1}{2} \le t+1 \le s$, with $n \in \mathbb{N}$. Their $SO(4)$ content is given by the set of half-integers $(p,q)$ satisfying $t+1 \le \pm p \le s \le q$. The sign discerns lowest $(+)$ or highest $(-)$ weight discrete series UIRs. The representations are unbounded because $q$ can increase indefinitely. The eigenvalue of the quadratic Casimir for these UIRs is given by $\mathcal{C}_2 = (2+t)(t-1)+ s(s+1)$, and it is positive for $s>1$. Note that the quadratic Casimir of the $\nu=0$, spin-$s$ principal series UIR coincides with that of the $t=-\tfrac{1}{2}$ spin-$s$ discrete series. (This is only true for fermionic UIRs, for bosonic ones one would need an imaginary $\nu$.) Physically, we can view these fermionic UIRs as being furnished by single-particle states of a (partially massless) fermionic field carrying half-integer spin $s$.\footnote{The exception to this is the $s=\tfrac{1}{2}$ case, which corresponds to a massless fermion. While the massless fermion corresponds to the $\nu=0$ limit of the principal series, it is further reducible into two bounded UIRs depending on the helicity structure. In fact, the massless fermion on dS$_2$ is reducible into a highest and lowest weight UIR of the double cover of $SO(2,1)$. A similar situation holds for massless fermions of higher spin, which are again reducible into two UIRs, namely $\mathcal{D}^\pm_{s,-\frac{1}{2}}$.} The totally massless gauge field corresponds to the case $t=s-1$.  We will explore the field theoretic interpretation of $\mathcal{D}^\pm_{s,s-1}$ and its shortcomings  in section \ref{section 4}. The discretuum of UIRs labeled by the allowed values of $t$ for fixed spin $s$ is the group theoretic counterpart of the partially massless spin-$s$ gauge fields appearing in de Sitter space \cite{Deser:1983mm}. The generalised version of the UIR corresponding to the Higuchi bound \cite{Higuchi:1986py} is given by $\mathcal{D}^\pm_{s,-\frac{1}{2}}$.
\newline\newline
It is interesting to note that the nature of the fermionic UIRs depends on the dimension of the de Sitter group \cite{ottoson1968classification,schwarz1971unitary,Schaub:2023scu, Letsios:2022tsq, Letsios:2023qzq}. For instance, the groups $SO(5,1)$ and $SO(3,1)$ only permit principal series fermionic representations, whilst the groups $SO(2,1)$ and $SO(4,1)$ admit both principal and discrete series fermionic representations. In the latter case, by fermionic we mean carrying half-integer labels of the $SO(2)$ subgroup of $SO(2,1)$. For $SO(D,1)$ with $D\ge 6$, one can have exceptional fermionic representations whose $SO(D)$ content necessarily becomes more elaborate than that of a totally symmetric spatial tensor.

\subsection{Tensor products}

Having listed the various fermionic UIRs of the double cover of $SO(4,1)$, we would like to discuss briefly their tensor product structure. This has been analysed in work of Martin \cite{martin1981tensor,martin1984tensor}. Generally speaking the tensor product of continuous UIRs can contain both continuous and discrete UIRs. 
\newline\newline
Specifically, from Theorem 4 of \cite{martin1981tensor} we have that the tensor product of two principal series UIRs satisfies
\begin{equation}
\mathcal{D}_{s_1,\nu_2} \otimes \mathcal{D}_{s_2,\nu_2} =  \int_{\mathcal{C}_\nu} \mathcal{D}_{s_3,\nu}  \oplus_\nu \mathcal{D}^+_{s_3,t}  \oplus \mathcal{D}^-_{s_3,t}~,
\end{equation}
where $s_1 + s_2 + s_3 \equiv 0 \, \text{mod} \, 1$, and $0$ or $\tfrac{1}{2} \leq t \leq \text{min} (s_3,s_1+s_2)-1$. On the right hand side, the whole set of principal series of values $\nu \in \mathbb{R}$ appear weighted by the Plancherel measure, and $\mathcal{C}_\nu$ is the principal series contour. Interestingly, for certain choices of $s_1$ and $s_2$ we can generate non-trivial discrete series UIRs in the tensor product. For instance, if we take $s_1 = \tfrac{1}{2}$ and $s_2 = 1$, then the discrete series UIRs with $s_3 = n + \tfrac{1}{2}$ with $n\in \mathbb{N}$ can appear. More physically, this would correspond to the tensor product of the single particle Hilbert space of a massive spinor and a massive vector field in dS$_4$. Moreover, if $s_1 = s_2 = 0$ or $s_1=0$ and $s_2=\tfrac{1}{2}$ only principal series UIRs will appear in the decomposition.
\newline\newline
The tensor product of a principal series UIR with a general UIR is given in Theorem 5 of \cite{martin1984tensor}. Again, here, one can obtain both principal and discrete series UIRs in the decomposition. As a concrete example,
\begin{equation}
\mathcal{D}_{\frac{1}{2},\nu} \otimes \mathcal{D}^\pm_{1,0} =  \int_{\mathcal{C}_\nu}  \mathcal{D}_{s_\nu,\nu}  \oplus_\nu \mathcal{D}^\pm_{s,t}~,
\end{equation}
where $s_\nu \ge  \tfrac{1}{2}$, $\tfrac{1}{2} + 1 + s \equiv 0 \, \text{mod} \, 1$, and $\tfrac{1}{2} \le t \le s-1$. The decomposition includes the $\mathcal{D}^\pm_{\frac{3}{2},\frac{1}{2}}$ UIRs. More prosaically, the tensor product of the single particle Hilbert space of the standard model electron and photon contains the single-particle Hilbert space of a de Sitter gravitino.

\subsection{Harish-Chandra characters}

In addition to classifying the UIRs, one can also compute group characters, as was done in early work of Hirai \cite{hirai1965characters}. Throughout our discussion, we are interested in fermionic representations of  the four-dimensional de Sitter group $SO(4,1)$. More specifically, we are interested in the reduced group character
\begin{equation}
    \chi_{\mathcal{D}}(t) \equiv \text{tr}_{\mathcal{D}} \,e^{-i H t} ~,
\end{equation}
where $H$ is the Hermitian generator of the $SO(1,1)$ subgroup of $SO(4,1)$ and $\mathcal{D}$ labels the UIR of interest. This particular character plays an central role in expressions for the Euclidean partition function of Euclidean de Sitter space \cite{Anninos:2020hfj}. We are interested here in the fermionic counterpart of these expressions. For the principal series $\mathcal{D}_{s,\nu}$, and defining $q\equiv e^{-t}$, one has
\begin{equation}
\chi_{\mathcal{D}_{s,\nu}} (t) \equiv (2s+1) \,\frac{q^{ \Delta } + q^{\bar{\Delta}}}{\left(1-q\right)^3}~, \quad\quad \Delta \equiv \tfrac{3}{2} + i \nu = 3-\bar{\Delta}~.
\end{equation}
For the maximal depth discrete series $\mathcal{D}^\pm_{s,s-1}$, which will be our main interest, one has
\begin{equation}\label{chiD}
\chi_{\mathcal{D}^\pm_{s,s-1}} (t) \equiv \frac{(2s+1)q^{s+1}-(2s-1)q^{s+2}}{\left(1-q\right)^3}~.
\end{equation}
We present some details of the derivation of the above character in appendix \ref{appchar}.

\section{Massive Rarita-Schwinger field}\label{sec:mrs}

In this section, we consider the classical massive spin-$\tfrac{3}{2}$ Rarita-Schwinger equation\footnote{In fact, spin-$\tfrac{3}{2}$ particles were considered earlier than Rarita and Schwinger's classic two-page paper \cite{PhysRev.60.61}, in work of Fierz and Pauli \cite{fierz1939relativistic} where their interaction with electromagnetic fields was considered, as well as in early work of Dirac \cite{dirac1936relativistic}. In what follows, however, we will refer to half-integer spin fermionic fields of arbitrary mass as Rarita-Schwinger fields as we follow their treatment most closely. Rarita and Schwinger also noted in their paper that for the massless case, their equations exhibit a gauge redundancy.} in a four-dimensional de Sitter (dS$_4$) background and its quantisation. Previous literature on this problem includes \cite{Metsaev:2006zy,Metsaev:2006ui,Zinoviev:2007ig,Anguelova:2003kf,Redkov:2011nes,Pejhan:2019ech,Nolland:2000fx,Letsios:2022tsq, Letsios:2023qzq, Boulanger:2023lgd}. Part of our focus will lie on the properties of the classical and quantum field at early and late times. The Rarita-Schwinger equation governs a spin-$\tfrac{3}{2}$ complex fermionic field $\Psi_{\mu\alpha}$, carrying a spinor index $\alpha$ and spacetime vector index $\mu$. A potential physical context where the massive Rarita-Schwinger field can appear is in theories of supergravity with de Sitter vacua \cite{Bergshoeff:2015tra}. For those de Sitter vacua that spontaneously break the supersymmetry, one finds a non-linear realisation of the supersymmetric transformations. The gravitino is rendered massive, with a mass scale related to the symmetry breaking scale. 

\subsection{Field equation}

We will describe our background de Sitter geometry in the planar coordinate system
\begin{equation}\label{planar}
\frac{ds^2}{\ell^2} = \frac{-d\eta^2 + d\bold{x}^2}{\eta^2} ~, 
\end{equation}
where $\bold{x} = \{x^1,x^2,x^3\} \in \mathbb{R}^3$ represents coordinates on $\mathbb{R}^3$, and $\eta\in \mathbb{R}^-$ is a conformal time coordinate. The late time boundary resides at $\eta=0^-$ and is denoted by $\mathcal{I}^+$. The de Sitter isometries $SO(4,1)$ act as Euclidean conformal maps at $\mathcal{I}^+$. We will set $\ell=1$ in what follows unless otherwise specified. We will often suppress the spinor index, and our general conventions are delineated in appendix \ref{conventions}. Our considerations will be for a Dirac massive Rarita-Schwinger gauge field, but similar considerations would apply for the Majorana case subject to the appropriate reality conditions. 
\newline\newline
The Rarita-Schwinger field equation is given by
\begin{equation} \label{eq: RS Dirac-type eqn generic m}
   \gamma^{\mu \rho \sigma } \left(  \nabla_{\rho} + \frac{m}{2}  \gamma_{\rho}   \right)   \Psi_{\sigma}   =   0 ~,
\end{equation}
where $m$ is a real valued mass parameter. By straightforward manipulations the transverse-traceless conditions follow directly from the above equation
\begin{equation} \label{eq: TT conditions}
    \nabla^{\mu}  \Psi_{\mu} = 0 = \gamma^{\mu}  \Psi_{\mu} ~.
\end{equation}
Given these constraints, it is immediate to conclude that the original sixteen off-shell degrees of freedom reduce to four complex on-shell degrees of freedom. An equivalent form of the field equation is
\begin{align} \label{eq: full RS eqn generic mass}
 \left(   \gamma^\nu {\nabla}_\nu     +  m \right) \Psi_{\mu} = 0 ~, \quad\quad \text{together with} \quad\quad \nabla^{\mu}  \Psi_{\mu} = \gamma^{\mu}  \Psi_{\mu}=0~.
\end{align}
Upon going to the planar coordinate system (\ref{planar}), the equation  for the $\eta$-component of the Rarita-Schwinger equation reads
\begin{align}\label{eq: Dirac op on psi_eta vec-spinor}
    -\Bigg( \eta \left( \gamma^{\underline{0}}\partial_{\eta} + \gamma^{\underline{1}}\partial_{1} + \gamma^{\underline{2}}\partial_{2} +\gamma^{\underline{3}}\partial_{3}  \right)- \frac{3}{2} \gamma^{\underline{0}}  \Bigg) \Psi_{\eta} + \frac{1}{ \eta}   \gamma^{\mu}   \Psi_{\mu}  =   - m   \Psi_{\eta} ~,
\end{align}
while the equation governing the spatial components is
\begin{align}\label{eq: Dirac op on psi_j vec-spinor}
    -\Bigg( \eta \left( \gamma^{\underline{0}}\partial_{\eta} + \gamma^{\underline{1}}\partial_{1} + \gamma^{\underline{2}}\partial_{2} +\gamma^{\underline{3}}\partial_{3}  \right)- \frac{3}{2} \gamma^{\underline{0}}  \Bigg) \Psi_{j} + \frac{1}{ \eta}   \gamma^{\eta}   \Psi_{j} + \frac{1}{ \eta}   \gamma^{i}  \delta_{ij}\Psi_{\eta}   =   - m   \Psi_{j}~.
\end{align}
As in Minkowski space, the transverse-traceless spin-$\tfrac{3}{2}$ field can be decomposed into $\pm \frac{1}{2}$ and $\pm \frac{3}{2}$ helicity sectors. The helicity corresponds to the eigenvalue of $\Psi_\eta^{(\pm\frac{1}{2})}$ and $\Psi_j^{(\pm\frac{3}{2})}$ with respect to the spinorial matrix
\begin{equation}
   S_{\bold{k}} =  \frac{1}{k} \begin{pmatrix}
        {\boldsymbol{\sigma}\cdot\bold{k}} & 0 \\
        0 & {\boldsymbol{\sigma}\cdot\bold{k}}
    \end{pmatrix} ~.
\end{equation}
The helicity of the solutions will be encoded in the building blocks, either  $\chi^{(\pm\frac{1}{2})}(\bf k)$ or $\chi_j^{(\pm\frac{3}{2})}(\bf k)$, which are eigenspinors of the $\mathbb R^3$ helicity operator. The  $\chi^{(\pm\frac12)}({\bold{k}})$ are complex spinors satisfying  
\begin{equation} \label{chi spinor eq.}
\frac{\boldsymbol{\sigma}  \cdot \bold{k} }k \, \chi^{(\pm\frac12 )}({\bold{k}}) =  \pm   \chi^{(\pm\frac12)}  ({\bold{k}})~,
\end{equation}
together with the completeness relation
\begin{equation}
\sum_{ s\in\{\pm\frac12\} } \chi^{( {s})}({\bold{k}}) \chi^{( {s})}({\bold{k}})^\dag = \mathbb{I}_2 ~,
\end{equation}
where $\mathbb{I}_2$ is the two-dimensional spinorial identity matrix. The $\chi^{( {\pm\frac32})}_{ {j}}$ are transverse-traceless vector-spinors  on $\mathbb{R}^{3}$
satisfying 
\begin{align} 
\label{transverse-traceless vector-spinor}
    \frac{{\boldsymbol{\sigma}   \cdot    \bold{k}}}k  \chi^{( {\pm\frac32})}_{ {j}} (  {\bold{k}} ) = \pm   \,  \chi^{(\pm\frac32)}_{ {j}}   (  {\bold{k}})~, \quad\quad \sigma^{j}\chi^{(\pm\frac32)}_{ {j}}({\bold{k}}) = 0 = k^{ {j}} \chi^{(\pm\frac32)}_{ {j}}({\bold{k}}  ) ~,
\end{align}
as well as the completeness relation
\begin{equation} \label{eq: completeness relation again}
    \underset{ {s} \in \{\pm \frac32 \}}{\sum} \, \chi_j^{({s})}(\bold{k}) \, \chi_l^{({s})}(\bold{k})^\dagger = \Pi_{jl}(\bold{k}) \equiv \delta_{jl}\mathbb{I}_2 - \frac{3}{2}\frac{k_j k_l}{k^2} \mathbb{I}_2 - \frac{\sigma_j \sigma_l}{2}   
     + \frac{1}{2}\frac{k_j}{k}\frac{\bold{k}\cdot\boldsymbol{\sigma}}{k}\sigma_l + \frac{1}{2}\frac{\sigma_j k_l}{k} \frac{\bold{k}\cdot\boldsymbol{\sigma}}{k} ~.
\end{equation}
We note that the projector $\Pi_{jl}(\bold{k})$ is transverse, $k^j \Pi_{jl}(\bold{k}) = 0 =  \Pi_{jl}(\bold{k}) k^l$, and $\sigma$-traceless, $\sigma^j \Pi_{jl}(\bold{k}) = 0 = \Pi_{jl}(\bold{k}) \sigma^l$.
\newline\newline
We now analyse the two types of solutions admitted by the massive Rarita-Schwinger field equations. The AdS$_4$ counterpart of these solutions were analysed, for instance, in \cite{Volovich:1998tj} (see equations (11) and (12) therein) and our solutions are related by a simple analytic continuation.

\subsection{Helicity$-\frac{1}{2}$ solutions} 

We start considering solutions of  helicity-$\frac{1}{2}$, which we refer to as type I solutions. We represent them in components form as
\begin{equation}
\Psi^{(s)}_{\mu} = \left( \Psi^{(s)}_{\eta}, \Psi^{(s)}_{j}    \right) , \quad\quad s = \pm \frac{1}{2}~.
\end{equation} 
Using the gamma-tracelessness of $\Psi^{(s)}_{\mu}$,  equation (\ref{eq: Dirac op on psi_eta vec-spinor}) reduces to
\begin{align}
    -\Bigg( \eta \left( \gamma^{\underline{0}}\partial_{\eta} + \gamma^{\underline{1}}\partial_{1} + \gamma^{\underline{2}}\partial_{2} +\gamma^{\underline{3}}\partial_{3}  \right)- \frac{3}{2} \gamma^{\underline{0}}  \Bigg) \Psi^{(s)}_{\eta} =   - m   \Psi^{(s)}_{\eta}~.
\end{align}
The above equation shows that $\Psi_\eta^{(s)}$ satisfies the Dirac equation for a massive spin-$\tfrac{1}{2}$ field in dS$_4$. The  positive frequency solution (in the Bunch-Davies sense) is  given by
\begin{align} \label{eq: pos freq spin-3/2 type-I psi_eta ; helicities +-}
    \Psi^{(s)}_{\eta} = \frac{1}{(2\pi)^{\frac{3}{2}}}  c^{(\text{I})}_{k}\,  i\sqrt{\frac{\pi}{2}}   \,  \eta^2  \,\begin{pmatrix}
        \tilde{s} \, e^{- \tfrac{ m \pi }{2}}\,   H^{(1)}_{\frac{1}{2}   +i m }(-k \eta)   ~ \chi^{({s})}({\bold{k}}) \\ \\
        e^{ \tfrac{m \pi }{2}}\,  H^{(1)}_{\frac{1}{2}   -i m }(-k \eta)   ~  \chi^{({s})}({\bold{k}}) 
    \end{pmatrix} 
  e^{i\bf{k}\cdot\bf{x}}~,
\end{align}
where $\tilde{s}\equiv\tfrac{s}{|s|} \in \{\pm \}$ is the sign of the helicity, $k=|\bf k|$ and $H^{(1)}_{\nu}(z)$ is the Hankel function of the first kind. The helicity-$\frac12$  character of the solution stems from   $ \chi^{({\pm \tfrac{1}{2}})}({\bold{k}})$.
The time component of the type I modes has thus been fully determined, up to the overall normalisation factor.
\newline\newline
Our task now is to find the spatial components $\Psi_j$. We take the following ansatz  
\begin{align} \label{Ansatz: pos freq spin-3/2 type-I psi_j ; helicities +-}
    \Psi^{( s)}_{j} = \frac{1}{(2\pi)^{\frac{3}{2}}} c^{(\text{I})}_{k}\,  i\sqrt{\frac{\pi}{2}}   \,   
    \begin{pmatrix}
        \left( A^{( {s})}_{k}(\eta) \, k_{ j}\, \mathbb I_2  +  B^{( {s})}_{k}(\eta) \, \sigma_{ {j}}  \right)    \chi^{( {s})}({\bold{k}}) \\ \\
        \left( C^{( {s})}_{k}(\eta) \, k_{j} \, \mathbb I_2    +  D^{( {s})}_{k}(\eta) \, \sigma_{ {j}}  \right)    \chi^{( {s})}({\bold{k}})
    \end{pmatrix}  e^{i\bf{k}\cdot\bf{x}} .
\end{align}
where $k_j$ denote the momentum components, $\sigma_j$ are the Pauli matrices, and $A^{( {s})}_{k}(\eta)$, $B^{( {s})}_{k}(\eta)$, $C^{( {s})}_{k}(\eta)$, $D^{( {s})}_{k}(\eta)$ are $\eta$-dependent functions that must be determined. The   tracelessness condition $\gamma^\mu\Psi_\mu=0$ allows to write the temporal component $\Psi_\eta$ as
\begin{align}\label{eq: condition from gamma_eta x gamma trace}
    \Psi_{\eta} = - \gamma_{\eta} g^{ij}\gamma_{i}  \Psi_{j} ~, 
\end{align}
which inserted in \eqref{eq: Dirac op on psi_j vec-spinor} simplifies the equation. Explicitly, the above equation takes the form (see appendix \ref{conventions} for conventions)
\begin{align}\label{eq: condition from gamma trace type-I spin-3/2}
    \Psi^{(s)}_{\eta} =-\frac{1}{ \eta^2} \begin{pmatrix}
         0 &  \sigma_i\\
         \sigma_i  &   0
     \end{pmatrix} \,  g^{ij} \Psi^{(s)}_{j}~.
\end{align}
Substituting (\ref{eq: pos freq spin-3/2 type-I psi_eta ; helicities +-}) and (\ref{Ansatz: pos freq spin-3/2 type-I psi_j ; helicities +-}) into (\ref{eq: condition from gamma trace type-I spin-3/2}), using \eqref{chi spinor eq.} and $\sigma_{ {i}}   \sigma^{ {i}} = 3\,\mathbb{I}_2$, we find the following relations 
\begin{align}
    e^{- \tfrac{m \pi }{2}}\,  H^{(1)}_{\frac{1}{2}   +i m }(-k \eta) = -\frac{1}{\eta^{2}}    \left(   k \, C^{( {s})}_{k}(\eta)  +  3  \tilde{s} \, D^{( {s})}_{k}(\eta)    \right)~,
\end{align}
\begin{align}
   e^{ \tfrac{m \pi }{2}}\,  H^{(1)}_{\frac{1}{2}   -i m }(-k \eta) =- \frac{1}{\eta^{2}}    \left(  \tilde{s} k \, A^{( {s})}_{k}(\eta)   +  3 \, B^{( {s})}_{k}(\eta)    \right)~.
\end{align}
Additional relations arise from imposing the transversality condition $    \nabla^{\mu}  \Psi_{\nu} = 0$, which, using   \eqref{eq: condition from gamma_eta x gamma trace}, we write as
\begin{align} \label{eq: condition from div freedom and gamma trace}
   -\eta^{2}  \partial_{\eta}  \Psi_{\eta}   + \frac{5}{2}  \eta   \Psi_{\eta} + g^{ij}  \partial_{i}  \Psi_{j}=0 ~.
\end{align}
Substituting (\ref{eq: pos freq spin-3/2 type-I psi_eta ; helicities +-}) and (\ref{Ansatz: pos freq spin-3/2 type-I psi_j ; helicities +-}) into (\ref{eq: condition from div freedom and gamma trace}), one obtains the following relations,
\begin{eqnarray}
    e^{-   \tfrac{m   \pi }{2} }\eta^{3}\left( - \eta \, \partial_{\eta}   + \frac{1}{2} \right) H_{\frac{1}{2}+im}^{(1)}(-k \eta) &=& \eta^{2} i k \, \left(  - \tilde{s}  k  A_{k}^{( {s})}(\eta)   -   B_{k}^{( {s})}(\eta) \right) ~, \\
     e^{  \tfrac{ m   \pi }{2} }\eta^{3}\left( - \eta \, \partial_{\eta}   + \frac{1}{2} \right) H_{\frac{1}{2}-im}^{(1)}(-k \eta) &=& \eta^{2}  i k\, \left( -  k  C_{k}^{( {s})}(\eta)   -\tilde{s}    D_{k}^{( {s})}(\eta)  \right) ~.
\end{eqnarray}
After the dust settles, the solutions are found to be
\begin{eqnarray} \label{A value}
    A^{( {s})}_{k}(\eta)  &=& -\tilde{s} \frac{\eta}{2k}  \Big( 2\eta e^{ \tfrac{m \pi }{2}} \, H^{(1)}_{\frac{1}{2}  -  i m }(- k \eta)  + 3e^{- \tfrac{ m \pi }{2}} \frac{1+im}{ik} \, H^{(1)}_{\frac{1}{2}  + i m }(- k \eta)  \Big) ~, \\ \label{B value}
    B^{( {s})}_{k}(\eta)  &=&   \frac{\eta}{2} e^{- \tfrac{m \pi }{2}} \frac{1+i m}{ik} \, H^{(1)}_{\frac{1}{2}  + i m }(- k \eta)  ~, \\
    C^{( {s})}_{k}(\eta)  &=& \tilde{s} \left(A^{( {s})}_{k}(\eta) \right)_{m \rightarrow  -m}~, \\
     D^{( {s})}_{k}(\eta) &=& \tilde{s} \left(B^{( {s})}_{k}(\eta) \right)_{m \rightarrow  -m}~.
\end{eqnarray}
We have thus fully determined the spatial components of the type I modes in (\ref{Ansatz: pos freq spin-3/2 type-I psi_j ; helicities +-}), up to the normalisation factor $c^{(\text{I})}_{k}$. In order to fix it, we use the standard Dirac inner product
\begin{align} \label{eq: Dirac inner product vector-spinor real general}
   \left( \Psi , \Psi'  \right) = i \int_{\mathbb{R}^{3}} d^{3}x \sqrt{-g} ~g^{\mu \nu}   \, \overline{\Psi}_{\mu} \,  \gamma^{\eta} \, \Psi'_{\nu}  ~,
\end{align}
where we are assuming that the mass $m$ has a real value. 
Recalling that $\overline{\Psi}_{\mu} =\Psi^{\dagger}_{\mu} \, i \, \gamma^{\underline{0}}$ and $\gamma^\eta=-\eta\,\gamma^{\underline 0}$, the inner product reduces to
\begin{align} \label{eq: Dirac inner product vector-spinor real specific}
    \left( \Psi , \Psi'  \right) = -\frac{1}{\eta} \int_{\mathbb{R}^{3}}  d^{3}x \, \left( - \Psi^\dagger_0  \Psi_0' + \Psi^\dagger_j  \Psi_j' \right) ~,
\end{align}
and consequently
\begin{align} \label{eq: inner product of 1/2 modes un-normalised}
  \left( \Psi^{(s)} , \, \Psi^{(s')} \right) = |c^{(\text{I})}_{k}|^{2} (m^2+1) \frac{3}{k^3}  \delta^{(3)}(\bold{k}-\bold{k}'  ) \, \delta_{s s'}~,~~~ \text{for} ~~ s, s' =\pm\tfrac{1}{2}  ~.
\end{align}
We then choose the normalisation factor $c^{(\text{I})}_{k}$ to be  
\begin{align} \label{eq: normalisation constant of dS massive type I modes}
    |c^{(\text{I})}_{k}| = \sqrt{\frac{k^3}{3(m^2+1)}}~~~\implies~~~\left( \Psi^{(s)} , \, \Psi^{(s')} \right) =   \delta^{(3)}(\bold{k}-\bold{k}'  ) \, \delta_{s s'}\,.
\end{align}
The singularity  at $m = \pm i$ is to be expected, as for those values  the theory acquires a gauge redundancy. Concomitantly, for $m = \pm i$ the helicity-$\tfrac{1}{2}$ sector becomes unphysical.

\subsection{Helicity$-\frac{3}{2}$ solutions} 

We now consider helicity-$\frac{3}{2}$ solutions which we refer to as type II solutions. These have the structure 
$$\Psi^{(s)}_{\mu} = \left( 0, \Psi^{(s)}_{j}    \right),~~~~  s = \pm \tfrac{3}{2}$$ 
Notice that type I solutions have $\Psi _{\eta} \neq  0$, whereas type II  ones have $\Psi_{\eta} = 0$. We make the following ansatz for the spatial components:
\begin{align} \label{Ansatz: pos freq spin-3/2 type-II psi_j ; helicities +-}
 \Psi^{(s)}_{j}  = \frac{1}{(2\pi)^{\frac{3}{2}}} c^{(\text{II})}_{k}\,  i\sqrt{\frac{\pi}{2}} \,    \begin{pmatrix}
        \tilde{s} \, g^{(\uparrow)}_{k}(\eta)   ~  \,\chi^{( {s})}_{ {j}}({\bold{k}}) \\ \\
        g^{(\downarrow)}_{k}(\eta)   ~  \,\chi^{( {s})}_{ {j}}({\bold{k}})
    \end{pmatrix}  e^{i\bf k\cdot x} ~,
\end{align}
where $\tilde{s}\in\{\pm\}$ is again the sign of the helicity. The helicity-$\frac32$ character of the solution is manifest in the building blocks $\chi^{( {s})}_{ {j}}   ( {\bold{k}}  )$.
\newline\newline
The coefficients $g^{(\uparrow)}_{k}(\eta)$, and $g^{(\downarrow)}_{k}(\eta)$ are fixed by inserting the ansatz \eqref{Ansatz: pos freq spin-3/2 type-II psi_j ; helicities +-} in  \eqref{eq: Dirac op on psi_j vec-spinor}. One finds the following system of equations:
\begin{equation} \label{eq: spatial component Dirac eq vector-spinor, system}
    \begin{cases}
         & \eta \partial_\eta g^{(\uparrow)}_{k}(\eta) -\left( \displaystyle\frac{1}{2} - im \right) g^{(\uparrow)}_{k}(\eta) + i\eta k g^{(\downarrow)}_{k}(\eta) = 0~, \\ \\
         & \eta \partial_\eta g^{(\downarrow)}_{k}(\eta) - \left( \displaystyle\frac{1}{2} + im \right) g^{(\downarrow)}_{k}(\eta) + i\eta k g^{(\uparrow)}_{k}(\eta) = 0 ~.
    \end{cases}
\end{equation}
The corresponding solutions are 
\begin{equation} \label{eq: g coefficients vector-spinor}
    g^{(\uparrow)}_{k}(\eta) = i e^{-\tfrac{m\pi}{2}} \eta \, H_{\frac{1}{2}+im}^{(1)}(-k\eta) ~, ~~~ g^{(\downarrow)}_{k}(\eta) = e^{-\tfrac{m\pi}{2}} \eta \, H_{-\frac{1}{2}+im}^{(1)}(-k\eta) ~,
\end{equation}
where we have adhered to the standard normalisation of positive frequency modes. We thus find
\begin{align} \label{Final:pos freq spin-3/2 type-II psi_j ; helicities +-}
   \Psi^{(s)}_{j}  = \frac{1}{(2\pi)^{\frac{3}{2}}} c^{(\text{II})}_{k}\,  i\sqrt{\frac{\pi}{2}}   \, \eta    
    \begin{pmatrix}
        i \tilde{s} \, e^{-\tfrac{m\pi}{2}}   \, H_{\frac{1}{2}+im}^{(1)}(-k\eta)   ~  \,\chi^{({s})}_{ {j}}({\bold{k}}) \\ \\
        e^{-\tfrac{m\pi}{2}}   \, H_{-\frac{1}{2}+im}^{(1)}(-k\eta)   ~  \,\chi^{( {s})}_{ {j}}({\bold{k}})
    \end{pmatrix}  e^{i\bf k\cdot x},~~~~  s = \pm \tfrac{3}{2}~.
\end{align}
\newline\newline
Finally, let us normalise type II mode functions using the Dirac inner-product (\ref{eq: Dirac inner product vector-spinor real general}). Taking into account that the temporal component vanishes, the expression \eqref{eq: Dirac inner product vector-spinor real specific} reduces to
\begin{align} \label{eq: Dirac inner product vector-spinor real specific 2}
    \left( \Psi , \Psi'  \right) = -\frac{1}{\eta} \int_{\mathbb{R}^{3}}  d^{3}x \, \Psi^\dagger_j  \Psi_j' ~,
\end{align} 
Substituting (\ref{Final:pos freq spin-3/2 type-II psi_j ; helicities +-}) into (\ref{eq: Dirac inner product vector-spinor real specific 2}) one finds
\begin{align} \label{eq: inner product of 3/2 modes un-normalised}
    \left(  \Psi^{(s)} , \,  \Psi^{(s')} \right) =   |c^{(\text{II})}_{k}|^{2}  \,  \frac{2}{k} \delta^{(3)}(    \bold{k} -\bold{k}') \, \delta_{s  s'}~,~~~ \text{for} ~~ s, s' =\pm\tfrac{3}{2}  ~.
\end{align}
We therefore choose the normalisation factor $c^{(\text{II})}_{k}$ to be  
\begin{align}
    |c^{(\text{II})}_{k}| = \sqrt{\frac{k}{2}} ~~~\implies~~~ \left( \Psi^{(s)} ,  \Psi^{(s')} \right) = \delta^{(3)}(\bold{k}-\bold{k}') \, \delta_{s  s'}    ~,
\end{align}
so as to guarantee the standard normalisation of the modes  \eqref{Final:pos freq spin-3/2 type-II psi_j ; helicities +-}. We note that unlike the case of helicity-$\tfrac{1}{2}$, the helicity-$\tfrac{3}{2}$ sector permits the limit $m = \pm i$. 
\newline\newline
Having obtained the solutions to the massive Rarita-Schwinger equation, we now discuss the late and early time behaviour.

\subsection{Late and early time behaviour}

The early time (or Minkowski) limit is given by taking $k\eta\to -\infty$, such that the proper wavelength of the modes becomes parametrically small. The late time limit is accordingly given by $k\eta \to 0^-$. It is useful to uncover the behaviour of the solutions in both limits, so as to understand the effect of time evolution on initially positive frequency modes.
\newline\newline
{\textbf{Early time behaviour.}} To asses the early time behaviour, whereby $k\eta\to-\infty$, it is useful to recall the asymptotic formula
\begin{equation}
    H_{\nu}^{(1)}(-k\eta) \approx e^{-\tfrac{i\pi}{2}(\nu+\frac{1}{2})} e^{-i k \eta} \sqrt{\frac{2}{-\pi k \eta}}~.
\end{equation}
The above implies that the mode solutions (\ref{eq: pos freq spin-3/2 type-I psi_eta ; helicities +-}), (\ref{Ansatz: pos freq spin-3/2 type-I psi_j ; helicities +-}), and (\ref{Final:pos freq spin-3/2 type-II psi_j ; helicities +-}) are indeed positive frequency modes. For the helicity-$\tfrac{1}{2}$ modes we have
\begin{equation}
    \Psi^{(\pm \frac{1}{2})}_\eta  \approx \frac{\beta_k}{(2\pi)^{\frac{3}{2}}} \frac{k (-\eta)^{\frac{3}{2}}}{\sqrt{3(m^2+1)}} \begin{pmatrix}
        \pm \chi^{(\pm\frac12)}({\bold{k}}) \\
        ~\chi^{(\pm\frac12)}({\bold{k}})
    \end{pmatrix} e^{-ik\eta+i\bf k\cdot x} ~,
\end{equation}
\begin{equation}
    \Psi^{(\pm \frac{1}{2})}_j  \approx \frac{\beta_k}{(2\pi)^{\frac{3}{2}}} \frac{k_j (-\eta)^{\frac{3}{2}}}{\sqrt{3(m^2+1)}} \begin{pmatrix}
        \pm \chi^{(\pm\frac12)}({\bold{k}}) \\
        ~\chi^{(\pm\frac12)}({\bold{k}})
    \end{pmatrix} e^{-ik\eta+i\bf k\cdot x} ~,
\end{equation}
where $c^{(\text{I})}_{k} \equiv \beta_k |c^{(\text{I})}_{k}|$. Similarly, the helicity-$\tfrac{3}{2}$ modes behaviour at early times is given by
\begin{equation}
    \Psi^{(\pm \frac{3}{2})}_{j}  \approx \frac{\alpha_k}{(2\pi)^{\frac{3}{2}}} \, i \sqrt{\frac{-\eta}{2}} \begin{pmatrix}
        \pm \chi^{(\pm\frac32)}_{j}({\bold{k}}) \\
        ~\chi^{(\pm\frac32)}_{j}({\bold{k}})
    \end{pmatrix} e^{-ik\eta+i\bf k\cdot x}~,
\end{equation}
where $c^{(\text{II})}_{k}   \equiv \alpha_k |c^{(\text{II})}_{k}|$. 
\newline\newline
{\textbf{Late time behaviour.}  We now expand (\ref{eq: pos freq spin-3/2 type-I psi_eta ; helicities +-}), (\ref{Ansatz: pos freq spin-3/2 type-I psi_j ; helicities +-}), and (\ref{Final:pos freq spin-3/2 type-II psi_j ; helicities +-}) in the limit $k\eta \rightarrow 0^-$. The relevant asymptotic behaviour obeyed by the Hankel functions is  given by 
\begin{eqnarray} \label{eq: Hankel asympt behaviour 3}
     H^{(1)}_{\frac{1}{2} \pm im}(-k \eta) &\approx &  \frac{-i \, 2^{\frac{1}{2} \pm i m} \Gamma(\frac{1}{2} \pm im)}{\pi}(-k\eta)^{-\frac{1}{2} \mp im}~,  \\
     H^{(1)}_{-\frac{1}{2} +im}(-k \eta) &\approx &   \frac{2^{\frac{1}{2} - i m} (1+\tanh{m\pi})}{\Gamma(\frac{1}{2}+im)}(-k\eta)^{-\frac{1}{2} + im} ~.  
\end{eqnarray}
\noindent This leads to the late time behaviour of the type I solutions
\begin{eqnarray} 
\Psi^{(\pm \frac{1}{2})}_{\eta} &\approx& \, \frac{\beta_k}{(2\pi)^2} \sqrt{\frac{2}{3(1+m^2)}} \, k (-\eta)^{\frac{3}{2}} \begin{pmatrix} 
    \pm e^{-\frac{m\pi}{2}} \left(\frac{2}{-k\eta}\right)^{i m} \, \Gamma(\frac{1}{2} + i m ) \, \chi^{(\pm\frac12)}({\bold{k}})  \\ \\
    e^{\frac{m\pi}{2}} \left(\frac{2}{-k\eta}\right)^{- i m} \, \Gamma(\frac{1}{2} - i m) \, \chi^{(\pm\frac12)}({\bold{k}})
    \end{pmatrix}e^{i\bf k\cdot x}, \nn\\ \\  \nonumber
   \Psi^{(\pm \frac{1}{2})}_{j}  &\approx& \, \frac{\beta_k}{(2\pi)^2} \sqrt{\frac{1}{6(1+m^2)}} \, (-\eta)^{\frac{1}{2}} \\
   &&\times\begin{pmatrix} 
    (i-m) \, e^{- \frac{m\pi}{2}} \left(\frac{2}{-k\eta}\right)^{i m} \, \Gamma(\frac{1}{2} + i m ) \left( \sigma_j \mp 3\frac{k_j}{k} \right) \, \chi^{(\pm\frac12)}({\bold{k}}) \\ 
    \pm (i+m) \, e^{\frac{m\pi}{2}} \left(\frac{2}{-k\eta}\right)^{- i m} \, \Gamma(\frac{1}{2} - i m) \left( \sigma_j \mp 3\frac{k_j}{k} \right) \, \chi^{(\pm\frac12)}({\bold{k}})
    \end{pmatrix} e^{i\bf k\cdot x}.
\end{eqnarray}
Note that the appearance of a projector in the spatial components of the field  is necessary for the field to satisfy $\gamma^\mu \Psi_\mu = 0$ at late times. We can also confirm that $\nabla^\mu \Psi_\mu = 0$ in (\ref{eq: condition from div freedom and gamma trace}) is satisfied by the late time expression above.
\newline\newline
Similarly, the late time behaviour for the type II solutions is
\begin{equation}
    \Psi^{(\pm \frac{3}{2})}_{j}  \approx - \alpha_k \frac{i e^{-\frac{m\pi}{2}}}{4\pi}  (-\eta)^{\frac{1}{2}} \begin{pmatrix}
        \pm \frac{\Gamma(\frac{1}{2}+im)}{\pi} \left(\frac{2}{-k\eta}\right)^{i m} \, \chi^{(\pm\frac32)}_{j}({\bold{k}}) \\ \\
        \frac{(1+\tanh m\pi)}{\Gamma(\frac{1}{2}+im)} \left(\frac{2}{-k\eta}\right)^{-i m} \, \chi^{(\pm\frac32)}_{j}({\bold{k}})
    \end{pmatrix}e^{i\bf k\cdot x}~. 
\end{equation}
Even though the solutions originate from a purely positive frequency mode {in the far past}, their late time behaviour encodes both positive and negative frequency modes. This is characteristic of the behaviour of fields in a de Sitter background, and is a classical avatar of cosmological particle creation. We note, further, that the negative frequency modes that appear at late times are suppressed (at large mass) by a Boltzmann type factor $e^{-m \pi  \ell  }$. Recalling the inverse de Sitter temperature, $\beta_{\text{dS}} = 2\pi\ell$, the suppression is an expression of the thermal nature of de Sitter space. 

\section{Quantisation of massive Rarita-Schwinger field} \label{section 4}

In this section, we quantise the free massive Rarita-Schwinger field in four-dimensional de Sitter space. A complementary approach to this problem was elegantly considered previously in \cite{Pejhan:2019ech}. The massive Rarita-Schwinger equation (\ref{eq: full RS eqn generic mass}) can stem from a variety of actions, as already noted in the original work \cite{PhysRev.60.61}. A particularly simple form, that requires no additional auxiliary fields, is given by
\begin{equation}
    S = - \int d^4x \, \sqrt{-g} \, \bar{\Psi}_\mu \gamma^{\mu\rho\sigma} \left(\nabla_\rho + \frac{m}{2} \gamma_\rho\right) \Psi_\sigma~.
\end{equation}
From the above action, one can straightforwardly obtain the Poisson anti-brackets for the odd Grassmann variables ${\Psi}_\mu$ and $\Pi_\nu$, namely
\begin{align}
    \{ {\Psi}_{\mu\alpha}(\eta,\mathbf{x}) , \, {\Pi}_{\nu\beta}(\eta,\mathbf{y}) \}  = g_{\mu\nu} \, \delta_{\alpha\beta} \, \delta(\mathbf{x}-\mathbf{y}) ~.
   \label{PB}
\end{align}
The canonical conjugate momenta are $$ \Pi^0 = 0 ~, ~~~ \Pi^j = \left( \frac{\delta L}{\delta(\nabla_0 \Psi_j)} \right)_R = -i\left(-\Psi^{0 \, \dagger} \gamma^{\underline{0}} \gamma^j + \Psi^{j \, \dagger} \gamma^{\underline{0}} \gamma^0 \right) ~,$$ along with their conjugates. Here $_R$ indicates that the direction of the derivative is from right to left. Consequently, we have the primary constraints \cite{PhysRevD.16.307}
\begin{equation}
\vartheta_j = \Pi_j + i(-\Psi^{0 \, \dagger} \gamma^{\underline{0}} \gamma_j + \Psi_j^{\, \dagger} \gamma^{\underline{0}} \gamma^0 )~, \quad \vartheta_0 =  \Pi^0~,
\end{equation}
along with their conjugates. Requiring a vanishing Poisson bracket between $\Pi_0$ and the Hamiltonian  gives rise to a set of secondary constraints. Consequently, one has to introduce Dirac brackets in order to properly quantise the theory \cite{Baaklini:1978qa}. Upon doing so, the term involving $\Psi_0$ drops out of the Hamiltonian. As a consequence, the $\Psi_j$ are the only dynamical variables of the theory. Indeed, the time component of helicity-$\frac{3}{2}$ modes vanishes whilst the time component of helicity-$\frac{1}{2}$ modes can be expressed in terms of the spatial ones since $\gamma^\mu \Psi_\mu=0$.

\subsection{Minkowski quantisation review} 

It is useful, at this stage, to recall the quantisation of a massive spin-$\tfrac{3}{2}$ field in Minkowski space. Such an analysis was performed in \cite{PhysRevD.80.104027}. {Taking  the line element of Minkowski spacetime to be
\begin{equation}
ds^{2}=-dt^{2}+d\bold{X}^2~,
\end{equation}
the equal-time anti-commutators determining the operator algebra for the spatial components of the massive spin-$\tfrac{3}{2}$ field are found in \cite{PhysRevD.80.104027} to be given by
\begin{align} \label{full anticom}
    \{\hat{\Psi}_j(t,\bold{X}), \hat{\Psi}_l(t,\bold{X}')^\dagger\} & = \{\hat{\Psi}_j(t,\bold{X}), \hat{\Psi}_l(t,\bold{X}')^\dagger\}_{\text{long }} + \{\hat{\Psi}_j(t,\bold{X}), \hat{\Psi}_l(t,\bold{X}')^\dagger\}_{\text{TT }}  \nonumber \\ 
    & = \left[ \delta_{jl}\mathbb{I}_4 - \frac{1}{3}\gamma_j\gamma_l + \frac{2}{3m^2}\mathbb{I}_4 \,\partial_j\partial_l + \frac{1}{3m}\left( \gamma_j\partial_l + \gamma_l\partial_j \right) \right] \delta(\bold{X}-\bold{X'}) ~,
\end{align}
where `long' and `TT' stand for `longitudinal' and `transverse-traceless' respectively, and the latter indicates that both the spatial $\gamma$-trace and spatial divergence of the `TT part' vanish. The longitudinal contribution stems from the sum of helicity-$\frac{1}{2}$ modes. We can further decompose it in three terms
\begin{equation} \label{long anticom}
    \{\hat{\Psi}_j(t,\bold{X}), \hat{\Psi}_l(t,\bold{X'})^\dagger\}_{\text{long }}  = \overset{3}{\underset{i=1}{\sum}} \, \{\hat{\Psi}_j(t,\bold{X}), \hat{\Psi}_l(t,\bold{X}')^\dagger\}^{(i)}_{\text{long}} ~,
\end{equation}
with
\begin{align}
   \{\hat{\Psi}_j(t,\bold{X}), \hat{\Psi}_l(t,\bold{X}')^\dagger\}^{(1)}_{\text{long}} = \frac{1}{3m} \left( \gamma_j\partial_l + \gamma_l\partial_j \right) \delta(\bold{X}-\bold{X'}) ~,
   \label{long1}
\end{align}
\begin{align}
  \{\hat{\Psi}_j(t,\bold{X}), \hat{\Psi}_l(t,\bold{X}')^\dagger\}^{(2)}_{\text{long}} = \frac{2}{3m^2} \mathbb{I}_4 \, \partial_j \partial_l \delta(\bold{X}-\bold{X'}) ~,
\end{align}
and
\begin{align} \label{long anticom 3}
  \{\hat{\Psi}_j(t,\bold{X}), \hat{\Psi}_l(t,\bold{X}')^\dagger\}^{(3)}_{\text{long}} = -\frac{3}{2} \left(\mathbb{I}_4\,\partial_j - \frac{1}{3}\gamma_j\gamma^{n}{\partial}_{n} \right) \frac{\delta(\bold{X}-\bold{X'})}{{\partial}^{n}\partial_{n}} \left(\overset{\leftarrow}{\partial}_{l} \mathbb{I}_4 - \frac{1}{3}\overset{\leftarrow}{{\partial}}_{n} \gamma^{n} \gamma_l \right) ,
\end{align}
where $j,l,n$ are spatial indices. The longitudinal anti-commutator (\ref{long anticom}) has been split into three terms for later convenience; in particular, (\ref{long anticom 3}) encodes the spatially $\gamma$-traceless contribution of helicity-$\tfrac{1}{2}$ modes to (\ref{full anticom}). The TT part is instead given by the sum of helicity-$\frac{3}{2}$ modes and reads
\begin{multline}
    \{\hat{\Psi}_j(t,\bold{X}), \hat{\Psi}_l(t,\bold{X'})^\dagger\}_{\text{TT}}  = \left(\delta_{jl} \mathbb{I}_4- \frac{1}{3}\gamma_j\gamma_l \right) \delta(\bold{X}-\bold{X'}) \, + \\
    \frac{3}{2} \left(\mathbb{I}_4\,\partial_j - \frac{1}{3}\gamma_j\gamma^{n}{\partial}_{n} \right) \frac{\delta(\bold{X}-\bold{X'})}{{\partial}^{n} \partial_{n}} \left(\overset{\leftarrow}{\partial}_{l} \mathbb{I}_4 - \frac{1}{3}\overset{\leftarrow}{{\partial}}_{n} \gamma^{n} \gamma_l \right) .
\end{multline}
The equal-time anti-commutators involving the time component of the Rarita-Schwinger field, $\hat{\Psi}_{t}(t, \bold{X}) $, can be found from the traceless condition by contracting the  anti-commutation relations of the spatial components with $\gamma^{j}$ and using $\hat{\Psi}_{t}(t, \bold{X}) = \gamma^{0}\gamma^{j} \hat{\Psi}_{j}(t, \bold{X})$.

\subsection{de Sitter quantisation} 

We now proceed to quantise the Dirac massive Rarita-Schwinger field in de Sitter space. We define the quantum field operator as
\begin{equation} \label{eq: quantum field op}
    \hat{\Psi}_j(\eta,\bold{x}) =  \underset{s \in \{\pm\frac{1}{2}, \, \pm \frac{3}{2} \} }{\sum} \int d^3 \bold{k}  \left( \hat{a}^s_{\bold{k}} \,\, \Psi^{(s)}_{j}   + \hat{b}^{s \, \dagger}_{\bold{k}} \, \Psi^{(s) \, C}_{j}  \right) ~,
\end{equation}
where $\Psi_j^C = \gamma^{\underline{2}}\Psi_j^*$.\footnote{The charge conjugate spinor  $\Psi^{C} = B^{-1} \Psi^{*}$ is defined with the matrix $B$ satisfying  $B \gamma^{\underline{a}} B^{-1} = +\left( \gamma^{\underline{a}} \right)^{*}$. This is the appropriate choice for real mass $m$. See appendix \ref{conventions}.}  We have decomposed the field operator (\ref{eq: quantum field op}) in terms of creation and annihilation operators $\hat{a}^s_{\bold{k}}$ and $\hat{b}^s_{\bold{k}}$ satisfying the standard operator algebra
\begin{equation} \label{eq: op algebra}
    \{\hat{a}^s_{\bold{k}}, \hat{a}^{s' \, \dagger}_{\bold{k'}}\} = \delta^{s s'} \delta(\bold{k}-\bold{k'}) = \{\hat{b}^{s \, \dagger}_{\bold{k}}, \hat{b}^{s'}_{\bold{k'}}\}~.
\end{equation}
The above operator algebra can be viewed as following from the requirement that the quantum field operator tends to the Minkowksian one in the limit $k \eta \to -\infty$. The Bunch-Davies state, denoted by $|0\rangle$, is annihilated by all the $\hat{a}^{s}_{\bold{k}}$ and $\hat{b}^{s}_{\bold{k}}$. The equal-time anti-commutators are given by
\begin{multline}
    \{\hat{\Psi}_j(\eta,\bold{x}), \hat{\Psi}_l(\eta,\bold{y})^\dagger\}  = \eta^2 \Biggl[ -\frac{1}{\eta} \{\hat{\Psi}_{\underline{j}}(\eta,\bold{x}), \hat{\Psi}_{\underline{l}}(\eta,\bold{y})^\dagger\}_{\text{TT }} - \eta \, \{\hat{\Psi}_{\underline{j}}(\eta,\bold{x}), \hat{\Psi}_{\underline{l}}(\eta,\bold{y})^\dagger\}^{(2)}_{\text{long}}
    \\
    - \frac{1}{\eta} \, \{\hat{\Psi}_{\underline{j}}(\eta,\bold{x}), \hat{\Psi}_{\underline{l}}(\eta,\bold{y})^\dagger\}^{(3)}_{\text{long}} 
    + \frac{m\mathbb{I}_4 - \gamma^{\underline{0}}}{3(m^2+1)} \left( \gamma_{\underline{j}}\partial_{\underline{l}} + \gamma_{\underline{l}}\partial_{\underline{j}} \right) \delta(\bold{x}-\bold{y}) \Biggr]~,
\end{multline}
where we observe that $m^2 \rightarrow m^2+1$ for the longitudinal bracket $\tiny{(2)}$ as compared to the Minkowski case \eqref{long1}. Indeed, $m=\pm i$ is a special value for which the Rarita-Schwinger equation acquires an additional gauge invariance, as we will analyse in section \ref{section 5}.
\newline\newline
We fixed the operator algebra (\ref{eq: op algebra}) by requiring compatibility with the flat space limit whereby $k\eta\to -\infty$. For the helicity-$\tfrac{1}{2}$ sector, this statement requires extra care. Taking into account the massive Dirac equations in planar coordinates, (\ref{eq: Dirac op on psi_eta vec-spinor}) and (\ref{eq: Dirac op on psi_j vec-spinor}), in the $k\eta \to -\infty$ limit one is left with the equations satisfied by massless Minkowskian modes. Thus, in the $k\eta \to -\infty$ limit, lower helicity massive modes of the field in de Sitter space reduce to the lower helicity massless modes in Minkowski space. But these are modes of a Minkowskian  Rarita-Schwinger gauge field. Similarly to (\ref{eq: inner product of 1/2 modes un-normalised}), the lower helicity modes of the massless Rarita-Schwinger field in Minkwoski space have vanishing norm. Indeed, in the $k\eta \to -\infty$ regime, with $k$ fixed, the norm of helicity-$\frac{1}{2}$ massive modes is only non-vanishing due to a sub-leading contribution of order $(-\eta)^{\frac{1}{2}}$. For the Rarita-Schwinger field in de Sitter space, the corresponding norm vanishes at the special value  $m=\pm i$ for which the theory acquires a gauge redundancy.

\subsection{Late time two-point function}

We have now assembled the relevant pieces to compute the equal-time two-point function of $\hat{\Psi}_j(\eta,\bold{x})$ in the Bunch-Davies vacuum, $|0\rangle$, namely
\begin{equation}
    G_{jl}(\eta;\bold{x},\bold{y}) \equiv \langle0| \hat{\Psi}_j(\eta, \bold{x}) \hat{\Psi}_l(\eta, \bold{y})^\dagger |0\rangle~.
\end{equation}
The contribution to $G_{jl}$ arising from helicity-$\frac{3}{2}$ modes gives
\begin{multline}
    G^{(\pm \frac{3}{2})}_{jl}(\eta;\bold{x},\bold{y}) = \eta^2 \underset{{s} \in \{\pm \frac{3}{2}\}}{\sum} \int \frac{d^3\bold{k}}{(2\pi)^3}  \frac{\pi k}{4} \, e^{i\bold{k} \cdot (\bold{x} - \bold{y})} \,  \times \\
      \begin{pmatrix}
         e^{-m \pi} H^{(1)}_{\frac{1}{2}   +i m }(-k \eta) \, H^{(2)}_{\frac{1}{2}   -i m }(-k \eta) \,\chi^{(s)}_j \, \chi^{{(s)} \, \dagger}_l & \tilde{s} H^{(1)}_{\frac{1}{2}   +i m }(-k \eta) \, H^{(2)}_{\frac{1}{2}   +i m }(-k \eta) \, \chi^{({s})}_j \, \chi^{{({s})} \, \dagger}_l \\\\
         \tilde{s} H^{(1)}_{\frac{1}{2}   -i m }(-k \eta) \, H^{(2)}_{\frac{1}{2}   -i m }(-k \eta) \, \chi^{({s})}_j \, \chi^{{({s})}\, \dagger}_l & e^{m \pi} H^{(1)}_{\frac{1}{2}   -i m }(-k \eta) \, H^{(2)}_{\frac{1}{2}   +i m }(-k \eta) \, \chi^{({s})}_j \, \chi^{{({s})} \, \dagger}_l
    \end{pmatrix}~.
\end{multline}
Using the completeness relation (\ref{eq: completeness relation again}), in the late time limit $k\eta\rightarrow0^-$ we find
\begin{multline}
   \lim_{k\eta\to 0^-} G^{(\pm \frac{3}{2})}_{jl}(\eta;\bold{x},\bold{y})  \approx  -\frac{\eta}{2} \int \frac{d^3 \bold{k}}{(2\pi)^3} \, e^{i\bold{k} \cdot (\bold{x} - \bold{y})} \,     \\~~~
      \times\begin{pmatrix}
         e^{-m \pi} \, \text{sech}m\pi \, \Pi_{jl}(\bold{k}) &  \frac{1}{\pi} \left(-\frac{2}{k\eta} \right)^{2im}  \Gamma(\frac{1}{2}+im)^2 \, \frac{\bold{k} \cdot \boldsymbol{\sigma}}{k} \Pi_{jl}(\bold{k}) \\
         \frac{1}{\pi} \left(-\frac{2}{k\eta} \right)^{-2im} \Gamma(\frac{1}{2}-im)^2 \, \frac{\bold{k} \cdot \boldsymbol{\sigma}}{k} \Pi_{jl}(\bold{k}) & e^{m \pi} \, \text{sech}m\pi \, \Pi_{jl}(\bold{k})
    \end{pmatrix}.
\end{multline}
A similar treatment can be performed for the two-point function  $G^{(\pm \frac{1}{2})}_{jl}(\eta;\bold{x},\bold{y})$ of the helicity-$\tfrac{1}{2}$ sector. Combining the helicity-$\tfrac{1}{2}$ and  helicity-$\tfrac{3}{2}$  contributions, one finds that the off-diagonal components of the late time momentum space two-point function read
\begin{equation}\label{G12}
   \lim_{k\eta\to 0^-} {G}^{(\text{tot, off-diag})}_{jl}(\eta,\bold{k}) \propto k^{\mp2im}\Bigg(\frac{\boldsymbol{\sigma} \cdot \bold{k}}{k}\Pi_{jl}(\bold{k})+   \frac{1\pm im}{6(1\mp im)} P_{j}(\bold{k}) ~\frac{\boldsymbol{\sigma} \cdot \bold{k}}{k}~P_{l}(\bold{k}) \Bigg)~,
\end{equation}
up to an overall normalisation factor. Here $P_j(\bold{k}) \equiv \sigma_j - 3 \tfrac{k_j}{k} \tfrac{\boldsymbol{\sigma}\cdot \bold{k}}{k}$ is the projector onto the $\sigma$-traceless sector.
\newline\newline
As expected, the two-point function is invariant under spatial translations, and acquires a scaling behaviour under $\eta\to\lambda \eta$. The late time two-point function (\ref{G12}) takes the form of a two-point function of a vector-spinor primary field in a three-dimensional conformal field theory of scaling dimension $\Delta = \tfrac{3}{2}\pm i m$, projected onto the $\sigma$-traceless sector (see \cite{Sotkov:1976xe,Osborn:1993cr} for a general treatment, and equation (21) of \cite{Volovich:1998tj} for the momentum space expression). This is a reflection of the fact that the $SO(4,1)$ de Sitter isometries act as conformal Killing vectors of $\mathbb{R}^3$ at late times. 

\subsection{General massive half-integer spin, briefly}

We can generalise the above discussion to the case of general half-integer spin. The spin-$s$ fermionic fields, with $s=\frac{3}{2},\tfrac{5}{2},\ldots$, will now carry a more general tensorial structure, $\Psi_{\mu_1\ldots\mu_s\alpha}$, and satisfy the field equation
\begin{equation}
\left(\gamma^\mu \nabla_\mu +m\right) \Psi_{\boldsymbol{\mu}} = 0~,
\end{equation}
where ${\boldsymbol{\mu}}\equiv \mu_1\mu_2\ldots\mu_{s-\tfrac{1}{2}}$, along with the generalised transverse and traceless conditions
\begin{equation}
\nabla^{\mu_i} \Psi_{\boldsymbol{\mu}} = 0 = \gamma^{\mu_i} \Psi_{\boldsymbol{\mu}}~, \quad\quad i =1,2,\ldots,s-\tfrac{1}{2}~.
\end{equation}
We have again supressed the spinor index. Altogether, these equations yield $(2s+1)$ on-shell degrees of freedom. One can decompose the equations into an increasing tower of helicities $\{\pm \tfrac{1}{2},\ldots, \pm s \}$. The spinor $\Psi_{\boldsymbol{\eta}}^{(\pm r)}$ with $r \in \{\pm \tfrac{1}{2},\ldots, \pm (s-1) \}$ and $\mu_1 = \mu_2 = \ldots = \mu_{s-\tfrac{1}{2}} = \eta$ will obey the Dirac equation for spin-$\tfrac{1}{2}$ fields, which has the same form as (\ref{eq: pos freq spin-3/2 type-I psi_eta ; helicities +-}). $\Psi_{\boldsymbol{\mu}}^{(\pm s)}$ vanishes if at least one the $\mu_i$ indices is equal to $\eta$. We expect each tower of helicity states to be governed by equations similar to the case of spin-$\tfrac{3}{2}$. Taking an ansatz for the highest helicity states $\Psi_{\boldsymbol{j}}^{(\pm s)}$ with $\mu_1, \mu_2, \ldots, \mu_{s-\tfrac{1}{2}} \in \{ i,j,k \}$ to have the same structure as that of the type II case (\ref{Ansatz: pos freq spin-3/2 type-II psi_j ; helicities +-}), we anticipate the slightly modified equations (see also equation (14) of \cite{Deser:2001xr})
\begin{equation} 
    \begin{cases}
         & \eta \partial_\eta g^{(\uparrow)}_{k}(\eta) + \left( s-2 + im \right) g^{(\uparrow)}_{k}(\eta) + i\eta k g^{(\downarrow)}_{k}(\eta) = 0~, \\ \\
         & \eta \partial_\eta g^{(\downarrow)}_{k}(\eta) - \left( 2-s + im \right) g^{(\downarrow)}_{k}(\eta) + i\eta k g^{(\uparrow)}_{k}(\eta) = 0 ~.
    \end{cases}
\end{equation}
The corresponding late time behaviour would then be $\Psi^{(\pm s)}_{\boldsymbol{j}} \approx (-\eta)^{2-s} \begin{pmatrix}
    (-k\eta)^{-im} \\ (-k\eta)^{im}
\end{pmatrix}e^{i\bf k\cdot x}$.

\section{Rarita-Schwinger gauge field} \label{section 5}

In this section we consider the spin-$\tfrac{3}{2}$ Rarita-Schwinger gauge field in dS$_4$ (see  \cite{Letsios:2023qzq,Letsios:2022tsq, Higuchi-Letsios, Letsios:2023awz} for a recent analysis, and \cite{Fang:1979hq,vasiliev1988free,metsaev1998fermionic,Deser:2001us,Deser:2001xr,Deser:2003gw,Zinoviev:2007ig,Metsaev:2006zy} for previous analyses). The motivation stems from the existence of unitary irreducible representations in the discrete series with half-integer spin. For the case of integer spin,  the discrete series UIR is furnished by the single particle Hilbert space of a partially massless gauge field. The simplest case is that of a Maxwell field in four-dimensional de Sitter space, as was recently discussed in \cite{RiosFukelman:2023mgq}. Similarly, the discrete series UIR $\mathcal{D}^\pm_{2,1}$ describes the single particle graviton states about a de Sitter background \cite{Higuchi:1991tn}, while the single particle content of a spin-$s$ Fronsdal field \cite{Fronsdal:1978rb} is encoded in $\mathcal{D}^\pm_{s,s-1}$. We might then expect that single-particle Hilbert spaces for partially massless gauge fields of half-integer spin are associated to the corresponding discrete series UIRs. We will generally set $\ell=1$ unless otherwise specified.

\subsection{Field equations}

We begin by considering the spin-$\tfrac{3}{2}$ Rarita-Schwinger field equations for the special case of $m\ell=i$. The field equations read
\begin{align} \label{eq: massles RS field eq}
   \gamma^{\mu \rho \sigma} \left(  \nabla_{\rho}+\frac{i}{2}\gamma_{\rho} \right) \Psi_{\sigma}=0 ~,
\end{align}
The reason for selecting such a peculiar value for the mass parameter is that the above equation acquires a gauge redundancy \cite{Deser:2001xr}. More specifically, (\ref{eq: massles RS field eq}) is invariant under the gauge transformation
\begin{equation}
    \Psi_{\mu\alpha} \rightarrow \Psi_{\mu\alpha} + \nabla_\mu \varepsilon_\alpha + \frac{i}{2} (\gamma_{\mu})_\alpha{}^\beta \, \varepsilon_\beta ~,
\end{equation}
where $\varepsilon$ is a Dirac spinor-valued gauge parameter and we have reinstated the spinor index. The gauge-invariant field strength is given by
\begin{equation}\label{Fgrav}
\mathcal{F}_{\mu\nu} = \nabla_{[\mu} \Psi_{\nu]} + \frac{i}{2}  \gamma_{[\mu} \Psi_{\nu]}~.
\end{equation}
A straightforward calculation further shows that $\mathcal{F}_{\mu\nu}$ satisfies a massless Dirac-type equation
\begin{equation}\label{F32}
\gamma^{\rho}\nabla_{\rho}\mathcal{F}_{\mu\nu}=0~.
\end{equation}
For the special value $m\ell=i$, the constraints for the massive theory (\ref{eq: TT conditions}) do not hold automatically. We can however use the gauge redundancy to impose certain gauge conditions on $\Psi_\mu$. Selecting, for instance, the $\gamma$-traceless gauge choice $\gamma^\mu \Psi_\mu = 0$, the field equations become
\begin{equation}\label{RSmzero}
 \left(   \gamma^\nu {\nabla}_\nu    + i \right) \Psi_{\mu} = 0~.
\end{equation}
The $\gamma$-traceless gauge condition can always be locally enforced with a suitable choice of $\varepsilon$. We now unpack the temporal and spatial components of the field equation (\ref{RSmzero}), yielding
\begin{eqnarray} \label{eq: Dirac op on psi_eta vec-spinor imaginary m}
    -\Bigg( \eta \left( \gamma^{\underline{0}}\partial_{\eta} + \gamma^{\underline{1}}\partial_{1} + \gamma^{\underline{2}}\partial_{2} +\gamma^{\underline{3}}\partial_{3}  \right)- \frac{3}{2} \gamma^{\underline{0}}  \Bigg) \Psi_{\eta} - \gamma^{\underline{\mu}}   \Psi_{\mu}  &=&   - i   \Psi_{\eta} ~, \\
\label{eq: Dirac op on psi_j vec-spinor imaginary m}
    -\Bigg( \eta \left( \gamma^{\underline{0}}\partial_{\eta} + \gamma^{\underline{1}}\partial_{1} + \gamma^{\underline{2}}\partial_{2} +\gamma^{\underline{3}}\partial_{3}  \right)- \frac{3}{2} \gamma^{\underline{0}}  \Bigg) \Psi_{j} - \gamma^{\underline{0}}   \Psi_{j} - \gamma^{\underline{i}}  \delta_{ij}\Psi_{\eta}   &=&   - i   \Psi_{j} ~.
\end{eqnarray}
The equations admit helicity-$\tfrac{1}{2}$ as well as helicity-$\tfrac{3}{2}$  solutions. Of these, the helicity-$\tfrac{3}{2}$ carry two physically propagating degrees of freedom. 
\newline\newline
Similarly to the Minkowsian case, the helicity-$\tfrac{1}{2}$ modes become pure gauge (see however \cite{Valenzuela:2022gbk,Valenzuela:2023aoa} for a different perspective). We do not have an ordinary action formalism, so we guide our discussion from the group theoretic perspective, which prompts us to consider the helicity-$\tfrac{3}{2}$ sector alone.

\subsection{Helicity$-\frac{3}{2}$ and helicity$-\frac{1}{2}$ solutions} 

Inspecting the equations, and working in the $\gamma^\mu \Psi_\mu = 0$ gauge, we observe that $\Psi_\eta$ can be consistently set to zero. The equation governing the $\Psi_j$ component can then be solved straightforwardly. Solutions satisfying the Bunch-Davies condition are given by
\begin{align} 
\label{eq: pos freq spin-3/2 type-II psi_j, massless}
 \Psi^{(\pm \frac{3}{2})}_{j} =  \frac{c^{(\text{II})}_{k}}{(2\pi)^{\frac{3}{2}}} \,
    \begin{pmatrix}
        \mp i (-\eta)^{\frac{1}{2}} k^{-\frac{1}{2}} \, \chi^{(\pm\frac32)}_j ({\bold{k}}) \\ \\
        (1 + i k \eta)(-\eta)^{-\frac{1}{2}} k^{-\frac{3}{2}} \, \chi^{(\pm\frac32)}_j ({\bold{k}})
    \end{pmatrix} e^{- i  k \eta+i\bf k\cdot x}~,
\end{align}
where  $\chi^{(\pm\frac32)}_{{j}} (  {\bold{k}} ) $ are vector-spinors defined in \eqref{transverse-traceless vector-spinor}. We can compare the solution (\ref{eq: pos freq spin-3/2 type-II psi_j, massless}) to the analogous calculation in AdS$_4$ (see equation (2.30) of \cite{Corley:1998qg} with $\Lambda=2$), and the results are related by a simple analytic continuation. 
\newline\newline
In the early time limit $k\eta\to-\infty$, the  solution \eqref{eq: pos freq spin-3/2 type-II psi_j, massless} becomes
\begin{align} \label{eq: pos freq spin-3/2 type-II psi_j, massless, flat space limit}
    \Psi^{(\pm \frac{3}{2})}_{j} \approx -i \frac{c^{(\text{II})}_{k}}{(2\pi)^{\frac{3}{2}}}  \sqrt{\frac{(-\eta)}{k}}
    \begin{pmatrix}
        \pm \chi^{(\pm\frac32)}_j ({\bold{k}}) \\ \\
        \chi^{(\pm\frac32)}_j ({\bold{k}})
    \end{pmatrix} e^{-ik\eta+i\bf k\cdot x} ~,
\end{align}
confirming that $\Psi^{(\pm \frac{3}{2})}_{j}$ is indeed a positive frequency Bunch-Davies mode. Imposing the standard normalisation further yields $|c^{(\text{II})}_{k}|^2 = \tfrac{k}{2}$.
At late times, whereby $k\eta \to 0^-$, one instead has
\begin{align} \label{eq: pos freq spin-3/2 type-II psi_j, massless, flat space limit}
    \Psi^{(\pm \frac{3}{2})}_{j} \approx \frac{c^{(\text{II})}_{k}}{(2\pi)^{\frac{3}{2}}}
    \begin{pmatrix}
        \pm i (-\eta)^{\tfrac{1}{2}} k^{-\frac{1}{2}} \chi^{(\pm\frac32)}_j ({\bold{k}}) \\ \\
      (-\eta)^{-\tfrac{1}{2}} k^{-\frac{3}{2}}  \chi^{(\pm\frac32)}_j ({\bold{k}})
    \end{pmatrix} e^{-ik\eta+i\bf k\cdot x}~.
\end{align}
Notice that the lower component grows at late times. This is characteristic of higher-spin fields in de Sitter space, including the linearised spin-two field. Nevertheless, the growing mode does not immediately imply an infrared breakdown, since the field itself is not a gauge-invariant object. 
\newline\newline
One can also write down the helicity-$\tfrac{1}{2}$ modes. These take the form of a gauge transformation and are given by
\begin{equation}
 \Psi^{(\pm \frac{1}{2})}_{\mu} = \left(\nabla_{\mu} + \frac{i}{2}\gamma_{\mu} \right) \, \xi(\eta,{\bold{x}}) ~.
\end{equation}
Imposing, further, the $\gamma$-traceless condition we find
\begin{equation}
    \xi(\eta,{\bold{x}}) = i \,\upsilon_k\, \eta^2 \begin{pmatrix}
    \mp   \, H^{(1)}_{-\frac{3}{2}}(-k \eta) \, \chi^{(\pm\frac12)}({\bold{k}}) \\ \\
    -   H^{(1)}_{\frac{5}{2}}(-k \eta) \, \chi^{(\pm\frac12)}({\bold{k}}) 
    \end{pmatrix} e^{i\bf k\cdot x} ~,
\end{equation}
where $v_{k}$ is a normalisation factor. Massless helicity-$\tfrac{1}{2}$ modes cannot be normalised using the same inner product as for the massive ones. The helicity-$\tfrac{1}{2}$ solutions do not contribute to the field strength (\ref{Fgrav}).
\newline\newline
From the structure of the solutions, it was argued in \cite{Deser:2001xr} that despite the imaginary `mass' the Rarita-Schwinger gauge field propagates causally. The essential reason is that  the large momentum propagation is dominated by a plane wave solution. More concretely, this can be seen from a retarded Green's function analysis of the Bessel equation governing the transverse-traceless sector, which indeed has support inside the light-cone. Relatedly, the field strength equation (\ref{F32}) yields causal propagation of $\mathcal{F}_{\mu\nu}$. This becomes clear upon observing that the classical field-strength mode solutions for the massless gravitino in dS$_{4}$ are related to the flat-space modes by a Weyl rescaling.\footnote{This is aligned with the observation of  \cite{Penrose}, that the equations governing the linearised field strengths of massless gauge fields in four dimensions are Weyl covariant.} In particular, the helicity-$\tfrac{3}{2}$ gravitino Fourier modes of positive frequency in Minkowski spacetime are 
\begin{align}  \label{flat-space gravitino modes}
 \Psi^{(\text{flat},\pm \frac{3}{2})}_{\underline{0}}=0,~~~~~~~~~    \Psi^{(\text{flat},\pm \frac{3}{2})}_{\underline{j}} =\frac{1}{\sqrt{2(2 \pi)^3}}
    \begin{pmatrix}
        \pm \chi^{(\pm\frac32)}_{\underline{j}} ({\bold{k}}) \\ \\
        \chi^{(\pm\frac32)}_{\underline{j}} ({\bold{k}})
    \end{pmatrix} e^{-ik\eta+i\bf k\cdot x} ~.
\end{align}
These solve the massless Rarita-Schwinger equation $\gamma^{\underline{\mu}} \partial_{\underline{\mu}} \Psi_{\underline{\nu}}^{(\text{flat})}=0$ on Minkowski spacetime with line element given by $ds^2=-d\eta^2 + d \bold{x}^2$. Then, the flat space field strength modes are simply given by $\mathcal{F}^{(\text{flat},\pm)}_{ \underline{\mu \nu}} \equiv \partial_{[\underline{\mu}}  \Psi_{\underline{\nu}]}^{(\text{flat},\pm \frac{3}{2})}$, such that
\begin{equation}
\mathcal{F}^{(\text{flat},\pm )}_{ \underline{0 j}} =  -\frac{ik}{2}  \Psi^{(\text{flat},\pm \frac{3}{2})}_{\underline{j}} ,  \quad \mathcal{F}^{(\text{flat},\pm )}_{ \underline{jl}} = {i}~k_{[\underline{j}}  \Psi_{\underline{l}]}^{(\text{flat},\pm \frac{3}{2})}.
\end{equation}
From the dS$_{4}$ helicity-3/2 modes (\ref{eq: pos freq spin-3/2 type-II psi_j, massless}), it is straightforward to compute the field strength modes as  $\mathcal{F}^{(\pm )}_{\mu \nu} = (\nabla_{[\mu}   + \frac{i}{2}  \gamma_{[\mu}) \Psi^{(\pm \frac{3}{2})}_{\nu]}$ (see \cite{Higuchi-Letsios} for expressions in global coordinates). Working in the orthogonal basis, whereby $\mathcal{F}^{(\pm \frac{3}{2})}_{\underline{\mu \nu}} = e^{\mu}\,_{\underline{\mu}}   e^{\nu}\,_{\underline{\nu}} \mathcal{F}^{(\pm \frac{3}{2})}_{\mu \nu}$, to make the comparison with flat space clearer, we find
  \begin{align}
    \mathcal{F}^{(\pm )}_{\underline{\mu \nu}} = -i  \,  {(-\eta)^{5/2}} ~~\mathcal{F}^{(\text{flat},\pm )}_{ \underline{\mu \nu}}.
  \end{align}
Thus,  the gravitino field strength modes on dS$_{4}$ are mapped to the flat space field strength modes by an overall rescaling.

\subsection{Physical mode quantisation}

The inclusion of an imaginary mass parameter, which ensures a gauge-invariance, comes at a price: the differential operator acting on the field $\Psi_\mu$ is not Hermitian. Relatedly, the equations of motion with imaginary mass parameter do not stem from an action with ordinary reality or positivity properties, or at least we know of no such action. For instance, modifying the ordinary Rarita-Schwinger gauge field action by a $\gamma^5$, can yield a real action whose quantisation gives an indefinite inner product \cite{Higuchi-Letsios}. Alternatively, quantising with a non-real action can yield a time-dependent inner product due to a non-Hermitian Hamiltonian. Generally speaking, a quantum field theory (even if free) admitting no action formalism is not in itself an obstruction.\footnote{The self-dual five-form field strength of ten-dimensional type IIB supergravity, which stems as the low energy limit of the IIB superstring, is often used as an example of a theory with no covariant and Lorentz invariant action formulation. Nonetheless, \cite{Sen:2015nph,Sen:2019qit} presented a recent construction with auxiliary fields. A similar situation can hold for other self-dual fields also, including two-dimensional chiral bosons \cite{Hull:2025rxy}. The Vasiliev equations in anti-de Sitter space \cite{Vasiliev:1990en,Vasiliev:2003ev} are classical field equations that do not admit any known complete local spacetime action formalism in terms of the Fronsdal fields (see, however, \cite{Boulanger:2011dd,Boulanger:2015kfa} for progress in this direction). Upon invoking the AdS/CFT correspondence, it is believed that they can be consistently quantised.} It does however make the quantisation procedure somewhat unclear.
\newline\newline
On the other hand, the quantisation of the Rarita-Schwinger gauge field on a Minkowksi background has been undertaken in \cite{PhysRevD.16.307}. The system has second class constraints, and a Dirac bracket analysis is therefore necessary. The result of \cite{PhysRevD.16.307} is that the quantum commutation relations of $\hat{\Psi}_j$ and $\hat{\Pi}_l$ in Minkowski space are given by
\begin{align}\label{Mimkowskian cc relations coulomb gauge}
    \{\hat{\Psi}_j(t,\bold{X}),\hat{\Pi}_l(t,\bold{X}') \} & = \left(\delta_{jl}\, \mathbb{I}_4- \frac{1}{3}\gamma_j\gamma_l \right) \delta(\bold{X}-\bold{X}') \, +\nonumber\\
    &\frac{3}{2} \left(\mathbb{I}_4\,\partial_j - \frac{1}{3}\gamma_j\gamma^{m}{\partial}_{m} \right) \frac{\delta(\bold{X}-\bold{X}')}{{\partial}_{m}\partial^{m}} \left(\overset{\leftarrow}{\partial}_{l} \mathbb{I}_4 - \frac{1}{3}\overset{\leftarrow}{{\partial}}_{m} \gamma^{m} \gamma_l \right),
\end{align}
where $\mathbb{I}_4$ is the $4 \times 4$ identity matrix in spinor space and $j,l,m \in \{1,2,3 \}$ denote the standard spatial coordinates on $\mathbb{R}^3$. 
\newline\newline
As already mentioned, motivated by the existence of a candidate UIR, namely $\mathcal{D}^\pm_{\frac{3}{2},-\frac{1}{2}}$, we are prompted to consider the quantisation of the Rarita-Schwinger gauge field satisfying the  field equation (\ref{RSmzero}). The challenge is to consistently quantise both UIRs $\mathcal{D}^\pm_{\frac{3}{2},-\frac{1}{2}}$ in a unified Hilbert space picture. The issue is that though independently $SO(4,1)$ invariant, a priori, the UIRs $\mathcal{D}^\pm_{\frac{3}{2},-\frac{1}{2}}$ can have a relative sign in their respective inner products \cite{Letsios:2023qzq, Higuchi-Letsios, Letsios:2023awz}. This, in turn, would  render the combined Hilbert space (which carries both helicities) non-positive. For integer spin, that one can indeed combine the two discrete series UIRs of $SO(4,1)$ into the single particle Hilbert space of a massless higher spin gauge field \cite{Higuchi:1991tn,RiosFukelman:2023mgq}.
\newline\newline
To proceed, we will consider the problem in the gauge $\hat{\Psi}_\eta(\eta,\bold{x})=0$, as this makes the physical degrees of freedom particularly manifest. Let us assume the existence of a quantum field operator $\hat{\Psi}_{j}(\eta,\bold{x})$ built from appropriate creation and annihilation operators and mode functions obeying the Bunch-Davies condition. We will proceed to deduce the structure of the conjugate momentum operator $\hat{\Pi}_l(\eta,\bold{y})$ satisfying the canonical algebra
\begin{equation} \label{eq: initial anticommutator dS}
    \{\hat{\Psi}_j(\eta,\bold{x}), \, \sqrt{-g} ~e^{\eta}_{~\underline{0}}~ \hat{\Pi}_l(\eta,\bold{y}) \} 
    = {i }  \,  \, \bm{\delta}^{\text{TT}}_{\underline{j} \underline{l}}(\bold{x}-\bold{y}) \, e^{~\underline{j}}_{j} \, e^{~\underline{l}}_{l} 
     = \frac{i }{\eta^{2}}  \,
   \bm{\delta}^{\text{TT}}_{{j} {l}}(\bold{x}-\bold{y}) ~.
\end{equation}
The $\eta^2$ factor in the denominator can be understood as coming from the metric, as in (\ref{PB}). Equivalently 
\begin{equation} \label{eq: final anticommutator dS}
    \{\hat{\Psi}_j(\eta,\bold{x}), \, \hat{\Pi}_l(\eta,\bold{y})\} =  - i\eta
    \, \bm\delta^{\text{TT}}_{{j} {l}}(\bold{x}-\bold{y}) ~.
\end{equation}
Here $\bm\delta^{\text{TT}}_{{j} {l}}(\bold{x}-\bold{y})$ is a transverse, $\gamma$-traceless   delta function  which takes the form
$$  \bm\delta^{\text{TT}}_{ {j}  {l}}(\bold{x}-\bold{y})
\equiv
\begin{pmatrix}
\delta^{\text{TT}}_{ {j}  {l}}(\bold{x}-\bold{y}) & &  \mathbf{0} \\
    \mathbf{0} &  & \delta^{\text{TT}}_{ {j}  {l}}(\bold{x}-\bold{y})
\end{pmatrix},
$$ with $\delta^{\text{TT}}_{{j} {l}}(\bold{x}-\bold{y})
$ a transverse,  $\sigma$-traceless  delta function  given by  
$$\delta^{\text{TT}}_{jl}(\bold{x}-\bold{y}) \equiv \left(\delta_{jl} \mathbb{I}_2- \frac{1}{3}\sigma_j\sigma_l \right) \delta(\bold{x}-\bold{y}) + \frac{3}{2} \left(\mathbb{I}_2\,\partial_j - \frac{1}{3}\sigma_j~\sigma_{m}{\partial}_{m} \right) \frac{\delta(\bold{x}-\bold{y})}{\partial^2} \left(\overset{\leftarrow}{\partial}_{l} \mathbb{I}_2 - \frac{1}{3}\overset{\leftarrow}{{\partial}}_{m} \sigma_{m} \sigma_l \right)~.$$
Explicitly, it satisfies $\sigma^{j}\delta^{\text{TT}}_{jl}(\bold{x}-\bold{y})  = \partial^{j}\delta^{\text{TT}}_{jl}(\bold{x}-\bold{y})=0$. This delta function has a $2 \times 2$ spinorial matrix structure with two tensor indices. It can be defined via the completeness relation of   vector-spinors \eqref{transverse-traceless vector-spinor}  as
\begin{equation} \label{eq: super explicit completeness relation dS}
   \underset{{s} \in \{\pm \frac32 \} }{\sum}  \int \frac{d^3 \bold{k}}{(2\pi)^3} \, e^{i\bold{k}\cdot(\bold{x}-\bold{y})} \, \chi_j^{({s})}(\bold{k}) \,  \chi_l^{( {s})}(\bold{k})^\dagger = \delta^{\text{TT}}_{jl}(\bold{x}-\bold{y}) ~.
\end{equation}
Our task is now to determine the operators $\hat{\Psi}_j(\eta,\bold{x})$ and $\hat{\Pi}_l(\eta,\bold{y})$ such that the canonical anti-commutation relation is satisfied.
We express the field operator as
\begin{equation} \label{eq: field operator dS}
    \hat{\Psi}_j(\eta,\bold{x}) = \underset{s \in \{ \pm \frac{3}{2} \} }{\sum} \int d^3 \bold{k} \,  \left( \hat{a}^s_{\bold{k}} \Psi^{(s)}_{j}   + \hat{b}^{s \, \dagger}_{\bold{k}} \Psi^{(s) \, C}_{j}  \right)~,
\end{equation}
where the charge conjugate modes are defined as $\Psi_j^C = \gamma^{\underline{2}} \gamma^5 \Psi_j^*$. We note that the charge conjugation matrix $\tilde{B}$ now has to satisfy $\tilde{B} \gamma^{\underline{a}} \tilde{B}^{-1} = - \left( \gamma^{\underline{a}} \right)^{*}$ because the `mass' is imaginary, and it is again defined up to a phase. The mode functions (\ref{eq: pos freq spin-3/2 type-II psi_j, massless}) are accompanied by creation and annihilation operators, which are postulated to satisfy the algebra
\begin{equation}
    \{\hat{a}^s_{\bold{k}}, \hat{a}^{s' \, \dagger}_{\bold{k'}}\} = \delta^{s s'} \delta(\bold{k}-\bold{k'}) = \{\hat{b}^{s \, \dagger}_{\bold{k}}, \hat{b}^{s'}_{\bold{k}'}\}~. 
\end{equation}
We note here that the quantum field operator $\hat{\Psi}_j$ reduces precisely to the corresponding one in Minkowski space upon taking the limit $k\eta\to-\infty$. As such, it reproduces quantum field theory in Minkowski space for small regions of spacetime. 
\newline\newline
In Minkowski space, one can consistently quantise the Rarita-Schwinger massless gauge field, and there is a conjugate momentum field operator accompanying $\hat{\Psi}_j$. We would thus like to explore, within the Coulomb gauge we are working in, whether such an operator exists in de Sitter space. To do so, we introduce the  field operator
\begin{equation} \label{eq: momentum operator dS}
    \hat{\Pi}_l(\eta,\bold{x}) \equiv  \underset{s \in \{\pm \frac{3}{2}\}}{\sum} \int d^3 \bold{k}\, \left( \hat{a}^{s \, \dagger}_{\bold{k}}    \Phi^{(s) \, \dagger}_{l}   + \hat{b}^{s}_{\bold{k}} \, \tilde{\Phi}^{(s) \, C \, \dagger}_{l}  \right)~,
\end{equation}
with modes $\Phi^{(s) \, \dagger}_{l}$, $\tilde{\Phi}^{(s) \, \dagger}_{l}$ to be determined.
The field momentum anti-commutator is then equal to
\begin{multline} \label{eq: field momentum anticommutator dS}
    \{\hat{\Psi}_j(\eta,\bold{x}), \, \hat{\Pi}_l(\eta,\bold{y})\} =  \sum_{s,s' \in \{ \pm \frac{3}{2} \}} \int d^3 \bold{k} \int d^3 \bold{k}' \Bigl( \{\hat{a}^s_{\bold{k}}, \hat{a}^{s' \, \dagger}_{\bold{k}'}\} \, \Psi^{(s)}_{j} \, \Phi^{(s') \, \dagger}_{l} 
    \\  
     + \{\hat{b}^{s \, \dagger}_{\bold{k}}, \hat{b}^{s'}_{\bold{k'}}\} \, \Psi^{(s) \, C}_{j} \, \tilde{\Phi}^{(s') \, C \, \dagger}_{l} 
     \Bigr)~.
\end{multline}
If we further define
\begin{equation} \label{pos freq spin-3/2 type-II phi_j, massless}
   \Phi^{(\pm \frac{3}{2})}_{l} \equiv \mp i \, \gamma^5 \, \Psi^{(\pm \frac{3}{2})}_{l} = \frac{c^{(\text{II})}_{k}}{(2\pi)^{\frac{3}{2}}} \sqrt{\frac{\pi}{2}} \, \eta
    \begin{pmatrix}
        \mp i \, H_{-\frac{3}{2}}^{(1)}(-k\eta) \,\chi^{(\pm\frac32)}_{j}({\bold{k}}) \\ \\
        H_{-\frac{1}{2}}^{(1)}(-k\eta) \,\chi^{(\pm\frac32)}_{j}({\bold{k}})
    \end{pmatrix}e^{i\bf k\cdot x} ~,
    \end{equation}
    and
\begin{align}
    \tilde{\Phi}^{(\pm \frac{3}{2})C}_{l} \equiv -\Phi^{(\pm \frac{3}{2})C}_{l} ~,
\end{align}
we can use the identity (\ref{eq: super explicit completeness relation dS}) to show that (\ref{eq: final anticommutator dS}) is satisfied. The outcome is that we can write the momentum operator as
\begin{equation} \label{eq: momentum operator dS 2.0}
    \hat{\Pi}_l(\eta,\bold{x}) = i \underset{s \in \{ \pm \frac{3}{2}\}}{\sum} \int d^3 \bold{k} \, \left( \hat{a}^{s \, \dagger}_{\bold{k}} \,\, \Psi^{(s) \, \dagger}_{l} 
    - \hat{b}^{s}_{\bold{k}} \, \Psi^{(s) \, C\,\dagger}_{l}  
    \right)  \tilde{s} \, \gamma^{5}~.
\end{equation}
We thus observe that at the level of the free gauge-fixed operators, there exists a candidate operator $\hat{\Pi}_l$. In the early time limit $k\eta\to-\infty$, $\hat{\Pi}_l$ reduces to the conjugate momentum of the flat space Rarita-Schwinger gauge field in the Coulomb gauge.
\newline\newline
As a final remark, we note that the anti-commutator of the gauge invariant field strength (\ref{Fgrav}) will share the same anti-commutation relations as the Minkowskian field strength, up to overall factors of $\eta$. In particular, it vanishes outside the light cone.

\subsection{Conformal operator basis}\label{confbasis}

To make the conformal properties of the field more manifest at late times, it is instructive to further express the quantum field operator (\ref{eq: field operator dS}) in the following form
\begin{equation} \label{eq: new field operator dS}
    \hat{\Psi}_j(\eta,\bold{x}) = \frac{\sqrt{\pi}}{2} \eta \int \frac{d^3\bold{k}}{(2\pi)^{\frac{3}{2}}} \, e^{i \bold{k}\cdot\bold{x}} \sqrt{k} \left[ \hat{\alpha}_j(\bold{k}) \begin{pmatrix}
        J_{-\frac{1}{2}}(-k\eta)  \\
        J_{-\frac{3}{2}}(-k\eta) 
    \end{pmatrix} + \hat{\beta}_j(\bold{k}) \begin{pmatrix}
        Y_{-\frac{1}{2}}(-k\eta)  \\
        Y_{-\frac{3}{2}}(-k\eta) 
    \end{pmatrix} \right] ~,
\end{equation}
where we have defined the time-independent operators 
\begin{equation} \label{eq: operator alpha}
    \hat{\alpha}_j(\bold{k}) \equiv \underset{s \in \{\pm\frac{3}{2}\}}{\sum} \left[ \hat{a}^s_{\bold{k}} \begin{pmatrix}
        i \tilde{s} \chi^{( {s})}_j(\bold{k}) & 0 \\
        0 & \chi^{( {s})}_j(\bold{k})
    \end{pmatrix} + \hat{b}^{s \, \dagger}_{-\bold{k}} \begin{pmatrix}
        i\sigma_2 & 0 \\
        0 & -i\sigma_2
    \end{pmatrix} \begin{pmatrix}
        -i \tilde{s} \chi^{( {s})}_j(-\bold{k})^* & 0 \\
        0 & \chi^{( {s})}_j(-\bold{k})^*
    \end{pmatrix} \right] ~,
\end{equation}
as well as
\begin{equation} \label{eq: operator beta}
    \hat{\beta}_j(\bold{k}) \equiv i\underset{s\in \{ \pm\frac{3}{2}\}}{\sum} \left[ \hat{a}^s_{\bold{k}} \begin{pmatrix}
        i \tilde{s} \chi^{({s})}_j(\bold{k}) & 0 \\
        0 & \chi^{({s})}_j(\bold{k})
    \end{pmatrix} - \hat{b}^{s \, \dagger}_{-\bold{k}} \begin{pmatrix}
        i\sigma_2 & 0 \\
        0 & -i\sigma_2
    \end{pmatrix} \begin{pmatrix}
        -i \tilde{s} \chi^{( {s})}_j(-\bold{k})^* & 0 \\
        0 & \chi^{({s})}_j(-\bold{k})^*
    \end{pmatrix} \right] ~.
\end{equation}
In the late time limit $k \eta \rightarrow 0^-$, the field operator is approximately given by
\begin{equation} \label{eq: late time limit of new field operator dS}
    \hat{\Psi}_j(\eta,\bold{x}) \approx \frac{1}{\sqrt{2}} \int \frac{d^3\bold{k}}{(2\pi)^{\frac{3}{2}}} \, e^{i\bold{k}\cdot\bold{x}} \left[ \hat{\alpha}_j(\bold{k}) \begin{pmatrix}
       \mathcal{O}(\eta^{\frac{1}{2}}) \\
        \displaystyle -\frac{1}{k} (-\eta)^{-\frac{1}{2}} 
    \end{pmatrix} + \hat{\beta}_j(\bold{k}) \begin{pmatrix}
        k(-\eta)^{\frac{3}{2}}  \\
        \displaystyle \mathcal{O}(\eta^{\frac{5}{2}}) 
    \end{pmatrix} \right] ~.
\end{equation}
The non-vanishing anti-commutation relations of the new operators are given by
\begin{eqnarray} \label{eq: anticommutators of alpha and beta}
    \{ \hat{\alpha}_j(\bold{k}), \hat{\alpha}_l(\bold{k}')^\dagger \} &=& 2 M_{jl}  (\bold{k}, \bold{k}') = \{ \hat{\beta}_j(\bold{k}), \hat{\beta}_l(\bold{k}')^\dagger \} ~,
\end{eqnarray}
where 
$$M_{jl}(\bold{k},\bold{k}') \equiv \delta(\bold{k}-\bold{k}') \underset{ {s}\in\{\pm\frac32\}}{\sum} 
\begin{pmatrix}
    \chi_j^{({s})}(\bold{k}) \,\chi_l^{({s})}(\bold{k}')^\dagger & 0 \\
    0 & \chi_j^{({s})}(\bold{k}) \, \chi_l^{({s})}(\bold{k}')^\dagger
\end{pmatrix} ~. $$
Thus, the late time behaviour is split into two operators
\begin{equation}\label{fermioncp}
\hat{\boldsymbol{\alpha}}_j(\bold{x}) \equiv \int \frac{d^3\bold{k}}{(2\pi)^{\frac{3}{2}}} e^{i \bold{k}\cdot \bold{x}} \, \frac{\hat{\alpha}_j(\bold{k})}{k}  \begin{pmatrix}
        0 \\
        \displaystyle  1
    \end{pmatrix}~,  \quad\quad \hat{\boldsymbol{\beta}}_j(\bold{x}) \equiv \int \frac{d^3\bold{k}}{(2\pi)^{\frac{3}{2}}} {e^{i \bold{k}\cdot \bold{x}}} \, k \, \hat{\beta}_j(\bold{k}) \begin{pmatrix}
        1 \\
        \displaystyle 0
    \end{pmatrix}~.
\end{equation}
These transform as spin-$\tfrac{3}{2}$ conformal primaries of weight $\bar{\Delta}=\tfrac{1}{2}$ and ${\Delta}=\tfrac{5}{2}$ respectively, and have the content of a Dirac two-spinor. This is directly analogous to the structure of higher-spin gauge fields with integer spin (see, for instance, section 2 of \cite{Anninos:2017eib}). 
\newline\newline
We anticipate that for a totally symmetric tensor with half-integer spin $s=n+\tfrac{1}{2}$, the structure in (\ref{eq: new field operator dS}) will generalise to 
\begin{equation} \label{fhss}
    \hat{\Psi}_{\boldsymbol{j}}(\eta,\bold{x}) = \frac{\sqrt{\pi}}{2} \eta \int \frac{d^3\bold{k}}{(2\pi)^{\frac{3}{2}}} \, e^{i \bold{k}\cdot\bold{x}} \sqrt{k} \left[ \hat{\alpha}_{\boldsymbol{j}}(\bold{k}) \begin{pmatrix}
        J_{-n+\frac{1}{2}}(-k\eta) \\
        J_{-n-\frac{1}{2}}(-k\eta)
    \end{pmatrix} + \hat{\beta}_{\boldsymbol{j}}(\bold{k}) \begin{pmatrix}
        Y_{-n+\frac{1}{2}}(-k\eta) \\
        Y_{-n-\frac{1}{2}}(-k\eta)
    \end{pmatrix} \right] ~,
\end{equation}
where $\boldsymbol{j} \equiv \{j_1,\ldots,j_n\}$, and the spatial tensor structure is totally symmetric, transverse, and traceless. The late time behaviour is now split into two conformal operators $\hat{\boldsymbol{\beta}}_{\boldsymbol{j}}(\bold{k})$ and  $\hat{\boldsymbol{\alpha}}_{\boldsymbol{j}}(\bold{k})$ akin to (\ref{fermioncp}). These transform as the Fourier transform of spin-$s$ supercurrents and shadows thereof, of respective weights $\Delta=1+s$ and $\bar{\Delta}=2-s$.

\subsection{Late time two-point function}

We now compute the late time structure of the equal-time two-point function of $\hat{\Psi}_i$ in the Bunch-Davies vacuum
\begin{equation}
    G_{jl}(\eta;\bold{x},\bold{y})  \equiv  \langle0| \hat{\Psi}_j(\eta, \bold{x}) \hat{\Psi}_l(\eta, \bold{y})^\dagger |0\rangle~.
\end{equation}
By construction, $|0\rangle$ is the state annihilated by all $\hat{a}^s_{\bold{k}}$ and $\hat{b}^s_{\bold{k}}$. Such a two-point function is of interest when studying fluctuations in inflationary cosmology. Here we compute the type of pattern that would be produced by a Rarita-Schwinger gauge field during inflation.
\newline\newline
The explicit form of the two-point function is given by
\begin{equation} \label{eq: field 2pt function}
   G_{jl}(\eta;\bold{x},\bold{y}) = \frac{1}{2}  \,\underset{ {s}\in\{\pm\frac32\}}{\sum} \int \frac{d^3k}{(2\pi)^3} \, e^{i\bold{k} \cdot (\bold{x} - \bold{y})} 
   \begin{pmatrix}
        (-\eta)\, \chi^{({s})}_j \, \chi^{({s}) \, \dagger}_l & - i \tilde{s} \frac{1-ik\eta}{k} \chi^{({s})}_j \, \chi^{({s}) \, \dagger}_l \\\\
        i \tilde{s} \frac{1+ik\eta}{k} \chi^{( {s})}_j \, \chi^{({s}) \, \dagger}_l & \frac{1+k^2\eta^2}{(-\eta)k^2} \chi^{({s})}_j \, \chi^{({s}) \, \dagger}_l
    \end{pmatrix}~.
\end{equation}
In the late time limit $k\eta \rightarrow 0^-$, the above reduces to
\begin{equation}\label{eq: field 2pt function late time}
 G_{jl}(\eta;\bold{x},\bold{y})  \approx   \begin{pmatrix}
        0 & -i G^{\text{const}}_{jl}(\bold{x},\bold{y}) \\\\
        i G^{\text{const}}_{jl}(\bold{x},\bold{y}) & \frac{1}{(-\eta)} G^{\text{leading}}_{jl}(\bold{x},\bold{y})
    \end{pmatrix} ~.
\end{equation}
In terms of the projector $\Pi_{jl}(\bold{k})$ defined in (\ref{eq: completeness relation again}),
\begin{equation}
    G^{\text{leading}}_{jl}(\bold{x},\bold{y}) = \frac{1}{2} \int \frac{d^3\bold{k}}{(2\pi)^3} \, e^{i\bold{k} \cdot (\bold{x} - \bold{y})} \frac{1}{k^2} \Pi_{jl}(\bold{k}) ~.
\end{equation}
More explicitly,
\begin{eqnarray} \label{eq: fourier transform of bottom right entry} \nonumber
    G^{\text{leading}}_{jl}(\bold{x},\bold{y}) &\equiv& \frac{1}{32\pi} \frac{1}{|\bold{x}-\bold{y}|} \Bigg( \mathbb I_2\,\delta_{jn} - \frac{1}{3}\sigma_j\sigma_n \Bigg) \Bigg[ \delta^n_{l} + 3\frac{(x-y)^n(x-y)_l}{|\bold{x}-\bold{y}|^2} - (x-y)^n \frac{\boldsymbol{\sigma}\cdot(\bold{x}-\bold{y})}{|\bold{x}-\bold{y}|^2} \sigma_l \Bigg] ~,\\ \nonumber
    G^{\text{const}}_{jl}(\bold{x},\bold{y}) &\equiv& -i \sigma_n \partial^n \, G^{\text{leading}}_{jl}(\bold{x},\bold{y}) ~.
\end{eqnarray}
As a consistency check, it is straightforward to verify that the above expressions are $\sigma$-traceless and divergence free.
\newline\newline
The late time two-point function $G^{\text{leading}}_{jl}(\bold{x},\bold{y})$ is conformally invariant with respect to conformal transformations of the boundary coordinates. The corresponding scaling dimension is $\Delta=3-\tfrac{5}{2}=\tfrac{1}{2}$. This is precisely the conformal weight of an operator given by the shadow transform of a spin-$\tfrac{3}{2}$ conserved supercurrent in a three-dimensional superconformal field theory.} It is also worth noting that the correlator decays slower than the two-point function of an Abelian gauge-field at late times, but faster than the two-point function of a linearised graviton which is logarithmic. On the other hand, for spin-$\tfrac{5}{2}$ and higher, we expect the two-point function to be infrared dominant over spin-two. 
\newline\newline
It would be interesting to see, as in  \cite{Anninos:2019nib}, if such infrared enhanced fermionic higher-spin fluctuations could have an imprint in inflationary cosmology. Any putative patterns imprinted in the CMB would be governed by (\ref{eq: field 2pt function late time}) and its higher-spin avatars.

\section{Euclidean path integral}\label{EPI}

In this section, we consider the Euclidean path integral, $\mathcal{Z}$, of the fermionic higher-spin fields on a four-dimensional sphere. Given that the round four-sphere is the Euclidean continuation of the static patch of dS$_4$, it has been argued that the sphere path integral \cite{Gibbons:1977mu,Gibbons:1976ue} plays an important role in the computation of the entropy of the de Sitter horizon. At a technical level, we will follow the treatment of \cite{Anninos:2020hfj}, but now focused on half-integer spin fields. As we shall see, though Euclidean, the partition functions are naturally built from group characters of the Lorentzian UIRs associated to the fermionic fields. In this sense, the section complements and adds to the Lorentzian analysis of the previous sections. 

\subsection{Sphere path integral of massive higher-spin field}

We first consider the four-sphere path integral, $\mathcal{Z}_m$ for a massive spin-$s$ fermionic field of the form $\Psi_{\mu_1\mu_2\ldots\mu_{s-\frac{1}{2}}\alpha}$, which was  analysed in \cite{Anninos:2020hfj}. The spatial index structure of  $\Psi_{\mu_1\mu_2\ldots\mu_{s-\frac{1}{2}}\alpha}$ is taken to be totally symmetric.  The Euclidean action is generally rather elaborate \cite{singh1974lagrangian} and involves the introduction of  several auxiliary fields. For spin-$\tfrac{3}{2}$, however, the situation is considerably simpler, and the action reads
\begin{equation}\label{SE32m}
    S_E[\Xi_\mu, \Psi_\sigma] =  \int d^4 x \sqrt{g} \, \Xi^{T}_\mu \gamma^{\mu\rho\sigma} \left( \nabla_\rho + \frac{m}{2} \gamma_\rho \right) \Psi_\sigma ~,
\end{equation}
where $\Xi_\mu$ and $\Psi_\mu$ are viewed as independent, unconstrained Dirac spin-$\tfrac{3}{2}$ fields, and $T$ indicates spinorial transposition. To compute the Euclidean path integral, we have to select an integration contour over the space of fields $\Psi_\mu$ and $\Xi_\mu$. Our choice is $\Xi_\mu = \Psi^*_\mu$. 
\newline\newline
A discussion of the fermionic path integral for spin-$\tfrac{3}{2}$ is given in appendix \ref{apppathintmassive}, where it is shown that an important contribution to the path integral stems from the transverse and $\gamma$-traceless eigenfunctions of the spin-$\tfrac{3}{2}$ Dirac operator.
More generally, the corresponding degeneracies for the spectrum of a transverse-traceless spin-$s$ Dirac operator were computed in \cite{rubin1984eigenvalues}, and are given by
\begin{equation}
    D_{n,s}^5 = \frac{1}{3} ( n+2) (2 s+1) \left(n-s+\frac{3}{2}\right) \left(n+s+\frac{5}{2}\right)~, \quad\quad n = s-\frac{1}{2},s+\frac{1}{2},s+\frac{3}{2}, \ldots~,
\end{equation}
while the eigenvalues are given by
\begin{equation}\label{specmrs}
    \lambda_{n,s}= m  \pm i \left(n+2\right)~, \quad\quad n = s-\frac{1}{2},s+\frac{1}{2},s+\frac{3}{2}, \ldots~.
\end{equation}
Following the general conjecture in equation (4.16) of \cite{Anninos:2020hfj}, the sphere path integral resulting from a heat kernel regularisation scheme is given by
\begin{equation}\label{Zmsum}
    \log \mathcal{Z}_m = -\sum_{n\ge -1}  \int_{\mathbb{R}^+} \frac{d\tau}{\tau}   D_{n,s}^5 e^{-\frac{\varepsilon^2}{4\tau} - \lambda_{n,s} \tau}~.
\end{equation}
The sum is regularised in a heat-kernel scheme, where the ultraviolet cutoff at length scale is given by $\varepsilon \equiv \frac{2e^{-\gamma}}{\sqrt{\Lambda_{\text{u.v.}}}}$, where $\Lambda_{\text{u.v.}}$ is some ultraviolet scale with units of inverse length squared (see appendix \ref{apppi}). 
\newline\newline
The extension of the sum (\ref{Zmsum}) beyond the physical range of $n$ in (\ref{specmrs}) stems from a careful treatment of the off-shell field content of the path integral. We analyse the case of a massive spin-$\tfrac{3}{2}$ field from a path integral perspective in appendix \ref{apppathintmassive}. The extended sum is  crucial to obtain the same logarithmic ultraviolet divergence as the path integral computation in appendix \ref{apppathintmassive}. Moreover, the extended mode sum ensures that there is no logarithmic ultraviolet divergence in odd spacetime dimensions, as necessitated by locality.
\newline\newline
The sum (\ref{Zmsum}) can be recast  in the following form  
\begin{equation}\label{Zchar}
\log \mathcal{Z}_m = - \int_{\mathbb{R}^+} \frac{dt}{t} \frac{e^{-\tfrac{t}{2}}}{1-e^{-t}} \left(2 \chi_{\mathcal{D}_{s,m}}(t) - \chi_{\text{edge}}(t) \right)~,
\end{equation} 
where we have formally set $\varepsilon=0$. The character $\chi_{\mathcal{D}_{s,m}}(t)$ is an $SO(4,1)$ group character for the $\mathcal{D}_{s,m}$ principal series UIR
\begin{equation}
\chi_{\mathcal{D}_{s,m}}(t) =  (2s+1) \frac{e^{-t(\frac{3}{2}+im) } + e^{-t(\frac{3}{2}-im )}}{(1-e^{-t})^3}~.
\end{equation}
The character $\chi_{\text{edge}}(t)$ is proportional to a principal series $SO(2,1)$ character
\begin{equation}
\chi_{\text{edge}}(t) = \frac{1}{12} \left(2s-{1} \right) \left(2s+{1}\right) \left(2s+{3}\right) \frac{e^{-t \left(\frac{1}{2}+i m\right)} + e^{-t \left(\frac{1}{2}-i m\right)}}{1-e^{-t}}~.
\end{equation}
The contribution from $\chi_{\text{edge}}(t)$, that appears for all $s\ge 1$, has been interpreted as an edge type contribution given that it adds ultraviolet divergences of codimension two. This interpretation has been further analysed in \cite{Law:2020cpj,Grewal:2022hlo,Law:2025ktz,Anninos:2021ihe}. We note that the edge mode contribution to (\ref{Zchar}) is itself fermionic, as can be seen by the form of (\ref{Zchar}). 

\subsection*{Coefficient of logarithmic divergence}

The logarithmic divergence of the sphere path integral can be obtained from a small-$t$ expansion of the integrand (\ref{Zchar}), and particularly the $\sim\frac{1}{t}$ term. The coefficient splits into a bulk and edge contribution, $\alpha_{\text{bulk}}$ and $\alpha_{\text{edge}}$, that read
\begin{equation}
\alpha_{\text{bulk}} = - (2 s+1)\frac{11+ 30 \left(m^2+2\right) m^2  }{180}~,  \quad\quad \alpha_{\text{edge}} = - (2 s-1) (2 s+1) (2 s+3)\frac{1+6 m^2 }{72}~.
\end{equation}
We note that the $s$-dependent coefficient of the edge contribution grows faster than that of the bulk contribution. For the specific case of spin-$\frac{3}{2}$ field, the overall coefficient of the logarithmic divergence is 
\begin{equation}
    {\mathcal{A}}_{S^4;\frac{3}{2}} = -\frac{41}{45} -\frac{2}{3} m^2(8+m^2) ~.
\end{equation}
We can compare the above coefficient to the one stemming from a direct path integral treatment in appendix \ref{apppathintmassive}. We find agreement.

\subsection{Sphere path integral for Rarita-Schwinger gauge field} \label{sec: massless euclidean p.i.}

We now move on to the sphere path integral, $\mathcal{Z}$, of the spin-$\tfrac{3}{2}$ fermionic gauge field. 
We take the Euclidean action to be given by
\begin{equation}\label{SE32}
    S_E[\Xi_\mu, \Psi_\sigma] =  \int d^4 x \sqrt{g} \, \Xi^{T}_\mu \gamma^{\mu\rho\sigma} \left( \nabla_\rho + \frac{i}{2}\gamma_\rho \right) \Psi_\sigma ~,
\end{equation}
where $g_{\mu\nu}$ is the standard metric on the round four-sphere. Upon variation of the fields, the above action yields the Euclidean continuation of the Rarita-Schwinger equations. The action admits the following gauge-redundancy
\begin{equation}
    \Psi_{\mu\alpha} \rightarrow \Psi_{\mu\alpha} + \nabla_\mu \varepsilon_\alpha + \frac{i}{2} (\gamma_{\mu})_\alpha{}^\beta  \, \varepsilon_\beta ~,
\end{equation}
where we have momentarily reinstated the spinor index $\alpha$ for clarity and $\varepsilon$ is a general Euclidean Dirac spinor. Although we did not consider an action formalism in Lorentzian signature, we will do so in Euclidean. One reason for this is that in Euclidean signature, fermionic actions do not satisfy any immediate reality properties. In any case, one can simply take (\ref{SE32}) as the starting point and explore the resulting partition function.
\newline\newline
As for the massive case,  the fields $\Xi_\mu$ and $\Psi_\sigma$ are, a priori, independent complexified fields. To compute the Euclidean path integral, we have to select an integration contour over the space of fields $\Psi_\mu$ and $\Xi_\mu$. Our choice is $\Xi_\mu = \Psi^*_\mu$. As we shall see, for this choice, the resulting partition function can be expressed in terms of an $SO(4,1)$ UIR character.   \\\\
We follow the technology developed in section 5 of \cite{Anninos:2020hfj}. As before, we accompany this with a careful treatment of the path integral in appendix \ref{apppathint}. One begins by expressing the path integral in terms of a (naive) sum over the modes of a transverse-traceless physical spin-$\tfrac{3}{2}$ field and the modes of a corresponding spin-$\tfrac{1}{2}$ ghost field. This reads
\begin{equation}\label{logZ}
    \log \mathcal{Z} = - \int_{\mathbb{R}^+} \frac{d\tau}{\tau} \, e^{-\frac{\varepsilon^2}{4\tau}} \underset{n\geq-1}{\sum} \left( D^{5}_{n,\frac{3}{2}} e^{-\tau(|\lambda_{n,3/2})|^2-1)} - D^{5}_{n,\frac{1}{2}} e^{-\tau(|\lambda_{n,1/2})|^2-4)} \right) ~,
\end{equation}
where the degeneracies are given by
\begin{eqnarray} 
    D^{5}_{n,\frac{1}{2}} &=& \frac{2}{3} (n+1)(n+2)(n+3) ~, \quad\quad n = 0,1,2, \ldots~, \\
    D^{5}_{n,\frac{3}{2}} &=&  \frac{4}{3} n(n+2)(n+4) ~, \quad\quad n = 1,2,3, \ldots~,
\end{eqnarray}
while the eigenvalues are given by
\begin{equation}
\lambda_{n,s}= \pm i \left(n+2\right) ~, \quad\quad n = s-\frac{1}{2},s+\frac{1}{2},s+\frac{3}{2}, \ldots~.
\end{equation}
Once again, following \cite{Anninos:2020hfj}, we have extended the sum of the degeneracy factors to $n=-1$ to account for additional zero modes in the path integral. Expression (\ref{logZ}) can be manipulated into the form
\begin{equation} \label{eq: naive p.i.}
    \log \mathcal{Z} = - \int_{\mathbb{R}^+} \frac{dt}{t} {F}_{\frac{3}{2}}(e^{-t}) ~,
\end{equation}
where
\begin{equation} \label{eq: naive p.i. integrands}
    {F}_{\frac{3}{2}}(e^{-t}) =  \underset{n\geq-1}{\sum} D^{5}_{n,\frac{3}{2}} e^{-\frac{t}{2}} \left( e^{-t(n+\frac{1}{2})} + e^{-t(n+\frac{5}{2})} \right) - \underset{n\geq0}{\sum} D^{5}_{n,\frac{1}{2}} e^{-\frac{t}{2}} \left( e^{-t(n-\frac{1}{2})} + e^{-t(n+\frac{7}{2})} \right) ~,
\end{equation}
and we have formally set $\varepsilon=0$. Defining $q\equiv e^{-t}$, the integrand of (\ref{eq: naive p.i.}) can be expressed as follows
\begin{equation} \label{eq: naive integrand with characters}
    {F}_{\frac{3}{2}}(q)  = \frac{q^{\frac{1}{2}}}{1-q} \left( 8\hat{\chi}_{\text{bulk},\frac{3}{2}} - 4\hat{\chi}_{\text{edge},\frac{3}{2}} - 4\hat{\chi}_{\text{bulk},\frac{1}{2}} \right)~.
\end{equation}
In the above we have defined the naive characters
\begin{equation}
    \hat{\chi}_{\text{bulk},\frac{3}{2}}  \equiv \frac{q^{\frac{5}{2}}+q^{\frac{1}{2}}}{(1-q)^3}~, \quad  \hat{\chi}_{\text{edge},\frac{3}{2}} \equiv \frac{q^{\frac{3}{2}}+q^{-\frac{1}{2}}}{1-q}~, \quad \hat{\chi}_{\text{bulk},\frac{1}{2}} \equiv \frac{q^{\frac{7}{2}}+q^{-\frac{1}{2}}}{(1-q)^3} ~.
\end{equation}
Due to the presence of terms $\sim q^0$ in (\ref{eq: naive p.i. integrands}), the integral (\ref{eq: naive p.i.}) is infrared (i.e. large $t$) divergent. The problematic behavior is caused by the inclusion of zero modes of the Euclidean action in the path integral, and more specifically the $n=-1$ spin-$\frac{3}{2}$ and the $n=0$ spin-$\frac{1}{2}$ modes in (\ref{eq: naive p.i. integrands}). To fix this pathology, we need to refine the integrand ${F}_{\frac{3}{2}}(q)$ by subtracting the contributions from such modes, namely $${F}_{\frac{3}{2}}(q) \rightarrow {F}_{\frac{3}{2}}(q) - {F}_{\frac{3}{2}}^0(q)~,$$ where
\begin{equation}
    {F}_{\frac{3}{2}}^0(q) = D_{-1,\frac{3}{2}}^{5}\left(1+q^2\right) - D_{0,\frac{1}{2}}^{5}\left(1+q^4\right) = -4\left(2+q^2+q^4\right) ~.
\end{equation}
Removing zero modes by hand can result in a loss of locality. However, this is counteracted by the necessity to divide by the locally defined volume of gauge transformations, whose residual part is generated by the set of Killing spinors
\begin{equation} \label{eq: non naive p.i.}
    \mathcal{Z} = \frac{1}{\text{vol} \, \mathcal{G}_\text{KS}} \exp \left[ - \int_{\mathbb{R}^+} \frac{dt}{t} \left( {F}_{\frac{3}{2}}(q) - {F}_{\frac{3}{2}}^0(q) \right) \right] ~.
\end{equation}
Let us examine the expression (\ref{eq: naive integrand with characters}) more in detail: the naive characters $\hat{\chi}$ cannot be associated to UIRs of $SO(4,1)$ due to the presence of negative powers of $q$ in the numerators. These negative powers of $q$ precisely correspond to the problematic zero modes of the Euclidean action. In order to rewrite (\ref{eq: naive integrand with characters}) in terms of unitary characters, we implement a fermionic version of the `flipping procedure' introduced in \cite{Anninos:2020hfj}. Namely,
\begin{equation} \label{eq: flipping rule}
    \hat{\chi} = \underset{k}{\sum} c_k q^k \longrightarrow [\hat{\chi}] 
    = \hat{\chi}  - \underset{k\leq-\frac{1}{2}}{\sum} c_k \left( q^{-k}+q^k \right)~,
\end{equation}
meaning that we flip the terms $\propto q^{-k}$ for $k\leq-\frac{1}{2}$. The integrand can then be rewritten in terms of the new characters $[\hat{\chi}]$. One finds
\begin{equation} \label{eq: integrand with unitary characters}
    {F}_{\frac{3}{2}}(q)
     = \frac{2q^{\frac{1}{2}}}{1-q} \left( 4q^{\frac{5}{2}}\frac{(2-q)}{(1-q)^3} - 4\frac{q^{\frac{3}{2}}}{1-q} {-4q^{-\frac{1}{2}}(1+q)} \right) ~.
\end{equation}
We recognise the first term in the previous expression as the bulk character (\ref{chiD}) of the $\mathcal{D}^\pm_{\frac{3}{2},\frac{1}{2}}$ UIR of $SO(4,1)$. The second term is proportional to the character of a discrete series UIR of $SO(2,1)$ of weight $\Delta=\frac{3}{2}$. For the sake of the reader, we recall these characters here
\begin{equation} \label{eq: unitary characters}
    \chi_{\mathcal{D}^\pm_{\frac{3}{2},\frac{1}{2}}} = 2q^{\frac{5}{2}} \frac{(2-q)}{(1-q)^3} ~, \quad\quad \chi_{\mathcal{D}^\pm_{\frac{3}{2}}} = \frac{q^{\frac{3}{2}}}{1-q}~.
\end{equation}
The contribution from the $SO(2,1)$ character is denoted as an edge character due to its reduced dimensionality. Hence, we find that the edge mode contribution for the Rarita-Schwinger gauge field in dS$_4$ is a fermion in the discrete series of dS$_2$. 
\newline\newline
We thus land on the expression
\begin{equation} \label{eq: non naive p.i. new}
    \mathcal{Z} = \frac{\mathcal{Z}_{F^0}  \mathcal{Z}_{\text{char}}}{\text{vol} \, \mathcal{G}_\text{KS}}~,
\end{equation}
where we have defined
\begin{eqnarray}
  \log  \mathcal{Z}_{\text{char}} &\equiv&   - \int_{\mathbb{R}^+} \frac{dt}{t} \frac{e^{-\frac{t}{2}}}{1-e^{-t}} \left( 4\chi_{\mathcal{D}^\pm_{\frac{3}{2},\frac{1}{2}}}(t)- \,8 \chi_{\mathcal{D}^\pm_{\frac{3}{2}}}(t)  \right) ~,  \\
  \log   \mathcal{Z}_{F^0} &\equiv&  \int_{\mathbb{R}^+} \frac{dt}{t} \left( {F}_{\frac{3}{2}}^0(q) + {8\frac{1+q}{1-q}} \right)~.
\end{eqnarray}
This is the Harish-Chandra character representation of the Euclidean partition function of a massless spin-$\tfrac{3}{2}$ gauge field. 
\newline\newline
It is worth noting that despite the presence of imaginary masses, the resulting characters that appear are unitary and their pre-factors are real-valued. This mimics the behaviour found in the dS$_2$ supergravity model of \cite{Anninos:2023exn}.

\subsection*{Coefficient of logarithmic divergence} 

As for the massive case, we can expand the integrand of $\log \mathcal{Z}$ at small $t$ to obtain the coefficient of the logarithmic divergence. From the $\log\mathcal{Z}_{\text{char}}$ contribution, we find the small-$t$ integrand expansion
\begin{equation}
\mathcal{I}_{\text{char}} \approx -\frac{8}{t^5} - \frac{16}{t^2}+\frac{52}{3 t^3} + \left( \left(\frac{289}{90}\right)_\text{bulk} + \left(\frac{10}{3}\right)_{\text{edge}}\right)\frac{1}{t} + \ldots ~,
\end{equation}
where we have indicated the bulk-edge split of the logarithmic coefficient. Let us now examine $\mathcal{Z}_{F^0}$. One readily finds the logarithmic contribution
\begin{equation} \label{eq: F^0 path integral evaluated}
 \mathcal{I}_{F^0} \approx \frac{16}{t^2} - \frac{16}{t} + \ldots ~.
\end{equation} 
It is worth noting that the seemingly non-local $\sim \tfrac{1}{t^2}$ terms cancel between $\mathcal{I}_{\text{char}}$ and $\mathcal{I}_{F^0}$. Finally, from the volume of the residual gauge group generated by Killing spinors, see  appendix \ref{resvolume} and (\ref{eq: final gravitino path integral}) for details, one obtains an additional factor of ${\mathcal{B}}_\text{vol}=+16$ to the logarithmic coefficient. Summing all the contributions together leads to 
\begin{equation}
    {\mathcal{B}}_{S^4;\frac{3}{2}} = {\mathcal{B}}_{\text{char}} 
    + {\mathcal{B}}_{F^0}  + {\mathcal{B}}_\text{vol} = \frac{589}{90} - 16 + 16 =  \frac{589}{90}~.
\end{equation}
We can compare the above coefficient, to that stemming from the direct path integral treatment. This is performed in appendix \ref{apppathint}, and we find agreement (see, also, the seventh entry of Table 2 of \cite{Bobev:2023dwx} for the AdS$_4$ counterpart).

\subsection{General spin proposal \& higher spin sums}

Our result for the character contribution to the spin-$\tfrac{3}{2}$ Rarita-Schwinger gauge field path integral, (\ref{eq: unitary characters}), agrees with the prediction below formula (5.22) of \cite{Anninos:2020hfj} --- our bulk character agrees with (5.9) of \cite{Anninos:2020hfj} and our edge character with (4.16) of \cite{Anninos:2020hfj}. Based on this, it is worth reiterating the conjecture of \cite{Anninos:2020hfj} for the character contribution to the partition function of a spin-$s$ fermionic totally massless gauge field. Namely, we denote the character contribution to the path integral as
\begin{equation}
    \mathcal{Z}^{(s)}_{\text{char}} \equiv \exp \left[- \int_{\mathbb{R}^+} \frac{dt}{t} \frac{e^{-\frac{t}{2}}}{1-e^{-t}} \left( \chi^{(s)}_{\text{bulk}}(t) -\chi^{(s)}_{\text{edge}}(t) \right) \right]~,
\end{equation}
where we have formally set the cutoff parameter $\varepsilon = 0$. The fermionic bulk and edge higher-spin characters are given by
\begin{eqnarray} \nonumber
\chi^{(s)}_{\text{bulk}}  &\equiv&  4(2s+1)\frac{q^{s+1}}{\left(1-q\right)^3} -4(2s-1)\frac{q^{s+2}}{\left(1-q\right)^3}~, \\ \nonumber
\chi^{(s)}_{\text{edge}} &\equiv& 4 \left(s-\frac{1}{2}\right) \left(s+\frac{1}{2}\right) \left(s+\frac{3}{2}\right) \frac{q^s}{3 (1-q)}-4 \left(s-\frac{3}{2}\right) \left(s-\frac{1}{2}\right) \left(s+\frac{1}{2}\right) \frac{q^{s+1}}{3 (1-q)}~.
\end{eqnarray}
We note that $\chi^{(\frac{1}{2})}_{\text{edge}} = 0$, compatible with the absence of an edge mode contribution for the Dirac spinor. 
\newline\newline
If we imagine a theory that has an infinite tower of higher-spin fermionic gauge fields, including a massless Dirac fermion, the following cancellation holds
\begin{equation}
\frac{8 q^{\frac{3}{2}}}{(1-q)^3} + \sum_{s=\frac{3}{2}}^\infty \left( \chi^{(s)}_{\text{bulk}} - \chi^{(s)}_{\text{edge}} \right) = 0~.
\end{equation}
In other words, the edge character contribution, though subleading in its ultraviolet divergent structure, sums up to a term that competes and cancels with the bulk character contribution.
\newline\newline
If the fermionic fields are accompanied by their bosonic counterpart, as happens for supersymmetric higher-spin theories, there will be an additional contribution from the bosonic fields. For Vasiliev theories in dS$_4$ \cite{Vasiliev:1990en,Iazeolla:2007wt}, one has a tower of higher-spin Fronsdal fields along with a conformally coupled scalar. The character contribution from the spectrum of bosonic fields appearing in the Vasiliev theory almost, but not quite, cancels \cite{Anninos:2020hfj}:
\begin{equation}\label{chs}
\int_{\mathbb{R}^+} \frac{dt}{2t} \frac{1+q}{1-q}\left(\frac{q+q^2}{(1-q)^3} + \sum_{s=1}^\infty \left( \chi^{(s)}_{\text{bulk}} - \chi^{(s)}_{\text{edge}} \right) \right) = - \int_{\mathbb{R}^+} \frac{dt}{2t} \frac{1+q}{1-q} \frac{q}{(1-q)^2}~.
\end{equation}
If one sums over the even spin spectrum alone, which is the spectrum of the minimal Vasiliev theory, we instead obtain the relation
\begin{multline}\label{chs2}
\int_{\mathbb{R}^+} \frac{dt}{2t} \frac{1+q}{1-q}\left( \frac{q+q^2}{(1-q)^3} + \sum_{s=1}^\infty \left( \chi^{(2s)}_{\text{bulk}} - \chi^{(2s)}_{\text{edge}} \right) \right)   \\ = - \int_{\mathbb{R}^+} \frac{dt}{2t} \frac{1+q}{1-q}\frac{q}{(1-q)^2} +  2\int_{\mathbb{R}^+} \frac{dt}{2t} \frac{1+q}{1-q} \frac{q^{\frac{1}{2}} + q^{\frac{3}{2}}}{(1-q)^2}~.
\end{multline}
The above also agrees with the analogous result for AdS$_4$ (specifically (2.24) of \cite{Giombi:2013fka} or (8.15) of \cite{Sun:2020ame}) with the main difference being a factor of two and the additional term that also appears in (\ref{chs}). 
\newline\newline
It is notable that there is a reduction in the degree of ultraviolet divergence in the resulting sum, especially given that summing the bulk or edge characters alone would lead to an enhanced ultraviolet divergence. Perhaps interestingly, in the $\mathcal{N}=2$ versions of these models one has two towers of integer and half-integer higher-spin fields, such that the one-loop phase of the four-dimensional sphere path integral (namely (5.14) of \cite{Anninos:2020hfj}) vanishes.
 
\subsection{A Euclidean higher spin speculation} 

The right-hand side of (\ref{chs}) can be viewed as the character contribution to the one-loop partition function of a three-dimensional conformal higher-spin gauge theory on $S^3$ \cite{Giombi:2013yva,Anninos:2020hfj}.\footnote{In even spacetime dimensions, a conformal higher-spin gauge theory admits a local action formulation, at least at sufficiently low orders in the weak field expansion \cite{Tseytlin:2013jya,Beccaria:2014jxa}. In odd spacetime dimensions, we define it to be the non-local theory that is induced by integrating out the fields of a free CFT with a general higher-spin background source turned on for all the higher-spin currents. See also \cite{Vasiliev:conformal, Diaz:2024kpr, Diaz:2024iuz, Segal:2002gd, Grigoriev:2019xmp}.} The right hand side of (\ref{chs2}) has in addition a term proportional to {\it twice} the logarithm of the partition function of a free conformally coupled real scalar on $S^3$.  This might lead us to, optimistically, hypothesise that the full four-sphere path integral has the general form \cite{anninos2026ds4metamorphosis}
\begin{equation}\label{chsS4}
{\mathcal{Z}^{(N)}_{\text{h.s.}}[S^4] = \frac{1}{\text{vol}\, \mathcal{G}_{\text{CHS}}} \int [\mathcal{D}\mathcal{B} ] \left|\mathcal{Z}^{(-N)}_{\text{free}}[\mathcal{B}]\right|^2}~.
\end{equation}
Here, $\mathcal{Z}^{(N)}_{\text{free}}[\mathcal{B}]$ is the partition function of $N$ free conformally coupled real scalar fields, $\varphi_I$ with $I=1,2,\ldots,N$, in three spacetime dimensions. The $N\to-N$ in (\ref{chsS4}) can be achieved by switching the statistics of the free fields \cite{LeClair:2006kb,Anninos2011dS}.\footnote{Relatedly, in Einstein gravity with a $\Lambda$ term, the finite part of the renormalised classical action of Euclidean AdS$_4$, $S^{\text{AdS}}_E = \tfrac{3\pi}{2  G_N|\Lambda|}$ (see for example (8.6) of \cite{Anninos:2021ihe}), is minus one half of the on-shell action of the Euclidean four-sphere, $S^{\text{dS}}_E = -\tfrac{3\pi}{G_N|\Lambda| }$ (see for example (1.3) of \cite{Anninos:2025ltd}). Here $G_N$ denotes the Newton constant.} The quantity $\mathcal{B}$ represents a general higher-spin background sourcing the whole set of $O(-N)\equiv\mathrm{Sp}(N)$ \cite{Anninos2011dS} invariant single-trace operators, which consist of a scalar and conserved currents for each even-spin. We can view  $\mathcal{Z}^{(N)}_{\text{free}}[\mathcal{B}]$ as a type of induced non-local conformal higher-spin gauge theory in three spacetime dimensions. The coupling for this theory will go as $\sim N^{-\frac{1}{2}}$. The measure $[\mathcal{D}\mathcal{B} ]$ must be gauge-invariant under the conformal higher-spin gauge group $\mathcal{G}_{\text{CHS}}$, and will be divided by the volume of $\mathcal{G}_{\text{CHS}}$ normalised in terms of the coupling. For the case of $N$ complex valued scalars, which is the $U(N)$ vector model, one will have conformal higher-spin gauge fields for each value of the spin.
\newline\newline
More concretely, for the case of a free scalar $O(N)$ vector model we have
\begin{equation}\label{Zchs}
\mathcal{Z}^{(N)}_{\text{free}}[\mathcal{B}] = {\det}^{-\frac{N}{2}} \left(- \nabla_c^2 + \mathcal{B}\right) = {\det}^{-\frac{N}{2}}\left(- \nabla_c^2 \right) e^{-\frac{N}{2}\left(\text{tr}\log\left(\boldsymbol{1} {-} \nabla_c^{-2} \mathcal{B} \right) {+} \text{tr} \nabla_c^{-2}\mathcal{B}\right)}~,
\end{equation}
where $-\nabla_c^2 \equiv -\nabla^2 + \tfrac{1}{8}R$ is the conformal Laplacian on $S^3$. We are regularising the functional determinant so that the `one-point functions' with respect to the source $\mathcal{B}$ vanish. We note, then, that upon reinstating  the heat-kernel regulator $\varepsilon$, it follows from the general results of  \cite{Anninos:2020hfj} that
\begin{equation}\label{detD}
\log {\det}^{-\frac{N}{2}}\left(- \nabla_c^2 \right) = N\int_{\varepsilon}^\infty \frac{dt}{2\sqrt{t^2-\varepsilon^2}} \frac{1+e^{-t}}{1-e^{-t}} \frac{e^{-t+\frac{1}{2}\sqrt{t^2-\varepsilon^2}}+e^{-t-\frac{1}{2}\sqrt{t^2-\varepsilon^2}}}{(1-e^{-t})^2}~.
\end{equation}
An explicit expression for the above determinant can be found in equation (C.22) of \cite{Anninos:2020hfj} upon setting $\nu=\tfrac{i}{2}$. The divergence structure goes as $\sim\tfrac{1}{\varepsilon^3}$ and $\sim \tfrac{1}{\varepsilon}$, and can be adjusted with the two local counterterms built from the background metric. Based on its structure, upon formally setting $\varepsilon=0$, the second term in (\ref{chs2}) can indeed be viewed as shifting $N$ by one unit in (\ref{chsS4}). The remaining path integral in  (\ref{chsS4}) is over the conformal higher-spin gauge fields, collectively denoted by $\mathcal{B}$, one for each  even spin. The effective action of this theory has an overall prefactor that goes as $N$. This leads to a perturbative series in $\sim \frac{1}{N}$. At one-loop, one has to path-integrate over the quadratic fluctuations of $\mathcal{B}$ and the conformal higher-spin path integral (\ref{Zchs})  yields \cite{Giombi:2013yva,Anninos:2020hfj} the first term in the second line of (\ref{chs2}). 
\newline\newline
Finally, we point out that the conformal higher-spin group in three dimensions, $\mathcal{G}_{\text{CHS}}$, has the same dimensionality as the higher-spin extension of dS$_4$, occasionally referred to as $\text{hs}(SO(4,1))$; a quotient of the universal enveloping algebra of $SO(4,1)$ by a certain ideal (see for instance \cite{joung2014notes}). At the level of the algebras, the two are in fact isomorphic \cite{vasiliev1988extended,Eastwood:2002su}. The Euclidean de Sitter higher-spin algebra, $\text{hs}(so(5))$, can be viewed as a specific real form of the complexification of $\mathcal{G}_{\text{CHS}}$ (perhaps due to a complex path-integration contour of the three-dimensional Weyl mode \cite{Gibbons:1978ac}). This complexification might also lead to a complexified ultraviolet cutoff parameter $\varepsilon$. The gluing type expression (\ref{chsS4}) is reminiscent of expressions appearing in recent discussions on general relativity on manifolds with finite size boundaries, as applied to the Euclidean sphere \cite{Anninos:2022hqo,Anninos:2024wpy,Silverstein:2024xnr,Allameh:2025gsa}. It is moreover reminiscent of a Euclidean counterpart to the proposal in \cite{Alishahiha:2004md}, and also \cite{Cotler:2025gui}.
\newline\newline
If the conformal field theory is a free supersymmetric field theory, the induced superconformal higher-spin gauge theory will contain an infinite tower of fermionic higher-spin currents also. For instance, we can consider the $\mathcal{N}=2$ theory comprised of $N$ free complex scalars, $\varphi_I$, combined with $N$ free Dirac fermions $\Psi_I$. The $U(N)$ invariant spectrum will consist of a complex scalar, a Dirac fermion, and two towers of integer- and half-integer traceless, conserved currents. The resulting superconformal higher-spin gauge theory will have the corresponding spectrum of conformal higher-spin gauge fields. The one-loop character contribution for such a superconformal higher-spin gauge theory will vanish. The higher-spin character contributions all vanish identically, as they carry no propagating degrees of freedom, whilst the two scalars appear to contribute with equal and opposite sign.\footnote{The way to note this, at least formally, is to use the relation  between the conformal higher-spin one-loop partition function and the ratio of bulk AdS$_4$ partition functions with alternate boundary conditions \cite{Giombi:2013yva}. The bulk AdS$_4$ theory in question has two bulk scalars, which in AdS$_4$/CFT$_3$ correspond to a $\Delta=1$ and $\Delta=2$ operator. As such, the corresponding conformal higher-spin characters following from (9.16) and (9.17) of \cite{Anninos:2020hfj}, along with general results in \cite{Beccaria:2014jxa, Klebanov:2011gs, Giombi:2013yva}, appear with equal and opposite sign.} The supersymmetric version of (\ref{detD}), will be less divergent at small $\varepsilon$ due to the vanishing of the $\sim\tfrac{1}{\varepsilon^3}$ term. The finite part follows from the results of \cite{Klebanov:2011gs} (see the $d=3$ entries of Table 1 and Table 2 therein), and at least in some regularisation scheme, might lead to the logarithm of an integer for some $N \in \mathbb{Z}$ due to the cancellation of the $\frac{3\zeta(3)}{8\pi^2}$ pieces. Furthermore, as noted in the concluding remarks of \cite{Anninos:2020hfj}, the dimension of the three-dimensional conformal higher-spin supergroup is given by the weighted sum over the number of conformal Killing tensors and spinors. The sum can be regularised in a straightforward manner as
\begin{equation}
\text{dim} \, \mathcal{G}_{\text{CHS}} =  \lim_{\delta\to 0^+} \sum_{s=1,\frac{3}{2},\ldots}  2(-1)^{2s} \frac{s (4 s^2-1)}{3} e^{-s \delta} = \frac{1}{8}~.
\end{equation}
Nonetheless, one must still make sense of the volume of the higher-spin supergroup to have a complete understanding of (\ref{chsS4}), or alternatively add some finite feature that effectively removes the group volume contribution. 
\newline\newline
In any case, the analysis of such classes of conformal higher-spin gauge theories, and the possibility of implementing supersymmetric localisation methods, is left for future work. They may constitute a microscopic definition of four-dimensional Vasiliev theories on a Euclidean four-sphere.

\section{Outlook} \label{outlook}

It is of interest to consider an interacting theory that has the dS$_4$ Rarita-Schwinger gauge field. One candidate, considered some time ago, is a de Sitter version \cite{Pilch:1984aw} of four-dimensional cosmological supergravity \cite{Townsend:1977qa}. It was argued there that the appearance of an imaginary mass leads to a problematic theory. This was due to either a non-real action for the $\mathcal{N}=1$ case --- essentially the imaginary mass of the Rarita-Schwinger gauge field --- or a wrong-sign kinetic term for the graviphoton in the $\mathcal{N}=2$ case. As already emphasised \cite{Deser:2001xr,Deser:2003gw,Anninos:2023exn, Higuchi-Letsios}, it may well be worth revisiting these de Sitter supergravities for both Lorentzian and Euclidean signature in light of our discussion.
\newline\newline
Another candidate, discussed in \cite{Leigh:2003gk,Sezgin:2012ag,Chang:2012kt,Hertog:2017ymy,Lang:2024dkt}, is a de Sitter version of the higher-spin gravity built from a tower of both integer and half-integer higher-spin gauge fields. These theories have no known action formalism, but do admit classical equations of motion \cite{Vasiliev:1990en,Vasiliev:2003ev} (see also \cite{Skvortsov:2020gpn, Skvortsov:2022syz, Ponomarev:2016lrm,Krasnov:2021nsq} where chiral higher-spin gravity theories admitting complex action are discussed). The interaction strength is governed by a dimensionless parameter $G_N \Lambda$. The higher-spin gauge theories enjoy a large gauge symmetry which is the higher-spin extension of the set of super-diffeomorphisms that renders much of the particle content gauge dependent. Thus, in such theories, the peculiar features of the Rarita-Schwinger field in dS$_4$ may be under more control. In what follows, we imagine the properties of such a theory, whose $\mathcal{N}=2$ spectrum has complex fields of spin $s=\{0,\tfrac{1}{2},1,\tfrac{3}{2},2,\tfrac{5}{2},3,\tfrac{7}{2},4,\ldots\}$. In particular there are two `metrics', perhaps one playing a role of a reference frame. There is an analogous $\mathcal{N}=1$ higher-spin spectrum with only even integer spins accompanied by a tower of half-integer spins. We follow \cite{Anninos:2017eib}, where a proposal was made for the quantum mechanical completion of the classical higher-spin theory in dS$_4$ with a purely bosonic spectrum (see also \cite{Neiman:2017zdr} for an approach centered on the static patch). 

\subsection{Higher-spin building blocks}

The building blocks of the quantum mechanical model, which we refer to as the $\mathcal{N}=2$ model, are as follows.  We have a complex field operator, $\hat{\mathcal{Q}}^I(\bold{x})$, and Dirac spinor valued field operators, $\hat{\Theta}^{I}_{\alpha}(\bold{x})$ and $\hat{\Lambda}_{\alpha I}(\bold{x})$. These transform as $(\Delta_{{\mathcal{Q}}},s_{{\mathcal{Q}}})=(\tfrac{1}{2},0)$ and $(\Delta_{{\Theta}},s_{{\Theta}})= (\Delta_{{\Lambda}},s_{{\Lambda}})= (1,\tfrac{1}{2})$ conformal primaries of $SO(4,1)$ on $\mathbb{R}^3$. The point $\bold{x}=\{x^1,x^2,x^3\}\in\mathbb{R}^3$ is interpreted as a point on $\mathcal{I}^+$ in the planar patch (\ref{planar}). The $U(N)$ vector index $I$ ranges in $\{1,\ldots,N\}$, and it is in the (anti-)fundamental when (raised) lowered. 
\newline\newline
Alongside $\hat{\mathcal{Q}}^I(\bold{x})$, $\hat{\Theta}^I(\bold{x})$, and $\hat{\Lambda}_I(\bold{x})$ we have the conjugate momentum operator $\hat{\Pi}^\mathcal{Q}_{I}(\bold{x})$, as well as the Hermitian conjugates ${\hat{\Theta}}_I(\bold{x})^\dag$ and ${\hat{\Lambda}}^I(\bold{x})^\dag$. These are taken to satisfy the standard (anti)-commutation relations
\begin{equation}\label{mopa}
    [\hat{\mathcal{Q}}^I(\bold{x}), \hat{\Pi}^\mathcal{Q}_{J}(\bold{y})] = i \delta^{I}_{J} \delta(\bold{x}-\bold{y})~, \,\ \{ \hat{\Theta}^I(\bold{x}) , {\hat{\Theta}}_J(\bold{y})^\dag \} = \mathbb{I}_2 \, \delta^{I}_{J} \delta(\bold{x}-\bold{y}) = \{ \hat{\Lambda}_J(\bold{x}) , \hat{\Lambda}^{I}(\bold{y})^\dag \} ~.
\end{equation}
Here, $\mathbb{I}_2$ is once again the spinorial identity matrix, and we have suppressed spinor indices. 
\newline\newline
Extending \cite{Anninos:2017eib} to the case at hand, we build $U(N)$ invariant spin-$s$ normal ordered conformal operators
\begin{eqnarray}\label{bspin}
    \hat{\mathcal{B}}^{(s)}(\bold{x}) &\equiv& :\hat{\mathcal{Q}}_I(\bold{x})^\dag \mathcal{D}_{\bold{x}}^{(s)} \hat{\mathcal{Q}}^I(\bold{x}) :~, \\
    \hat{\Xi}^{(s)}(\bold{x}) &\equiv&  : \hat{\Lambda}_I(\bold{x}) \tilde{\mathcal{D}}_{\bold{x}}^{(s)} \hat{\Theta}^I(\bold{x}):~,\label{fspin} \\ 
    \hat{\Psi}^{(s)}(\bold{x}) &\equiv&  :   \hat{\mathcal{Q}}_I(\bold{x} )^\dag \mathcal{P}_{\bold{x}}^{(s)} \hat{{\Theta}}^I(\bold{x})    :~, \label{ffspin} \\
    \hat{\Phi}^{(s)}(\bold{x}) &\equiv&  : \hat{\Lambda}_I(\bold{x}) \tilde{\mathcal{P}}_{\bold{x}}^{(s)} \hat{\mathcal{Q}}^I(\bold{x} ) :~. 
\end{eqnarray}
Their scaling dimensions are given by $\Delta_{{\mathcal{B}}^{(s)}}  = \Delta_{{\Xi}^{(s)}} = \Delta_{{\Psi}^{(s)}} = \Delta_{{\Phi}^{(s)}} = s+1$. These include a pair of spin-0 and spin-$\tfrac{1}{2}$ $U(N)$ invariant operators
\begin{eqnarray}\label{scalars}
    \hat{{\mathcal{B}}}^{(0)}(\bold{x}) &\equiv& : \hat{\mathcal{Q}}_I(\bold{x})^\dag \hat{\mathcal{Q}}^I(\bold{x}) :~, \quad\quad \hat{\Xi}^{(0)}(\bold{x})  \equiv \,\,\, : \hat{\Lambda}_I(\bold{x}) \hat{\Theta}^I(\bold{x}): ~, \\
    \hat{\Psi}^{(\frac{1}{2})}(\bold{x}) &\equiv& :\hat{\mathcal{Q}}_I(\bold{x}) ^\dag \hat{{\Theta}}^I(\bold{x}):~, \quad\quad \hat{\Phi}^{(\frac{1}{2})}(\bold{x}) \equiv \,\,\, :\hat{\Lambda}_I(\bold{x}) \hat{\mathcal{Q}}^I(\bold{x} ): ~,
\end{eqnarray}
of scaling dimension $\Delta_{{\mathcal{{B}}}^{(0)}} = 1$, $\Delta_{{\Xi}^{(0)}}=2$, and $\Delta_{{\Psi}^{\left({1}/{2}\right)}} = \tfrac{3}{2} = \Delta_{{\Phi}^{\left({1}/{2}\right)}}$ respectively. For the integer spin quantities, we are contracting spinor indices. There are additional $U(N)$ invariants we can construct by combining with $\hat{\Pi}_I^{\mathcal{Q}}$,  $\hat{\Theta}^\dag_I$, $\hat{\Lambda}^{I\,\dag}$, and so on. The additional operator content is in contrast to the situation in AdS/CFT, and stems from the fact that in de Sitter space each bulk field comes appended with two late time conformal operators. We can render the operators Hermitian by adding to them their Hermitian conjugate. We have also introduced the differential operators $\mathcal{D}_{\bold{x}}^{(s)}$, $\tilde{\mathcal{D}}_{\bold{x}}^{(s)}$, ${\mathcal{P}}_{\bold{x}}^{(s)}$, and ${\tilde{\mathcal{P}}}_{\bold{x}}^{(s)}$, which appear in the construction of a conserved spin-$s$ current from a free massless complex scalar a free Dirac fermion theory in three-dimensions. For example,
\begin{eqnarray}
    \mathcal{D}_{\bold{x}}^{(2)} & = & -\frac{1}{3} \partial_j \partial_l - \frac{1}{3}\overset{\leftarrow}{\partial}_j \overset{\leftarrow}{\partial}_l + 2 \overset{\leftarrow}{\partial}_{(j} {\partial}_{l)} - \frac{2}{3}\delta_{jl}\overset{\leftarrow}{\partial}_n \,{{\partial}^n} ~, \\
    \tilde{\mathcal{D}}_{\bold{x}}^{(2)} &=& \gamma_j \overset{\leftrightarrow}{\partial}_l + \gamma_l \overset{\leftrightarrow}{\partial}_j ~,  \\  
    \mathcal{P}_{\bold{x}}^{(\frac{3}{2})} & = & \frac{1}{3} \overset{\leftarrow}{\slashed{\partial}} \gamma_j + \frac{1}{3} \overset{\leftrightarrow}{\partial}_j \mathbb{I}_2 - \frac{1}{3} \gamma_j {\slashed{\partial}} ~, \\
    \tilde{\mathcal{P}}_{\bold{x}}^{(\frac{3}{2})} & = & \frac{1}{3} \overset{\leftarrow}{\slashed{\partial}} \gamma_j - \overset{\leftrightarrow}{\partial}_j \mathbb{I}_2 - \frac{1}{3} \gamma_j {\slashed{\partial}} ~,
\end{eqnarray} 
where $\overset{\leftrightarrow}{\partial}_j \equiv \overset{\leftarrow}{\partial}_j - {\partial}_j$ and the $\gamma_j$ are the three-dimensional Euclidean $\gamma$-matrices satisfying $\{ \gamma_j,\gamma_l\} = 2\delta_{jl}$, expressed in a complex two-dimensional representation.
\newline\newline
The conformally invariant vacuum $|0\rangle$ is defined to be a quantum state governed entirely by the following non-vanishing two-point functions 
\begin{eqnarray}
    \langle 0 |\hat{\mathcal{Q}}^I(\bold{x}) \hat{\mathcal{Q}}_J(\bold{y})^\dag| 0 \rangle &\equiv& - \delta^{I}_{J}  \partial^{-2}  = \frac{\delta^{I}_{J}}{4\pi|\bold{x}-\bold{y}|}~,  \\
    \langle 0 | \hat{{\Theta}}^I(\bold{x}) \hat{{\Theta}}_J(\bold{y})^\dag| 0 \rangle &\equiv&  \delta^{I}_{J} \slashed{\partial}^{-1} =  \delta^{I}_{J} \frac{i \, \boldsymbol{\sigma} \cdot (\bold{x}-
    \bold{y})}{4\pi |\bold{x}-\bold{y}|^{3}}~, \label{thetac} \\
    \langle 0 | \hat{{\Lambda}}_I(\bold{x}) \hat{{\Lambda}}^J(\bold{y})^\dag| 0 \rangle &\equiv&  \delta_{I}^{J} \slashed{\partial}^{-1} =  \delta_{I}^{J} \frac{i \, \boldsymbol{\sigma} \cdot (\bold{x}-
\bold{y})}{4\pi |\bold{x}-\bold{y}|^{3}}~.
\end{eqnarray}
We are discarding contact terms in the above, but they will generally also be pertinent for a complete analysis. All other correlators are obtained by Wick contractions. As simple examples,
\begin{equation}
    \langle 0 | \hat{\mathcal{B}}^{(0)}(\bold{x}) \hat{\mathcal{B}}^{(0)}(\bold{y}) | 0 \rangle = \frac{N}{16\pi^2|\bold{x}-\bold{y}|^2}~, \quad\quad \langle 0 | \hat{\Xi}^{(0)}(\bold{x}) \hat{\Xi}^{(0)}(\bold{y})^\dag | 0 \rangle = -\frac{N}{8\pi^2|\bold{x}-\bold{y}|^4}~,
\end{equation}
for the spin-0 operators, and
\begin{equation}
 \langle 0 | \hat{\Psi}^{(\frac{1}{2})}(\bold{x}) \hat{{\Psi}}^{(\frac{1}{2})}(\bold{y})^\dag | 0 \rangle =   \frac{i \, N \,\boldsymbol{\sigma} \cdot (\bold{x}-\bold{y})}{16\pi^2 |\bold{x}-\bold{y}|^4} = \langle 0 | \hat{\Phi}^{(\frac{1}{2})}(\bold{x}) \hat{{\Phi}}^{(\frac{1}{2})}(\bold{y})^\dag | 0 \rangle ~,
\end{equation}
for the spin-$\tfrac{1}{2}$ operators. For the spin-$\tfrac{3}{2}$ operator $\hat{\Psi}_j^{(\frac{3}{2})}(\bold{x})$, we have
\begin{equation}
    \langle 0 | \hat{\Psi}_j^{(\frac{3}{2})}(\bold{x}) \hat{{\Psi}}_l^{(\frac{3}{2})}(\bold{y})^\dag | 0 \rangle = \frac{i \, N}{12\pi^2} \left( \delta_{jn} - \frac{1}{3} \sigma_j \sigma_n \right) \frac{\boldsymbol{\sigma} \cdot (\bold{x}-\bold{y})}{|\bold{x}-\bold{y}|^6} \left( \delta_{nl} - 2\frac{(x-y)_n (x-y)_l}{|\bold{x}-\bold{y}|^2} \right) ~.
\end{equation}
 The above takes the form of a supercurrent two-point function \cite{Osborn:1993cr,Corley:1998qg}, as expected.
\newline\newline
Being composite, the operators $\{\hat{\mathcal{B}}^{(s)}, \hat{\Psi}^{(s)}, \hat{\Phi}^{(s)}, \hat{\Xi}^{(s)}\}$ will also generally have non-trivial three- and higher-point functions. If we normalise the two-point function of the composite operators to be order one, the size of the higher-point correlations is suppressed by $\sim {N^{-\frac{1}{2}}}$.

\subsection{Toward a map for bulk fields}

In section \ref{confbasis}, we noted that the late time structure of free bulk higher-spin fields in dS$_4$ in the transverse-traceless gauge is captured by a collection of transverse-traceless conformal spin-$s$ operators, $\hat{\boldsymbol{\alpha}}_{\boldsymbol{i}_s}(\bold{x})$ and $\hat{\boldsymbol\beta}_{\boldsymbol{i}_s}(\bold{x})$, where $\boldsymbol{i}_s \equiv \{i_1,\ldots,i_s\}$. The scaling dimensions of the conformal operators are $\Delta_{\boldsymbol{\alpha}} = 2-s$ and $\Delta_{\boldsymbol{\beta}} = s+1$. The bosonic counterpart of such higher-spin operators can be found in section 2 of \cite{Anninos:2017eib}, and the structure is again similar.
\newline\newline
For the case of a purely bosonic tower of interacting higher-spin gauge fields in de Sitter space, it has been proposed \cite{Anninos:2011ui,Anninos:2017eib} that the underlying microphysical content is captured by a vastly reduced set of operators akin to the $\{\hat{\mathcal{Q}}^I(\bold{x}),\hat{\Pi}^{\mathcal{Q}}_I(\bold{x})\}$ of the previous section. The dominant part of the late time profiles of the gauge-fields are captured by $U(N)$ invariant composite operators obtained from the shadow transform of $\hat{\mathcal{B}}^{(s)}(\bold{x})$.\footnote{The shadow transform involves integrating these operators against a spin-$s$ two-point function \cite{Ferrara:1972kab}. Specifically, given a totally symmetric, traceless, spin-$s$ conserved current $\mathcal{O}_{\boldsymbol{i}}(\bold{x})$ of dimension $\Delta_s=s+1$, its shadow transform is given (somewhat schematically) by
\begin{equation}
\mathcal{S}_{\boldsymbol{i}}(\bold{x}) =   \int_{\mathbb{R}^3} \frac{d^3 \bold{y}}{(2\pi)^3} G^{(s)}_{\boldsymbol{i}\boldsymbol{j}}(\bold{x},\bold{y}) \mathcal{O}_{\boldsymbol{j}}(\bold{y})~.
\end{equation}
Here $G^{(s)}_{\boldsymbol{i}\boldsymbol{j}}$ is the two-point function of a spin-$s$, dimension $\bar{\Delta}_s = 2-s$, conformal operator, `gauge fixed' to be transverse and traceless. The output of the shadow transform, $\mathcal{S}_{\boldsymbol{i}}(\bold{x})$, is an operator which transforms locally as a linearised spin-$s$  conformal gauge field. A more detailed account can be found, for instance, in \cite{Metsaev:2008fs} and appendix A of \cite{Anninos:2017eib}. The reason for the gauge redundancy is that the two-point function of a conserved current, $G^{(s)}_{\boldsymbol{i}\boldsymbol{j}}(\bold{x},\bold{y})$, has a non-trivial kernel.} Using the quantum state $|0\rangle$, one can reproduce the late time tree-level correlations in the Bunch-Davies vacuum from the microscopic operators, and complete them to all orders \cite{Anninos:2017eib}. The structure also echoes  the speculative formula (\ref{chsS4}).
\newline\newline
Along the same vein, we imagine that the microscopic completion of a putative higher-spin gauge theory comprised of an infinite tower of effective fields with integer and half-integer spin has a similar completion. In particular, the dominant late time operators $\hat{\boldsymbol\alpha}_{\boldsymbol{i}_s}(\bold{x})$ are microscopically defined by the shadow transform of the microscopic operators $\{\hat{\mathcal{B}}^{(s)}(\bold{x}), \hat{\Psi}^{(s)}(\bold{x}), \hat{\Phi}^{(s)}(\bold{x}), \hat{\Xi}^{(s)}(\bold{x})\}$. Using the shadow transform, we can map the higher-spin operators $\{\hat{\mathcal{B}}^{(s)}(\bold{x}), \hat{\Psi}^{(s)}(\bold{x}), \hat{\Phi}^{(s)}(\bold{x}), \hat{\Xi}^{(s)}(\bold{x})\}$ to a collection of operators whose conformal structure mimics that of the late time profile of bulk higher-spin fields $\hat{\boldsymbol\alpha}_{\boldsymbol{i}_s}(\bold{x})$. A similar treatment follows for the spin-$0$ and spin-$\tfrac{1}{2}$ operators (see appendix \ref{appDirac} for a discussion of massless Dirac fermions in dS$_4$). The Hilbert space that these operators act on must be invariant under all higher-spin supersymmetries --- the higher-spin generalisation of the fact that the de Sitter isometries are gauged in gravity \cite{higuchi1991quantum}. From the perspective of a dS/CFT correspondence \cite{Strominger2001,Maldacena2003}, wavefunctionals will take the form of partition functions of superconformally invariant field theories \cite{Hertog:2017ymy}. They will solve the late time limit of the supersymmetric Wheeler-DeWitt equation, and thus be annihilated by the supercharges \cite{Teitelboim:1977fs,DEath:1992wve}. 
\newline\newline
The bulk operators $\hat{\boldsymbol{\alpha}}_{\boldsymbol{i}_s}(\bold{x})$ are accompanied by an additional tower of conformal higher-spin operators, $\hat{\boldsymbol{\beta}}_{\boldsymbol{i}_s}(\bold{x})$. Moreover, one has a non-trivial bulk operator algebra generated by the $\hat{\boldsymbol{\alpha}}_{\boldsymbol{i}_s}(\bold{x})$ and $\hat{\boldsymbol{\beta}}_{\boldsymbol{i}_s}(\bold{x})$. It is conceivable that in the limit $N\to\infty$ we can reconstruct this bulk operator algebra from the additional operators required to furnish the microscopic operator algebra,  (\ref{mopa}),  at least within some portion of $\mathcal{I}^+$. However, at finite $N$ the bulk operator algebra is at best an approximation \cite{Anninos:2017eib}.

\section*{Acknowledgements}

It is a great pleasure to acknowledge discussions with  Tarek Anous, Nikolay Bobev, Frederik Denef, Neil Lambert, Alan Rios Fukelman, Dami\'an Galante, Sean Hartnoll, Vladimir Schaub, Carlo Iazeolla,  Peter West, Mat\'\i as Semp\'e, Facundo Cruz, and especially Beatrix M\"uhlmann for insightful comments. D.A. is funded by the Royal
Society under the grant “Concrete Calculables in Quantum de Sitter”, the STFC Consolidated grant ST/X000753/1, and the KU Leuven C1 project C16/25/01. C.B. is funded by STFC under the grant reference
STFC/2887726.  V. A. L.  is funded by the ULYSSE Incentive Grant for Mobility in Scientific Research [MISU] F.6003.24, F.R.S.-FNRS, Belgium.  {In the early stage of this work, V. A. L.  was supported by a fellowship from the Eleni Gagon Survivor's Trust for research at the Department of Mathematics at King's College London. G.A.S. is funded by CONICET and ANPCYT grants. G.A.S. would like to thank the organizers of Physics Session Initiative 2025 workshop, held at Pollica, where part of this work was done.

\appendix

\section{Conventions and eigenspectra}\label{conventions}

In this appendix, we provide our choice of conventions and spell out the eigenspectra of the Dirac operators. We use a mostly plus metric in Lorentzian signature. Spatial coordinates are denoted by middle of the alphabet Latin indices $i,j,...$. Orthonormal frame indices are denoted by  $\underline a,\underline b,..$. Greek letters from the beginning of the alphabet $\alpha, \beta,...$ denote spinor indices. The four-dimensional de Sitter and Euclidean sphere metrics have curvature $R = 4\Lambda$, and we write $\Lambda = + \tfrac{3}{\ell^2}$. The Riemann tensor of the de Sitter and Euclidean sphere is given by $R_{\mu\nu\rho\sigma} = \tfrac{\Lambda}{3} \left(g_{\mu\rho}g_{\nu\sigma} - g_{\mu\sigma}g_{\nu\rho} \right)$.

\subsection{Vierbeins, spin connections and Christoffell symbols}

For Lorentzian signature we work  in the planar coordinate system $x^\mu=(\eta,x^i)$ of four-dimensional de Sitter space. The metric reads
\begin{equation}
\frac{ds^2}{\ell^2} = \frac{-d\eta^2 + d\bold{x}^2}{\eta^2}~,
\end{equation}
where $\eta \in \mathbb{R}^-$, and $\bold{x}\in\mathbb{R}^3$. The metric covers half of the full de Sitter manifold, with the future boundary $\mathcal{I}^+$ residing at $\eta=0^-$.
\newline\newline
The non-vanishing Christoffel symbols are  
\begin{equation}
{\Gamma}^\eta_{\eta\eta} = -\frac{1}{\eta}~, \quad {\Gamma}^\eta_{ij} = -\frac{\delta_{ij}}{\eta}~, \quad {\Gamma}^i_{\eta j} = {\Gamma}^i_{j \eta } = -\frac{{\delta}^i_j}{\eta},~~~~~i,j=1,2,3~.
\end{equation}
The non-vanishing vierbein and spin connection components are given by
\begin{eqnarray}
e_{\mu}\, ^{\underline{\nu}} &=& \displaystyle - \frac{1}{\eta} \delta_\mu^\nu  ~,~~\qquad
\omega_{i}{}^{\underline{0 j}} =   - \omega_{i}{}^{\underline{j0}}= -\frac1\eta\delta_{ i j}  ~.
\end{eqnarray}
Covariant derivatives on spinors $\nabla_\mu=\partial_\mu+\frac14\omega_\mu{}^{\underline{ab}}\gamma_{\underline{ab}}$ then read
$$\nabla_\eta=\partial_\eta,~~~~~\nabla_i=\partial_i-\frac1{2\eta}\gamma_{\underline{0i}}~.$$
The covariant derivative acting on a vector-spinor field $\Psi_{\mu\alpha}$ includes a Christoffel symbol and takes the form
\begin{equation}
\nabla_\mu \Psi_{\nu\alpha} = \partial_\mu \Psi_{\nu\alpha} + \frac{1}{4} \omega_\mu{}^{  \underline{ab}} (\gamma_{\underline{ab}})_\alpha{}^\beta  \Psi_{\nu\beta}  - \Gamma^\lambda_{\mu\nu}\Psi_{\lambda\alpha} ~.
\end{equation}

\subsection{Spinor conventions and $\gamma$-matrices}

\subsubsection*{Lorentzian properties}

For an orthonormal frame in Lorentzian signature, we have
\begin{equation}
\{ \gamma^{\underline{a}}, \gamma^{\underline{b}}  \} = 2 \eta^{\underline{a} \underline{b}}~.
\end{equation}
Our signature convention implies $$(\gamma^{\underline{0}})^{\dagger} = - \gamma^{\underline{0}}, ~~~~  (\gamma^{\underline{0}})^{2} = -1~~~~  
\text{and}~~~~ (\gamma^{\underline{j}})^{\dagger} = \gamma^{   \underline{j}    }~,   \quad (\gamma^{\underline{j}})^{2}=1~,
$$  
here $\underline{j}$ denotes spatial orthonormal frame indices. 
$\gamma$-matrices conjugate as follows
\begin{equation}
{\gamma^{\mu}}^{\dagger} =  \gamma^{\underline{0}}   \gamma^{\mu}    \gamma^{\underline{0}}~.
\end{equation}
The chiral matrix is defined as $\gamma^{5} \equiv i \gamma^{\underline{0}}   \gamma^{\underline{1}} \gamma^{\underline{2}}   \gamma^{\underline{3}}$. An explicit basis is given by
\begin{equation}
\label{basischoice}
    \gamma^{\underline{0}} = i \begin{pmatrix}
        \mathbb{I}_2 & 0 \\
        0 & -\mathbb{I}_2
    \end{pmatrix} ~, \quad \gamma^{\underline{j}} = \begin{pmatrix}
        0 & i\sigma_{ {j}} \\
        -i\sigma_{  {j}} & 0
    \end{pmatrix} ~, \quad \gamma^{5} = \begin{pmatrix}
        0 & \mathbb{I}_2 \\
        \mathbb{I}_2 & 0
    \end{pmatrix} ~.
\end{equation}
where $\sigma_i$ are the Pauli matrices, and $\mathbb{I}_2$ is the two-dimensional identity matrix. The Dirac conjugate is  defined as
\begin{equation}
\bar {\psi} \equiv \psi^{\dagger} i\gamma^{\underline{0}}~, 
\end{equation}
and the charge conjugate spinor as 
\begin{equation}
\Psi^C \equiv B^{-1} \Psi^* \,,
\end{equation}
where $B \gamma^{\mu} B^{-1} = \pm (\gamma^{\mu})^{*}$.
For the basis \eqref{basischoice}, one finds $B = \gamma^{\underline{2}}$ for the plus sign (this is relevant to the case of spinors with real mass $m$). Similarly, we have  $\tilde{B} = -\gamma^{\underline{2}} \gamma^5$ for the minus sign, which is relevant to spinors with pure imaginary mass $m$.

\subsubsection*{Euclidean properties}

In Euclidean signature, the gamma matrices are $\gamma^{\underline{1}}$, $\gamma^{\underline{2}}$, $\gamma^{\underline{3}}$, and $\gamma^{\underline{4}} = -i\gamma^{\underline{0}}$. They are all Hermitian,  
\begin{equation}
{\gamma^{\mu}}^{\dagger} =  \gamma^{\mu} ~.
\end{equation}
The Euclidean Clifford algebra reads
\begin{equation}
\{ \gamma^{\underline{a}}, \gamma^{\underline{b}}  \} = 2 \delta^{\underline{a} \underline{b}}~.
\end{equation}
We now define $\gamma^{5} \equiv \gamma^{\underline{1}} \gamma^{\underline{2}}   \gamma^{\underline{3}} \gamma^{\underline{4}} = \begin{pmatrix}
    0 & \mathbb{I}_2 \\
        \mathbb{I}_2 & 0
\end{pmatrix} $.

\subsubsection*{Curved space $\gamma$-matrices}

The Clifford algebra in curved space is given by
\begin{equation}
\{ \gamma^{\mu}, \gamma^{\nu}  \} = 2 g^{\mu \nu}~,
\end{equation}
where $\gamma^\mu \equiv e^\mu{}_{\underline a}\,\gamma^{\underline a}$.

\subsubsection*{$\gamma$-matrix identities}

We  define anti-symmetrised products with weight one and denote it by square brackets. In particular,
\begin{equation}
\gamma^{\mu \nu} \equiv \gamma^{[\mu}   \gamma^{\nu]}= \frac{1}{2} [\gamma^{\mu},  \gamma^{\nu}] = \gamma^{\mu}  \gamma^{\nu} - g^{\mu \nu}~.
\end{equation}
In Lorentzian signature, its Hermitean conjugate satisfies
\begin{equation}
(\gamma^{\mu \nu})^{\dagger} = \gamma^{  \underline{0}   }   \gamma^{\mu \nu}  \gamma^{\underline{0}}~.
\end{equation}
The anti-symmetrised triple product is given by
\begin{equation}
 \gamma^{\mu \nu \rho} \equiv \frac{1}{3!} (\gamma^{\mu}  \gamma^{\nu} \gamma^{\rho} -\gamma^{\mu}  \gamma^{\rho}  \gamma^{\nu} + \gamma^{\rho} \gamma^{\mu}   \gamma^{\nu}-\gamma^{\rho} \gamma^{\nu}   \gamma^{\mu}+ \gamma^{\nu} \gamma^{\rho}   \gamma^{\mu} - \gamma^{\nu} \gamma^{\mu}   \gamma^{\rho}  )~,
\end{equation}
or equivalently
\begin{equation}
\gamma^{\mu \nu \rho} \equiv \gamma^{[\mu}   \gamma^{\nu}   \gamma^{\rho]} = \gamma^{\mu}  \gamma^{\nu}   \gamma^{\rho}   -g^{\mu \nu}  \gamma^{\rho} - g^{\nu \rho} \gamma^{\mu} + g^{\mu \rho} \gamma^{\nu} = \gamma^{\mu} \gamma^{\nu \rho} -g^{\mu \nu} \gamma^{\rho} +g^{\mu \rho} \gamma^{\nu}~,
\end{equation}
The following relation also holds
\begin{equation}
\gamma^{\mu \nu \rho} = \frac{1}{3} (   \gamma^{\mu}   \gamma^{\nu \rho}   + \gamma^{\rho}   \gamma^{\mu \nu}  + \gamma^{\nu}   \gamma^{ \rho  \mu} )~.
\end{equation}

\subsection{Commutators and Killing spinors}

In de Sitter space, the commutator of  covariant derivatives acting on spinors gives
\begin{equation}
[\nabla_{\mu} , \nabla_{\nu}] \psi  = \frac{1}{4} R_{\mu \nu}{}^{\rho\sigma} \gamma_{\rho\sigma} \psi=\frac{\Lambda}{3} \frac{1}{2}  \gamma_{\mu\nu}\psi  ~.
\label{comp}
\end{equation}
Upon setting $\ell=1$, we have that $\frac{\Lambda}{3}$ is equal to +1 for dS$_4$ and $-1$ for AdS$_4$. For vector-spinor fields we have
$$[\nabla_\mu,\nabla_\nu]\Psi_\rho=\frac1{2 }\gamma_{\mu\nu} \Psi_\rho+g_{\mu\rho}\Psi_\nu- g_{\nu\rho}\Psi_\mu ~ .$$
A useful identity for spinors in arbitrary backgrounds is 
\begin{equation}
\nabla^2\psi = \slashed{\nabla}^2\psi + \frac R4\psi~.
\end{equation}
For $\gamma$-traceless vector-spinors the identity reads
$$\nabla^2\Psi_\mu = \left(\slashed{\nabla}^2  + \frac R4+1\right)\Psi_\mu.$$
A Killing spinor, $\lambda$, is a (generally complex) spinor satisfying 
\begin{equation}\label{KSE}
\nabla_\mu \lambda = \alpha \gamma_\mu \lambda~,
\end{equation}
for some $\alpha\in\mathbb{C}$. Compatibility with \eqref{comp} requires $\alpha^2=-\tfrac{1}{4\ell^2}$ in de Sitter and Euclidean spheres.  
\newline\newline
A conformal Killing spinor  is a spinor, $\varepsilon$, satisfying
\begin{equation}
\nabla_\mu \varepsilon = \frac{1}{4} \gamma_\mu \slashed{\nabla} \varepsilon~.
\end{equation}
For a maximally symmetric spacetimes, the real and imaginary parts of the Killing spinor solve the conformal Killing spinor equation (see appendix C of \cite{Anous:2014lia} for more details).

\subsection{Fermionic spherical harmonics} \label{appsphharmonics}

We now introduce spinor spherical harmonics. These are spinor  eigenfunctions of the Dirac operator on $S^4$ \cite{Camporesi:1995fb}
\begin{equation} \label{eq: spinor spherical harmonics expansion}
    \slashed{\nabla}\psi_{\pm}^{n,l} = \pm \frac{i}{\ell} (n +2) \psi_{\pm}^{n,l} ~,
\end{equation}
Here $n=0,1,2,\dots$ parametrize the eigenvalue and $l$ labels the remaining quantum numbers. For each fixed value of $n$, both sets of spinor spherical harmonics $\{ \psi_{-}^{n,l} \}$ and $\{\psi_{+}^{n,l} \}$ independently form a unitary irreducible representation of $so(5)$. They are equivalent since they can be related as $\psi_{-}^{n,l}=\gamma^5\psi_{+}^{n,l}$, their highest weight vector is $f^{(s=\frac{1}{2})}_{n} = (n+\frac{1}{2} , \frac{1}{2})$. Its dimension, i.e. the degeneracy of the $n$-eigenvalue, is given by \cite{Camporesi:1995fb}
\begin{equation} \label{eq: spinor degeneracy}
    D^{5}_{n,\frac{1}{2}} = \frac{(3+n)!}{n!}\frac{4}{3!}  =\frac{2}{3} (n+1)(n+2)(n+3) ~.
\end{equation}
{\textbf{Killing spinors on $S^4$.}} For $n=0$, the spinor spherical harmonics reduce to two families of Killing spinors, $\psi_{+}^{n=0,l}$ and  $\psi^{n=0,l}_{-}$, satisfying
\begin{equation}
    \nabla_{\mu}\psi_{\pm}^{ n=0,l} = \pm \frac{i}{2 \ell} \gamma_{\mu}\psi_{\pm}^{ n=0,l} ~.
\end{equation}
Comparing to (\ref{KSE}) we recognize the Killing spinor equation, each with (complex) degeneracy $D^{5}_{0,\frac{1}{2}}=4$. This is in contrast to the case of AdS$_4$ in either Euclidean or Lorentzian signature. 
\newline\newline
The TT vector-spinor spherical harmonics  are vector-spinor eigenfunctions of the Dirac operator on $S^{4}$ satisfying \cite{Letsios:2022tsq}
\begin{align}
    \slashed{\nabla}\psi_{\pm ~\mu} ^{n,l} & = \pm \frac{i}{\ell} (n +2) \psi_{\pm~\mu}^{n,l} ~~~\text{and}~~~ \gamma^{\mu}\psi_{\pm ~\mu}^{n,l} = 0 = \nabla^{\mu}\psi^{n,l}_{\pm ~\mu} ~,
\end{align}
where $n=1,2,\dots$ and $l$ labels all the other relevant quantum numbers (note the value $n=0$ is not allowed). For each fixed value of $n$, both sets of vector-spinor spherical harmonics $\{ \psi_{-\mu}^{n,l} \}$ and $\{\psi_{+\mu}^{n,l} \}$ independently form a unitary irreducible representation of $so(5)$, although equivalent. Their highest weight vector is $f^{(s=\frac{3}{2})}_{n} = (n+\frac{1}{2} , \frac{3}{2})$ and their dimension \cite{Homma:2020has}
\begin{align} \label{eq: vector spinor degeneracy}
  D^{5}_{n,\frac{3}{2}} =  \frac{4}{3} n(n+2)(n+4) ~.
\end{align}

\section{Derivation of discrete series character} \label{appchar}

In this appendix, we provide some details for the derivation of a discrete series character of maximal depth, namely $\mathcal{D}^{\pm}_{s,s-1}$. These were originally computed by Hirai \cite{hirai}, and also discussed in  \cite{Basile:2016aen,Anninos:2020hfj}.

\subsection{Subgroup decomposition}

We consider unitary $so(4,1)$ representations in the decomposition
\begin{equation}
so(4,1) \supset so(4)  \supset so(3) \supset so(2)~,
\end{equation}
as in \citep{ottoson1968classification, schwarz1971unitary}. In a given representation with fixed $\Delta=s+1$, the anti-hermitian $so(4,1)$ generators,  $M_{AB}$,    act on a Hilbert space that is spanned by states $|{q,p;l;m}\rangle$, which carry labels (quantum numbers) of $so(4)$, $so(3)$ and $so(2)$. In particular, the labels $q$ and $p$ with $q \geq |p| $ are $so(4)$ quantum numbers. The label $l$, with $q \geq l \geq |p|$, is a $so(3)$ quantum number, while $m \in \{ -l, ...,0,...,l \}$ is a $so(2)$ quantum number. 
The Hilbert space is equipped with a positive-definite inner product such that  $$\langle{q',p';l';m' | q,p;l;m}\rangle = \delta_{qq'} 
 \delta_{pp'
 } \delta_{ll'} \delta_{mm'}~.
 $$
To be specific, there are two Hilbert spaces of interest corresponding to the representation spaces of the two discrete series, $\mathcal{D}^{+}_{s,s-1}$ and $\mathcal{D}^{-}_{s,s-1}$. The Hilbert space on which  $\mathcal{D}^{+}_{s,s-1}$ acts is spanned by the states:
\begin{align}
    \left \{|{q,p=s;l;m}\rangle \equiv |{q;l;m}\rangle_{+}~\text{with}~ q \geq  l \geq s~~ \text{and}~~l \geq m   \geq -l   \right\} ~.
\end{align}
Similarly, the Hilbert space on which  $\mathcal{D}^{-}_{s,s-1}$ acts is spanned by the states:
\begin{align}
    \left \{|{q,p=-s;l;m}\rangle \equiv |{q;l;m}\rangle_{-}~\text{with}~ q \geq  l \geq s~~ \text{and}~~l \geq m   \geq -l   \right\} ~.
\end{align}
It is easy to understand that states belonging to $\mathcal{D}^{+}$  UIRs are orthogonal to states in the
$\mathcal{D}^{-}$ UIRs, because the inner product is $so(4)$-invariant, and states that belong to different $so(4)$ representations are orthogonal to each other. 
\\ \\
If the spin $s$ is an integer, then the quantum numbers $q$, $l$, $m$ are also integers. If the spin $s$ is a half(-odd-)integer, then then the quantum numbers $q$, $l$, $m$ are themselves also half(-odd-)integers. At the level of the Lie algebra, $so(4,1)$, both integer and half-integer spins can be accommodated. However, at the group level, integer-spin representations of $so(4,1)$ are lifted to representations of $SO(4,1)$, while half-integer-spin representations are lifted to representations of the double cover $\text{Spin}(4,1)$.

\subsection{Infinitesimal matrix elements} 

The anti-hermitian generator $M_{04}$ acts on states as follows \cite{schwarz1971unitary, ottoson1968classification}:
\begin{align}\label{infinitesimal boost mtrx elements}
M_{04}  |{q;l;m}\rangle_{-} =& - \frac{i}{2}  \sqrt{(q-l+1)(q+l+2)}  ~ |{q+1;l;m}\rangle_{-} \nonumber \\
&- \frac{i}{2}  \sqrt{(q-l)(q+l+1)}  ~ |{q-1;l;m}\rangle_{-}~, \\
M_{04}  |{q;l;m}\rangle_{+} =& - \frac{i}{2}  \sqrt{(q-l+1)(q+l+2)}  ~ |{q+1;l;m}\rangle_{+} \nonumber \\
&- \frac{i}{2}  \sqrt{(q-l)(q+l+1)}  ~ |{q-1;l;m}\rangle_{+}~.  \nonumber
\end{align}
Note that states in the discrete series $\mathcal{D}^{+}$ do not mix with states in $\mathcal{D}^{-}$ under the action of any $so(4,1)$ generator. The  non-zero infinitesimal matrix elements, $_{\pm}\langle{q+1;l;m}| M_{04}|  {q;l;m}\rangle_{\pm}$ and $_{\pm}\langle{q-1;l;m}| M_{04}$  $|{q;l;m}\rangle_{\pm}$, do {not} depend on the spin $s$. This means that the  $M_{04}$-matrix elements for all strictly massless fields in the discrete series have the same form. Further, these matrix elements do {not} depend on $m$. They only depend on the $so(4)$ quantum number $q$ and on the $so(3)$ quantum number $l$. Moreover, the action of the boost generator shifts only the value of $q$, while it leaves $l$ and $m$ fixed.
We also observe that the aforementioned infinitesimal boost matrix elements  are equal for the two discrete series, namely
\begin{align}\label{def: mtrx element so(4,1) <q+1,1>}
A^{l}(q) =&  \,_{-}\langle{q+1;l;m}| M_{04}|  {q;l;m}\rangle_{-} = \,_{+}\langle{q+1;l;m} |M_{04}|  {q;l;m}\rangle_{+} = \nonumber \\ =& - \frac{i}{2}  \sqrt{(q-l+1)(q+l+2)} ~,
\end{align}
and
\begin{align}\label{def: mtrx element so(4,1) <q-1,1>}
A^{l}(q-1) =&   \,_{-}\langle{q-1;l;m}| M_{04}|  {q;l;m}\rangle_{-} = \,_{+}\langle{q-1;l;m}| M_{04}|  {q;l;m}\rangle_{+} = \nonumber\\
=& - \frac{i}{2}  \sqrt{(q-l)(q+l+1)} =  - A^{l}(q-1)^{*} ~.
\end{align}
The action of $M_{04}$ on states is then conveniently expressed as
\begin{align}\label{boost matrx element so(4,1) convenient}
M_{04}  |{q,l;m}\rangle_{\pm} = A^{l}(q)  ~ |{q+1;l;m}\rangle_{\pm} \nonumber +A^{l}(q-1) \,  |{q-1;l;m}\rangle_{\pm}~.
\end{align}
The action of the boost generator on the states determines the group matrix elements
\begin{align}
    \,_{\pm}\langle{q';l';m'}| e^{M_{04}\,t}|{q;l;m}\rangle_{\pm}~,
\end{align}
where the boost group element is expanded as $\displaystyle e^{M_{04}\,t} = \sum_{n=0}^{\infty} \frac{t^{n}}{n!} (M_{04})^{n}$. 

\subsection{Determination of character}

To determine the characters of the discrete series UIRs we focus on the diagonal group matrix elements 
\begin{align}\label{def: grp matrix element taylor so(4,1)}
    K_{q}^{l}(t) \equiv \,_{\pm}\langle{q;l;m} |e^{M_{04}\,t}|{q;l;m}\rangle_{\pm} = \sum_{n=0}^{\infty} \frac{t^{n}}{n!} ~\,_{\pm}\langle{q;l;m}|(M_{04})^{n}\,|{q;l;m}\rangle_{\pm}~.
\end{align}
As in the case of infinitesimal boost  matrix elements (\ref{infinitesimal boost mtrx elements}), it easy to understand that the group matrix elements $K_{q}^{l}(t)$ depend only on $q$ and $l$. The character of the UIR is given by
\begin{equation}
  \chi_{\mathcal{D}^{\pm}_{s,s-1}}(t) \equiv  \text{tr}_{\mathcal{D}^{\pm}_{s,s-1}}(e^{M_{04}\,t}) = \sum_{q=s}^{\infty}\sum_{l=s}^{q}  \sum_{m=-l}^{l} K_{q}^{l}(t)   =\sum_{q=s}^{\infty} \sum_{l=s}^{q}   (2l+1)K_{q}^{l}(t)~,
\end{equation}
where in the last equality we performed the summation over $m$ because the group matrix elements are independent of $m$. It is convenient to introduce new summation variables that we will denote as $\Delta_{so(2,1)}$ and $m_{so(2)}$ --- the reason for introducing this notation will become clear shortly. The old and the new summation variables are related to each other as follows: 
\begin{align}
 q=m_{so(2)}-1~,~~~~   l= \Delta_{so(2,1)}-1~.
\end{align}
Then, in terms of the new variables, we have
\begin{align}
  \chi_{\mathcal{D}^{\pm}_{s,s-1}} (t)
     =&\sum_{m_{so(2)}=s+1}^{\infty} ~\sum_{\Delta_{so(2,1)}=s+1}^{m_{so(2)}}   (2\Delta_{so(2,1)}-1)~K^{\Delta_{so(2,1)}\,-\,1}_{m_{so(2)}\,-\,1}(t) \\
     =& \sum_{\Delta_{so(2,1)}=s+1}^{\infty}~(2\Delta_{so(2,1)}-1)~ ~\sum_{m_{so(2)}=\Delta_{so(2,1)}}^{\infty}   ~K^{\Delta_{so(2,1)}\,-\,1}_{m_{so(2)}\,-\,1}(t) \label{before the final boss so(4,1) character}~.
\end{align}
The sum over $m_{so(2)}$ can be re-written in a form that corresponds to a character of discrete series UIRs of $so(2,1)$ with scaling dimension $\Delta_{so(2,1)}$. This can be shown easily by using well-known formulas of $so(2,1)$ discrete series matrix elements (see for example \cite{sugiura,Letsios-Fukelman, schwarz1971unitary, ottoson1968classification}) as discussed in the next subsection. Then, in the case of $\mathcal{D}^{+}_{s,s-1}$ we find
\begin{equation}\label{final char forml appendix group th}
  \chi_{\mathcal{D}^{+}_{s,s-1}} (t)
     = \sum_{\Delta_{so(2,1)}=s+1}^{\infty}~(2\Delta_{so(2,1)}-1) \chi^{so(2,1)}_{\mathcal{D}^{+}_{\Delta_{so(2,1)}}}(t)
     = \frac{(2s+1)q^{s+1}-(2s-1)q^{s+2} }{(1-q)^{3}} ~,
\end{equation}
with $q\equiv e^{-t}$, and $t>0$. In the above, we have made use of the expression for the $so(2,1)$ discrete series character \cite{sugiura}
\begin{equation}
\chi^{so(2,1)}_{\mathcal{D}^{+}_{\Delta_{so(2,1)}}}(t)= \frac{e^{-\Delta_{so(2,1)}t}}{(1-e^{-t})} ~.
\end{equation}
The result (\ref{final char forml appendix group th}) is in agreement with \cite{Anninos:2020hfj, Basile:2016aen, hirai1965characters}. 
\newline\newline
Working similarly, we find the same expression for $\chi_{\mathcal{D}^{-}_{s,s-1}}$
$$\chi_{\mathcal{D}^{-}_{s,s-1}} (t)  = \frac{(2s+1)q^{s+1}-(2s-1)q^{s+2} }{(1-q)^{3}}~.$$
%
Note that the expressions for $\chi_{\mathcal{D}^{+}_{s,s-1}}$ and $\chi_{\mathcal{D}^{-}_{s,s-1}}$ hold for discrete series UIRs of $so(4,1)$ with either integer or half-integer spin $s$.

\subsection{Review of $SO(2,1)$}

{
The algebra $so(2,1)$ is generated by the three generators $\{J_{02}, J_{01}, J_{21}\}$. The discrete series UIRs $\mathcal{D}^{+}_{\Delta_{so(2,1)}}$ of $so(2,1)$ are labelled by the $so(2,1)$ scaling dimension $\Delta_{so(2,1)} \in \{ 1,\frac{1}{2},2,...  \}$ \cite{ottoson1968classification, schwarz1971unitary}. These discrete series UIRs can be realised on a Hilbert space spanned by states $|{m_{so(2)}}\rangle$ with $m_{so(2)} \in \{ \Delta_{so(2,1)}, \Delta_{so(2,1)}+1,...  \}$, where the quantum number $m_{so(2)}>0$ parameterises the eigenvalue of $so(2)$ generator $J_{21}$, i.e. $ J_{21}  |{m_{so(2)}}\rangle = i m_{so(2)}|{m_{so(2)}}\rangle$.
The boost generator $J_{02}$ acts on these states as \cite{ottoson1968classification, schwarz1971unitary}
\begin{align}\label{infinitesimal boost mtrx elements 2D}
J_{02}  |{m_{so(2)}}\rangle =& - \frac{i}{2}  \sqrt{(m_{so(2)}-\Delta_{so(2,1)}+1)(m_{so(2)}+\Delta_{so(2,1)})}  ~ |{m_{so(2)}+1}\rangle \nonumber \\
&- \frac{i}{2}  \sqrt{(m_{so(2)}-\Delta_{so(2,1)})(m_{so(2)}+\Delta_{so(2,1)}-1)}  ~ |{m_{so(2)}-1}\rangle~.
\end{align}
The character for the $so(2,1)$ discrete series $\mathcal{D}^{+}_{\Delta_{so(2,1)}}$ is given by \cite{sugiura}\footnote{See also appendix A of \cite{Letsios-Fukelman} on details concerning how this sum can be performed and obtain the discrete series character for $\Delta=\tfrac{3}{2}$.}
\begin{equation}\label{def:so(2,1) character discrete}
\chi^{so(2,1)}_{\mathcal{D}^{+}_{\Delta_{so(2,1)}}}(t) \equiv  \text{tr}_{\mathcal{D}^{+}_{\Delta_{so(2,1)}}}(e^{J_{02}\,t})  = \frac{e^{-\Delta_{so(2,1)}|t|}}{1-e^{-|t|}} ~.
\end{equation}
The $so(2,1)$ matrix elements for the generator $J_{02}$ are related to the $so(4,1)$ matrix elements $A^l(q)$ defined in (\ref{def: mtrx element so(4,1) <q+1,1>}), (\ref{def: mtrx element so(4,1) <q-1,1>}) as follows
\begin{align}\nonumber
&\langle{m_{so(2)}+1|J_{02}|m_{so(2)}}\rangle=A^{\Delta_{so(2,1)}-1}(m_{so(2)}-1)=  \,_{+}\langle{q+1;l;m} |M_{04}  |{q;l;m}\rangle_{+}\Big|_{\parbox{2cm}{\centering\scriptsize $q \rightarrow m_{so(2)}-1,$ \\[0.5ex] $l\rightarrow \Delta_{so(2,1)}-1$}} ~,\\ \nonumber
&\langle{m_{so(2)}-1|J_{02}|m_{so(2)}}\rangle=A^{\Delta_{so(2,1)}-1}(m_{so(2)}-2)=  \,_{+}\langle{q-1;l;m} |M_{04}  |{q;l;m}\rangle_{+}\Big|_{\parbox{2cm}{\centering\scriptsize $q \rightarrow m_{so(2)}-1,$ \\[0.5ex] $l\rightarrow \Delta_{so(2,1)}-1$}} ~.
\end{align}
For  diagonal matrix elements of the form $\langle{m_{so(2)}|(J_{02})^{n}|m_{so(2)}}\rangle$, with $n =0,1,2,...$, the following holds
\begin{align}
&\langle{m_{so(2)}|(J_{02})^{n}|m_{so(2)}}\rangle=  \,_{+}\langle{q;l;m}| (M_{04})^{n}  |{q;l;m}\rangle_{+}\Big|_{\parbox{2cm}{\centering\scriptsize $q \rightarrow m_{so(2)}-1,$ \\[0.5ex] $l\rightarrow \Delta_{so(2,1)}-1$}} ~.
\end{align}
Both sides of this equation vanish for odd values of $n$. Multiplying both sides with $t^{n}/n!$, summing over $n$ and using (\ref{def: grp matrix element taylor so(4,1)}), we find a useful  relation between $SO(2,1)$ ($\text{Spin}(2,1)$) and $SO(4,1)$ ($\text{Spin}(4,1)$) finite boost matrix elements
\begin{align}\label{eq:key eqtn for computation of character 4D}
&\langle{m_{so(2)}|e^{J_{02}t}|m_{so(2)}}\rangle =  \,_{+}\langle{q;l;m}| e^{M_{04}t}  |{q;l;m}\rangle_{+}\Big|_{\parbox{2cm}{\centering\scriptsize $q \rightarrow m_{so(2)}-1,$ \\[0.5ex] $l\rightarrow \Delta_{so(2,1)}-1$}} = K_{m_{so(2)}-1}^{\Delta_{so(2,1)}-1}(t)~.
\end{align}
The finite boost matrix element $K^{l}_{q}(t)$ has been defined in (\ref{def: grp matrix element taylor so(4,1)}). The relation (\ref{eq:key eqtn for computation of character 4D})  between  finite $SO(2,1)$ and $SO(4,1)$ matrix elements, corresponding to the discrete series  $\mathcal{D}^{+}_{\Delta_{so(2,1)}}$ and  $\mathcal{D}^{+}_{s,s-1}$, respectively, can be easily shown to hold for the case of $\mathcal{D}^{-}_{\Delta_{so(2,1)}}$ and $\mathcal{D}^{-}_{s,s-1}$.
Note that (\ref{eq:key eqtn for computation of character 4D}) has been used to go from (\ref{before the final boss so(4,1) character}) to (\ref{final char forml appendix group th}).}

\section{Details for fermionic path integrals} \label{apppi}

In this appendix we provide some details on the computation of the Euclidean partition function for a free spin-$\frac{3}{2}$ field in four spacetime dimensions. The path integral takes the general form
\begin{equation}
\mathcal{Z} = \frac{1}{\text{vol} \, \mathcal{G}} \int \mathcal{D} \Psi \mathcal{D} \Psi^\dagger e^{-S_E[\Psi_\mu]}~.
\end{equation}
The division by the volume of the gauge redundancies, $\text{vol} \, \mathcal{G}$, is mandatory for the case of   gauge fields. Throughout this appendix we set $\ell=1$.

\subsection{Rarita-Schwinger gauge field}\label{apppathint}

We begin with the case of a spin-$\tfrac{3}{2}$ massless field, governed by the Euclidean action $S_E[\Psi_\mu]$ given in (\ref{SE32}), and reproduced here for convenience
\begin{equation}
    S_E[\Psi_\mu] =  \int d^4 x \sqrt{g} \, \Psi^{\dag}_\mu \gamma^{\mu\rho\sigma} \left( \nabla_\rho + \frac{i}{2}\gamma_\rho \right) \Psi_\sigma ~.
\end{equation}
We consider the decomposition
\begin{equation} \label{eq: first splitting of vector-spinor}
    \Psi_\mu = \varphi_\mu + \nabla_\mu \chi + \gamma_\mu \lambda ~, ~~~\text{with}~~~ {\nabla}^\mu\varphi_\mu = 0 = \gamma^\mu\varphi_\mu ~,
\end{equation}
where $\chi$ and ${\lambda}$ are independent Dirac spinors. The above decomposition is redundant whenever $\nabla_\mu \chi =  \gamma_\mu \chi'$, or equivalently $\gamma_\mu \lambda =  \nabla_\mu \lambda'$, and we must ensure that such configurations are not double counted. We can avoid such double counting, by removing all Killing spinor configurations from $\chi$. 
\newline\newline
Upon implementing the decomposition (\ref{eq: first splitting of vector-spinor}), the Euclidean action $S_E[\Psi_\mu]$ reduces to the sum $S_E[\varphi_\mu] + S_E[\chi,\lambda]$, where 
\begin{equation} \label{eq: varphi TT action}
    S_E[\varphi_\mu] = \int d^4x \sqrt{g} \, \varphi_\mu^\dagger \left( \slashed{\nabla} + i \right) \varphi^\mu ~,
\end{equation}
and
\begin{equation} \label{eq: chi lambda action}
    S_E[\chi,\lambda] = \frac{3}{2} \int d^4x \sqrt{g} \, \bigl( \chi^\dagger -2i{\lambda}^\dagger \bigr) \bigl( \slashed{\nabla} - 2i \bigr) \bigl( \chi + 2i{\lambda} \bigr) ~.
\end{equation}
If we further apply the field transformation  $\lambda = \tfrac{({\varepsilon}-\chi)}{2i}$, we can rewrite (\ref{eq: chi lambda action}) as
\begin{equation} \label{eq: epsilon lambda action}
    S_E[\varepsilon] = \frac{3}{2} \int d^4x \sqrt{g} \, {\varepsilon}^\dagger \bigl( \slashed{\nabla} - 2i \bigr) {\varepsilon} ~.
\end{equation}
The zero modes are spinors, $\varepsilon = \varepsilon^+$, satisfying $\left( \slashed{\nabla} - 2i \right) {{\varepsilon}}^+ = 0$, that is Killing spinors of the type $\nabla_\mu {\varepsilon}^+ - \frac{i}{2} \gamma_\mu {\varepsilon}^+ = 0$. This corresponds to a pure gauge configuration of $\Psi_\mu$. Let us then define 
\begin{equation} \label{eq: varphi TT partition function}
    \mathcal{Z}^{\text{TT}}_\varphi \equiv \int \mathcal{D} \varphi \mathcal{D} \varphi^\dagger e^{-S_E[\varphi_\mu]} \quad\text{and}\quad   \mathcal{Z}_{{\varepsilon}} \equiv \int \mathcal{D}' {{\varepsilon}} \mathcal{D}' {{\varepsilon}}^\dagger e^{-S_E[{\varepsilon}]} ~,
\end{equation}
where the prime in $\mathcal{D}' {{\varepsilon}}$ and $ \mathcal{D}' {{\varepsilon}}^\dagger$ designates the exclusion of $\varepsilon^+$ spinors from path integration in $\mathcal{Z}_{{\varepsilon}}$. We note that we must still path-integrate over the $\varepsilon^+$ configurations to compute the full $\mathcal{Z}$, as not doing so would render the problem non-local.
\newline\newline
We are now tasked with computing the measure of the path integral with care.

\subsubsection*{Jacobian factor} 

As our measure of path integration, we select the following local measure
\begin{equation} \label{eq: psi measure}
    1 = \int \mathcal{D} \Psi \mathcal{D} \Psi^\dagger \exp \left( - \Lambda_{\text{u.v.}}^{\frac{1}{2}} \int d^4 x \sqrt{g} g^{\mu\nu} \Psi^\dagger_\mu \Psi_\nu \right) ~,   
\end{equation}
where $\Lambda_{\text{u.v.}}$ is some reference high-energy cutoff scale with units of inverse length squared. When introducing the change of variables (\ref{eq: first splitting of vector-spinor}), we must preserve the above normalisation condition and therefore introduce a Jacobian factor. In other words, we need a factor $J$ such that
\begin{align} \label{eq: psi expanded measure}
    1 = & \int \mathcal{D} \Psi \mathcal{D} \Psi^\dagger \exp \left( - \Lambda_{\text{u.v.}}^{\frac{1}{2}} \int d^4 x \sqrt{g} g^{\mu\nu} \Psi^\dagger_\mu \Psi_\nu \right)  \nonumber \\  
     =& \int \mathcal{D} \varphi \mathcal{D} \varphi^\dagger \mathcal{D}'' \chi \mathcal{D}'' \chi^\dagger \mathcal{D} {\lambda} \mathcal{D} {\lambda}^\dagger \, J \, \exp \Biggl( - \Lambda_{\text{u.v.}}^{\frac{1}{2}} \int d^4 x \sqrt{g} g^{\mu\nu} \biggl[\varphi_\mu + \nabla_\mu \chi + \gamma_\mu {\lambda}\biggr]^\dagger  \nonumber \\
    & \biggl[\varphi_\nu + \nabla_\nu \chi + \gamma_\nu {\lambda}\biggr] \Biggr) ~,
\end{align}
where the double prime in $\mathcal{D}'' \chi$, and $\mathcal{D}'' \chi^\dagger$ denotes the exclusion of all Killing spinors from path integration. Expanding the argument of the exponential in the previous expression and integrating by parts, we find
\begin{multline} \label{eq: expanded exponential argument}
     \biggl[\varphi_\mu + \nabla_\mu \chi + \gamma_\mu {\lambda}\biggr]^\dagger \biggl[\varphi^\mu + \nabla^\mu \chi + \gamma^\mu {\lambda}\biggr] = \\
     \varphi^{\mu\dagger}\varphi_\mu  + 4{\lambda}^\dagger{\lambda} +  \left( \nabla^\mu \chi \right)^\dagger \nabla_\mu \chi + \left( \nabla^\mu \chi \right)^\dagger \gamma_\mu {\lambda} + {\lambda}^\dagger \gamma^\mu \, \nabla_\mu \chi ~.
\end{multline}
In order to remove the cross-terms mixing $\chi$ and ${\lambda}$, we consider the field transformation $\lambda = -\tfrac{1}{4} \left(  \slashed{\nabla} \, \chi - \zeta \right)$ and $\chi = \chi$. After integrating by parts, we obtain
\begin{multline} \label{eq: second expanded exponential argument}
     \biggl[\varphi_\mu + \nabla_\mu \chi + \gamma_\mu {\lambda}\biggr]^\dagger \biggl[\varphi^\mu + \nabla^\mu \chi + \gamma^\mu {\lambda}\biggr] 
     = \varphi^{\mu\dagger}\varphi_\mu + \frac{1}{4}\zeta^\dagger\zeta - \frac{3}{4}\chi^\dagger \left( \nabla^2 +1 \right) \chi ~.
\end{multline}
Once again, the change of variables is a shift in $\lambda$. As such, it does not affect the Jacobian. Thus, (\ref{eq: psi expanded measure}) simplifies to
\begin{align} \label{eq: new expanded measure} \nonumber
    1 = \int \mathcal{D} \varphi \mathcal{D} \varphi^\dagger \mathcal{D}'' \chi \mathcal{D}'' \chi^\dagger \mathcal{D} \zeta \mathcal{D} \zeta^\dagger \, J \, \exp \left( - \Lambda_{\text{u.v.}}^{\frac{1}{2}} \int d^4 x \sqrt{g} \left[\varphi^{\mu\dagger}\varphi_\mu + \frac{1}{4}\zeta^\dagger\zeta - \frac{3}{4}\chi^\dagger \left( \nabla^2 +1 \right) \chi \right] \right)~.
\end{align}
Defining the measures governing the fields in the decomposition of $\Psi_\mu$ as
\begin{eqnarray} \label{eq: varphi measure}
    1 &\equiv& \int \mathcal{D} \varphi \mathcal{D} \varphi^\dagger \exp \left( - \Lambda_{\text{u.v.}}^{\frac{1}{2}} \int d^4 x \sqrt{g} \, \varphi^{\mu\dagger} \varphi_\mu \right) ~,   \\ 
    1 &\equiv& \int \mathcal{D} \chi \mathcal{D} \chi^\dagger \exp \left( - \Lambda_{\text{u.v.}}^{\frac{3}{2}} \int d^4 x \sqrt{g} \, \chi^{\dagger} \chi \right) ~,  \\
    1 &\equiv& \int \mathcal{D} \zeta \mathcal{D} \zeta^\dagger \exp \left( - \Lambda_{\text{u.v.}}^{\frac{1}{2}} \int d^4 x \sqrt{g} \, \zeta^{\dagger} \zeta \right) ~,   
\end{eqnarray}
we readily find that $J = \tfrac{1}{Y_{\chi}Y_\zeta}$, where
\begin{eqnarray} \label{eq: Jacobian 1}
    Y_{\chi} &=& \int \mathcal{D}'' \chi \mathcal{D}'' \chi^\dagger \, \exp \left( - \frac{3}{4} \Lambda_{\text{u.v.}}^{\frac{1}{2}} \int d^4 x \sqrt{g} \, \chi^\dagger (-\nabla^2-1) \chi \right) ~, \\
    Y_{\zeta} &=& \int \mathcal{D} \zeta \mathcal{D} \zeta^\dagger \, \exp \left( - \frac{1}{4} \Lambda_{\text{u.v.}}^{\frac{1}{2}} \int d^4 x \sqrt{g} \, \zeta^\dagger \zeta \right) ~.
\end{eqnarray}
It is possible to rewrite the above in terms of determinants as
\begin{equation} \label{eq: Jacobian 1 det}
    Y_{\chi} = \text{det}''   \frac{3(-\nabla^2-1)}{4\Lambda_{\text{u.v.}}}   ~,  \quad\quad
    Y_{\zeta} = \text{local}~.
\end{equation}
We note that we must exclude from the $\chi$-determinant Killing spinor with either sign for the eigenvalue, and this is indicated by the double prime notation.

\subsubsection*{Residual gauge group `volume'} \label{resvolume}

Since the Euclidean action $S_E$ is gauge-invariant under $\Psi_\mu \rightarrow \Psi_\mu + \left(\nabla_\mu + \frac{i}{2}\gamma_\mu \right)\omega$, where $\omega$ is a generic spinor, we must divide the path integral by a corresponding gauge group `volume' $$\text{vol} \, \mathcal{G}  = \int \mathcal{D} \omega \mathcal{D} \omega^\dagger~.$$ It stems from the following locally defined metric for the gauge parameter $\omega$
\begin{equation} \label{eq: metric for gauge parameter}
    ds^2_\omega = \Lambda^{\frac{3}{2}}_{\text{u.v.}} \int d^4x \sqrt{g} \, \delta\omega^\dagger \delta\omega ~,
\end{equation}
and it cancels almost perfectly with $\int \mathcal{D}'' \chi \mathcal{D}'' \chi^\dagger$. What is left is the residual `volume'
\begin{equation} \label{eq: residual volume}
    \frac{\int \mathcal{D}'' \chi \mathcal{D}'' \chi^\dagger \int \mathcal{D}\varepsilon^+ \left(\mathcal{D}\varepsilon^{+}\right)^\dagger} {\text{vol}\,\mathcal{G}} \equiv \frac{\Lambda_{\text{u.v.}}^4}{\text{vol} \, \mathcal{G}_{\text{KS}}} ~.
\end{equation}
For the purposes of our analysis, we are interested in the $\Lambda_{\text{u.v.}}$ dependence of the residual volume, and we will not compute the full $\text{vol} \, \mathcal{G}_{\text{KS}}$. The $\Lambda_{\text{u.v.}}^4$ in the numerator stems from the fact that $\varepsilon$ and $\chi$ have different units, length$^{-3/2}$ and length$^{-1/2}$ respectively. As there are four distinct complex Killing spinors that generate $\mathcal{G}_{\text{KS}}$, we find $${\text{vol} \, \mathcal{G}_{\text{KS}}} = a \, \Lambda^{-6}_{\text{u.v.}} ~,$$ for some constant $a$. Note that the power of $\Lambda_{\text{u.v.}}$ gets an extra minus sign as opposed to the analogous computation for the bosonic case, and this is due to the fact that the mode measure for the fermionic field $\omega$ is $\underset{i}{\prod} \,\Lambda^{-\frac{3}{2}}_{\text{u.v.}} \ d\omega^\dagger_i d\omega_i$, where the power of $\Lambda_{\text{u.v.}}$ is flipped with respect to the metric (\ref{eq: metric for gauge parameter}).

\subsubsection*{Final result}

Thus, using Gaussian path integration formulas for fermionic fields, we  obtain the following contributions for the $\varphi^\mu$ and $\varepsilon$ sectors
\begin{equation} \label{eq: varphi TT partition function as a det}
    \mathcal{Z}^{\text{TT}}_\varphi = \text{det} \left( \frac{\slashed{\nabla}+i}{\Lambda_{\text{u.v.}}^{\frac{1}{2}}} \right) ~, \quad \text{and} \quad \mathcal{Z}_{\varepsilon} = \text{det}' \left( \frac{3}{2} \frac{\slashed{\nabla}-2i}{\Lambda_{\text{u.v.}}^{\frac{1}{2}}} \right)~.
\end{equation}
Combined with the Jacobian factor, and assembling all the remaining pieces together, we arrive at the final expression
\begin{equation} \label{eq: final gravitino path integral}
    \mathcal{Z} = a \, \Lambda^{6+4-2}_{\text{u.v.}}  \frac{\text{det}^{(\frac{3}{2})} \left( \frac{\slashed{\nabla}+i}{\Lambda_{\text{u.v.}}^{\frac{1}{2}}} \right)}{\text{det}{''}^{\,(\frac{1}{2})} \left( -\frac{\slashed{\nabla}+2i}{\Lambda^{\frac{1}{2}}_{\text{u.v.}}} \right) } ~,
\end{equation}
where overall constants have been reabsorbed in the dimensionless pre-factor $a$. The double prime indicates that we are removing all Killing spinor configurations from the spectrum.

\subsubsection*{Coefficient of the logarithmic divergent term}

We will now proceed to study the functional determinants appearing in (\ref{eq: final gravitino path integral}). Given the spectra and degeneracies (\ref{eq: spinor spherical harmonics expansion}) - (\ref{eq: vector spinor degeneracy}), let us begin with the determinant stemming from the TT component of the vector-spinor. Employing the methods of \cite{Anninos:2020hfj}, one finds
\begin{equation}
    \log \text{ det}^{(\frac{3}{2})} \left( \frac{\slashed{\nabla}+i}{\Lambda_{\text{u.v.}}^{\frac{1}{2}}} \right) = - \int_\varepsilon^\infty \frac{dt}{\sqrt{t^2-\varepsilon^2}} \, 8 \frac{2\sinh t +3\cosh t -2}{(e^{t}-1)^4} \left( e^{\sqrt{t^2-\varepsilon^2}} + e^{-\sqrt{t^2-\varepsilon^2}} \right) ~.
\end{equation}
If we formally set $\varepsilon=0$, taking the small-$t$ expansion of the integrand yields the $\frac{1}{t}$ coefficient
\begin{equation}
    \mathcal{B}_1 = -\frac{191}{45} ~,
\end{equation}
which is associated with the logarithmic divergence. Similarly, for the determinant related to the ${\lambda}$ factor of the vector-spinor decomposition, we find
\begin{equation}
    \log \text{ det}''^{\,(\frac{1}{2})} \left( -\frac{\slashed{\nabla}+2i}{\Lambda^{\frac{1}{2}}_{\text{u.v.}}} \right)  = - \int_\varepsilon^\infty \frac{dt}{\sqrt{t^2-\varepsilon^2}} \, 4 \frac{8\cosh t -6 -e^{-2t}}{(e^{t}-1)^4} \left( e^{2\sqrt{t^2-\varepsilon^2}} + e^{-2\sqrt{t^2-\varepsilon^2}} \right) ~.
\end{equation}
If we again set $\varepsilon=0$ and take the small-$t$ expansion of the integrand, one obtains the $\frac{1}{t}$ coefficient
\begin{equation}
    \mathcal{B}_2 = \frac{469}{90} ~.
\end{equation}
It is then possible to find the overall coefficient of the logarithmic divergent term in the path integral $\mathcal{Z}$ as
\begin{equation} \label{eq: coeff log div func dets}
    \mathcal{B}_{S^4;\frac{3}{2}} = \left( -\frac{191}{45} \right) - \left( \frac{469}{90} \right) +{8\times2}= \frac{589}{90} ~,
\end{equation}
where the final $\mathcal{B}_{\text{vol}}= {+8\times2}$ is due to a proper treatment of the path integration measure --- it stems from the $\Lambda_{\text{u.v.}}$ dependent prefactor in (\ref{eq: final gravitino path integral}).

\subsection{Massive fermionic path integral} \label{apppathintmassive}

We now proceed to perform the computation of the Euclidean partition function for a massive spin-$\frac{3}{2}$ field,
\begin{equation}
\mathcal{Z}_m = 
\int \mathcal{D} \Psi \mathcal{D} \Psi^\dagger e^{-S_E[\Psi_\mu]}~,
\end{equation}
where the Euclidean action is given in (\ref{SE32m}) and the path integration contour choice is $\Xi_\mu = \Psi^*_\mu$. We reproduce this below, for convenience
\begin{equation}\label{SE32m appendix}
    S_E[\Psi_\mu]  =  \int d^4 x \sqrt{g} \, {\Psi}^{\dagger}_\mu \gamma^{\mu\rho\sigma} \left( \nabla_\rho + \frac{m}{2} \gamma_\rho \right) \Psi_\sigma ~.
\end{equation}
We note that no auxiliary fields are necessary in the local formulation of the free massive spin-$\tfrac{3}{2}$ theory. This is no longer true for spins greater than $\frac{3}{2}$ \cite{singh1974lagrangian}, and even for spin-$\tfrac{3}{2}$ a Stueckelberg formulation is available, as discussed in the next subsection. It is convenient to decompose the massive vector-spinor as
\begin{equation} \label{eq: second splitting of vector-spinor massive}
    \Psi_\mu =\varphi_\mu + \left( \nabla_\mu - \frac{1}{4} \gamma_\mu \slashed{\nabla} \right) \chi+ \frac{1}{4} \gamma_\mu {\lambda}~, ~~~\text{with}~~~ \nabla^\mu\varphi_\mu  = 0 = \gamma^\mu\varphi_\mu ~,
\end{equation}
where $\chi$ and $\lambda$ are independent Dirac spinors, {and we remove all Killing spinor configurations from $\chi$ to avoid redundancies in the parameterisation}. The Euclidean action $S_E[\Psi_\mu]$ then reduces to $S_E[\varphi_\mu] + S_E[\chi,\lambda]$, where  
\begin{equation} \label{eq: varphi TT action massive}
    S_E[\varphi_\mu] = \int d^4x \sqrt{g} \, \varphi_\mu^\dagger \left( \slashed{\nabla} + m \right) \varphi^\mu ~,
\end{equation}
and
\begin{equation} \label{eq: chi lambda action massive}
    S_E[\chi,\lambda] = \frac{3}{8} \int d^4x \sqrt{g} \, \begin{pmatrix}
        \chi^\dagger & \lambda^\dagger
    \end{pmatrix} \begin{pmatrix}
        -(\slashed{\nabla}+2m)(\nabla^2+1) & (\nabla^2+1) \\
        -(\nabla^2+1) & (\slashed{\nabla}-2m)
    \end{pmatrix} \begin{pmatrix}
        \chi \\ \lambda
    \end{pmatrix} ~.
\end{equation}
Let us further define 
\begin{equation} \label{eq: varphi TT partition function massive}
    \mathcal{Z}^{\text{TT}}_\varphi \equiv \int \mathcal{D} \varphi \mathcal{D} \varphi^\dagger e^{-S_E[\varphi_\mu]} \quad\text{and}\quad   \mathcal{Z}_{\chi,\lambda} \equiv \int \mathcal{D}'' {\chi} \mathcal{D}'' {\chi}^\dagger \mathcal{D} \lambda \mathcal{D} \lambda^\dagger e^{-S_E[\chi,\lambda]} ~,
\end{equation}
where the double prime in $\mathcal{D}'' {\chi}$ and $ \mathcal{D}'' {\chi}^\dagger$ designates the exclusion of Killing spinors from path integration. The Jacobian analysis then follows in an identical way as that of the Rarita-Schwinger gauge field. Differently from the massless case, one no longer has to divide by any residual gauge volume. 
\newline\newline
In addition to the determinant over the transverse-traceless sector, essentially $\mathcal{Z}^{\text{TT}}_\varphi$ in  (\ref{eq: varphi TT partition function as a det}) but with a general mass parameter,  one ends up with a ratio of determinants, one coming from the Jacobian, as in (\ref{eq: Jacobian 1}), and the other from $\mathcal{Z}_{\chi,\lambda}$ in (\ref{eq: varphi TT partition function massive}). These determinants almost entirely cancel, but a residual term
\begin{equation}\label{mzm}
    \left( \frac{\Lambda_{\text{u.v.}}^{\frac{1}{2}}}{f(m)}  \right)^8 \times \text{det} \left( 4\frac{m^2+1}{\Lambda_{\text{u.v.}}} \right) = \left( \frac{\Lambda_{\text{u.v.}}^{\frac{1}{2}}}{f(m)}  \right)^8 \times \text{local}
\end{equation}
remains, with $f(m)$ a linear function in $m$. The overall $\Lambda_{\text{u.v.}}^4$ in the above expression stems from the omission of the eight complex Killing spinor modes from $\chi$.

\subsubsection*{Coefficient of the logarithmic divergent term}
 
Given the spectra and degeneracies in appendix \ref{appsphharmonics}, the determinant stemming from the transverse-traceless component of the massive vector-spinor in a heat-kernel regularisation is found to be
\begin{equation}
    \log \text{ det}^{(\frac{3}{2})} \left( \frac{\slashed{\nabla}+m}{\Lambda_{\text{u.v.}}^{\frac{1}{2}}} \right) = - \int_\varepsilon^\infty \frac{dt}{\sqrt{t^2-\varepsilon^2}} \, 8 \frac{2\sinh t +3\cosh t -2}{(e^{t}-1)^4} \left( e^{im\sqrt{t^2-\varepsilon^2}} + e^{-im\sqrt{t^2-\varepsilon^2}} \right) ~.
\end{equation}
Formally setting $\varepsilon=0$, and taking the small-$t$ expansion of the integrand yields the $\frac{1}{t}$ coefficient
\begin{equation}
    \mathcal{B}_1 = -\frac{401}{45} -\frac{2}{3} m^2(8+m^2) ~.
\end{equation}
The overall coefficient of the logarithmic divergent term for the massive path integral $\mathcal{Z}_m$ is thus
\begin{equation} \label{eq: coeff log div func dets}
    \mathcal{A}_{S^4;\frac{3}{2}} = -\frac{401}{45} -\frac{2}{3} m^2(8+m^2) + 4\times2 = -\frac{41}{45} -\frac{2}{3} m^2(8+m^2) ~,
\end{equation}
where the additional $\mathcal{B}_{\text{KS}}=+4\times2$ is due to the $\Lambda_{\text{u.v.}}$-dependent prefactor in (\ref{mzm}).

\subsection{Stueckelberg formulation of massive spin-$\tfrac{3}{2}$}

For the sake of completeness, we consider here the massive spin-$\tfrac{3}{2}$ theory as formulated with auxiliary fields \cite{buchbinder2006gauge,Zinoviev:2007ig,Boulanger:2023lgd}. The net effect is to introduce an additional Dirac spinor $\chi$ that couples quadratically to $\Psi_\mu$. In Lorentzian signature, the Stueckelberg action for the massive Dirac spin-$\frac{3}{2}$ field is
\begin{multline}
    S_{{L},\text{st}}[\Psi_\mu, \chi]=-\int d^4x \sqrt{-g} ~\Bigg( \bar{\Psi}_{\mu}\gamma^{\mu \rho \sigma}\left(\nabla_\rho +\frac{m}{2} \gamma_\rho   \right) \Psi_{\sigma} + \\\frac{3 \alpha_2}{2}\left( \bar{\Psi}_{\mu}\gamma^{\mu}\chi-\bar{\chi} \gamma^{\mu}\Psi_{\mu}  \right)+ \frac{3}{2} \bar{\chi}\,\gamma^\beta \nabla_\beta \chi  +3m \bar{\chi}\chi\Bigg)~.
\end{multline}
The above theory has a gauge redundancy
\begin{equation}\label{Stueck gauge sym 3/2}
\delta \Psi_\mu = \nabla_\mu \xi + \frac{m}{2}\gamma_\mu \xi~, \quad\quad \delta \chi = \alpha_2 \, \xi~, \quad\quad \alpha_2 \equiv {\pm} \sqrt{m^2+1}~.
\end{equation}
The Dirac spinor $\xi$ is a gauge parameter. The auxiliary spinor $\chi$ is the Stueckelberg `counterpart' of $\Psi_\mu$. 
\newline\newline
In Euclidean signature, we consider the following Stueckelberg action 
\begin{multline}
    {S}_{{E},\text{st}} = \int d^4x \sqrt{g} ~\Bigg( {\Xi}^{T}_{\mu}\gamma^{\mu \rho \sigma}\left(\nabla_\rho +\frac{m}{2} \gamma_\rho   \right) \Psi_{\sigma} +\frac{3 \alpha_2}{2} \left( {\Xi}^{T}_{\mu}\gamma^{\mu}\chi-{\zeta}^{T} \gamma^{\mu}\Psi_{\mu}  \right) + \\ \frac{3}{2} {\zeta}^{T}\,\gamma^\beta \nabla_\beta \chi + 3m \, \zeta^{T}\chi\Bigg)~,
\end{multline}
where $\alpha_2 = \pm\sqrt{m^2 +1} \in \mathbb{R}$. At this stage, the two Euclidean Dirac vector-spinors $\Xi_{\mu}$ and $\Psi_{\mu}$ are not related to each other, and the same is true for the two Dirac spinors $\chi$ and $\zeta$. We note that the Stueckelberg action is invariant under the following gauge transformations:
\begin{align} \label{complx stueck gauge sym Psi} 
    \delta \Psi_{\mu} = \left(\nabla_{\mu} + \frac{m}{2} \gamma_\mu\right) \xi ~, \quad\quad \delta \chi = \alpha_2 \, \xi~,
\end{align}
\begin{align} \label{complx stueck gauge sym Xi}
    \delta \Xi_{\mu} = \left(\nabla_{\mu} - \frac{m}{2} \gamma^{*}_\mu\right) \tilde{\xi} ~, \quad\quad \delta \zeta = \alpha_2 \, \tilde{\xi} ~,
\end{align}
where $\xi$, $\tilde{\xi}$ are independent spinor gauge parameters. It is also useful to rewrite the Stueckelberg action in a form that makes the aforementioned gauge symmetries manifest
\begin{multline}
    {S}_{{E},\text{st}} = \int d^4x \sqrt{g} ~\Bigg({\Xi}_{\mu}^*-\frac{1}{\alpha_2} \left(\nabla_{\mu}-\frac{m}{2}\gamma_{\mu}\right) \zeta^* \Bigg)^{\dagger} \gamma^{\mu \rho \sigma} \left(\nabla_\rho +\frac{m}{2} \gamma_\rho   \right) \\ \Bigg({\Psi}_{\sigma}-\frac{1}{\alpha_2} \left(\nabla_{\sigma}+\frac{m}{2}\gamma_{\sigma}\right) \chi \Bigg).
\end{multline} 
It is easy to verify that the  reality conditions $\Xi_{\mu}=\Psi_{\mu}^*$, $\zeta = \chi^*$ are {not} consistent with the gauge transformations (\ref{complx stueck gauge sym Psi}), (\ref{complx stueck gauge sym Xi}). Interestingly, the reality conditions $\Xi_{\mu}=\gamma^5\Psi^{*}_{\mu}$ and $\zeta = \gamma^5 \chi^{*}$ are instead compatible with the previous gauge transformations, alongside with the relation $\tilde{\xi} = \gamma^5 \xi^{*}$ (see for instance B.22 of \cite{Bobev:2023dwx} for the AdS$_4$ counterpart).

\section{Massless Dirac fermion in dS$_4$}\label{appDirac}

In this appendix we briefly review the massless Dirac fermion on a dS$_4$ background (\ref{planar}). Some general considerations of fermions in de Sitter were  recently studied in \cite{Pethybridge:2021rwf,Schaub:2023scu, Letsios:2020twa}. The Lorentzian action is given by
\begin{equation}\label{freedirac}
    S_L=  \int \frac{d\eta d\bold{x}}{\eta^4} \, \Psi^\dag i \gamma^{\underline{0}} \gamma^\mu {\nabla}_\mu \Psi~,
\end{equation}
which is real. Since the theory is conformally invariant we can map it to half of flat space by taking $\Psi \to (-\eta)^{\frac{3}{2}}\Psi$, and solve the problem in ordinary Minkowski space and map the final results back to de Sitter space. Following the standard quantisation procedure, we have the local field operator
\begin{equation}
    \hat{\Psi}(\eta,\bold{x}) = (- \eta)^{\frac{3}{2}} \sum_{s = \pm\frac{1}{2}} \int d^3\bold{k} \left( \hat{a}^{s}_{\bold{k}} \upsilon^{(s)} e^{i \bold{k}\cdot\bold{x}} + \hat{b}^{s\dag}_{\bold{k}}  {\upsilon}^{(s) \, C} e^{- i \bold{k}\cdot\bold{x}} \right) ~,
\end{equation}
with the standard anti-commutators
\begin{equation}
    \{\hat{a}^{s}_{\bold{k}}, \hat{a}^{s'\dag}_{\bold{k}'}\} = \delta^{ss'} \delta(\bold{k}-\bold{k}') = \{ \hat{b}^{s}_{\bold{k}}, \hat{b}^{s'\dag}_{\bold{k}'} \} ~.
\end{equation}
In terms of the local fields
\begin{equation}\label{DFlagebra}
\{ \hat{\Psi}_\alpha(\eta,\bold{x}) , \hat{\Psi}_\beta(\eta,\bold{y})^\dag \} = -\eta^3 \delta_{\alpha\beta} \delta(\bold{x}-\bold{y})~,
\end{equation}
where we have reinstated the four-spinor indices, and we recall that $\eta \in \mathbb{R}^-$. The four-spinor mode functions are given by
\begin{equation} \label{eq: chi explicit}
    \upsilon^{(\pm\frac{1}{2})} = \frac{1}{\sqrt{2(2\pi)^3}} \begin{pmatrix} 
        \pm \chi^{(\pm)}(\bold{k}) \\ 
        \chi^{(\pm)}(\bold{k}) 
    \end{pmatrix} e^{-ik\eta} ~, \quad 
\end{equation}
where the helicity two-spinor eigenstates satisfy (\ref{chi spinor eq.}) and, in the basis (\ref{basischoice}}), they are given by
\begin{equation}\label{eq: spinors explicit on R^3}
\chi^{(\pm)}(\bold{k}) = \frac{1}{\sqrt{2k(k\mp k_3)}} \begin{pmatrix}
        \pm (k_1-ik_2) \\
        k \mp k_3
    \end{pmatrix} ~.
\end{equation}
The vacuum state $|0\rangle$ is annihilated by all the $\hat{a}^{s}_{\bold{k}}$ and $\hat{b}^{s}_{\bold{k}}$. The four-spinors $\upsilon^{(s)}$ and ${\upsilon}^{(s) \, C}  \equiv \gamma^{\underline{2}} \upsilon^{(s)*}$ satisfy the massless Dirac equation in momentum space. They are normalised as
\begin{eqnarray}
    \sum_s {\upsilon}^{(s)} \,\overline{\upsilon^{(s)}} &=& \frac{i}{2k \, (2\pi)^3} \left( \gamma^{\underline{0}} k - \gamma^{\underline{j}} k_{\underline{j}} \right) ~,\\
    \sum_s {{\upsilon}}^{(s) \, C} \,\overline{\upsilon^{(s)  C}} &=& \frac{i}{2k \, (2\pi)^3} \left( \gamma^{\underline{0}} k - \gamma^{\underline{j}} k_{\underline{j}} \right) ~.
\end{eqnarray}
Up to a phase, the spatial parity operator $\hat{P}$ acts on the creation and annihilation operators as $\hat{P}\hat{a}^s_{\bold{k}}\hat{P}^{-1} = \hat{a}^s_{-\bold{k}}$ and so on. This yields $\hat{P}\hat{\Psi}(\eta,\bold{x})\hat{P}^{-1} \to \gamma^{\underline{0}} \hat{\Psi}(\eta,-\bold{x})$ up to a phase.

\subsection{Chiral and parity basis}

It is also convenient to go to a chiral basis. To do so we define the chiral projectors $P_L = \tfrac{1+\gamma^5}{2}$ and $P_R = \tfrac{1-\gamma^5}{2}$. For the massless Dirac theory, the chirality and the helicity agree. So we can project the operator onto $\hat{\Psi}_{L/R} \equiv P_{L/R} \hat{\Psi}$ which has the content of a two-spinor. The operator algebra is now
\begin{equation}
\{\hat{\Psi}_{L/R}(\eta,\bold{x}), \hat{\Psi}_{L/R}(\eta,\bold{y})^\dag \} = - \eta^3 \, P_{L/R} \, \delta(\bold{x}-\bold{y})~,
\end{equation}
with other anti-commutators vanishing. Under a parity transformation and up to a phase, $\Psi_L(t,\bold{x}) \to \Psi_R(t,-\bold{x})$. Thus, the late time surface $\eta=0^-$ is populated by four two-spinor operators. 
\newline\newline
We can also define parity even and parity odd parts of the fermion by projecting onto $P_\pm = \tfrac{(1\mp i \gamma^{\underline{0}})}{2}$. The parity even fermionic operator is $\Psi_+ \equiv \Psi_L + \Psi_R$, while the parity odd one is $\Psi_- \equiv \Psi_L - \Psi_R$. The corresponding non-vanishing operator algebra is now
\begin{equation}
\{\hat{\Psi}_{\pm}(\eta,\bold{x}), \hat{\Psi}_{\pm}(\eta,\bold{y})^\dag \} = - \eta^3 \, P_\pm \, \delta(\bold{x}-\bold{y})~. 
\end{equation}
Again, $\Psi_\pm$ have the content of a two-spinor.

\subsection{Constant time two-point functions}

The two-point function on a constant-$\eta$ surface is given by
\begin{equation}
\langle 0 | \hat{\Psi}(\eta,\bold{x}) \hat{\Psi}(\eta,\bold{y})^\dag| 0 \rangle = - \frac{\eta^3}{2} \mathbb{I}_4 \,\delta(\bold{x}-\bold{y}) - \frac{i \eta^3}{2\pi^2}  \frac{ \boldsymbol{\gamma} \cdot (\bold{x}-\bold{y}) }{|\bold{x}-\bold{y}|^4} \gamma^{\underline{0}}~. 
\end{equation}
In terms of the chiral basis, we have the non-vanishing two-point functions
\begin{eqnarray}
\langle 0 | \hat{\Psi}_L(\eta,\bold{x}) \hat{\Psi}_L(\eta,\bold{y})^\dag| 0 \rangle = -   \frac{\eta^3}{4} \mathbb{I}_2 \, \delta(\bold{x}-\bold{y}) - \frac{i \eta^3}{4\pi^2}  \frac{ \boldsymbol{\sigma} \cdot (\bold{x}-\bold{y}) }{|\bold{x}-\bold{y}|^4}   ~, \\
\langle 0 | \hat{\Psi}_R(\eta,\bold{x}) \hat{\Psi}_R(\eta,\bold{y})^\dag| 0 \rangle =  -  \frac{\eta^3}{4} \mathbb{I}_2 \, \delta(\bold{x}-\bold{y}) +  \frac{i \eta^3}{4\pi^2}  \frac{ \boldsymbol{\sigma} \cdot (\bold{x}-\bold{y}) }{|\bold{x}-\bold{y}|^4}~. 
\end{eqnarray}
The left-hand sides are understood as projected onto the appropriate chirality sector. We note that the non-local pieces of the two-point functions are those of a two-spinor conformal primary of weight $\Delta=\frac{3}{2}$ in a Euclidean three-dimensional conformal field theory. The contact terms, far from being innocuous, are directly related to the quantum field and the operator algebra (\ref{DFlagebra}).
\newline\newline
The constant-$\eta$ two-point functions in the parity basis are given by
\begin{eqnarray}
\langle 0 | \hat{\Psi}_+(\eta,\bold{x}) \hat{\Psi}_-(\eta,\bold{y})^\dag| 0 \rangle &=&  - \frac{i \eta^3}{2\pi^2}  \frac{ \boldsymbol{\sigma} \cdot (\bold{x}-\bold{y}) }{|\bold{x}-\bold{y}|^4}   ~, \\
\langle 0 | \hat{\Psi}_-(\eta,\bold{x}) \hat{\Psi}_+(\eta,\bold{y})^\dag| 0 \rangle &=& - \frac{i \eta^3}{2\pi^2}  \frac{ \boldsymbol{\sigma} \cdot (\bold{x}-\bold{y}) }{|\bold{x}-\bold{y}|^4}   ~, \\
\langle 0 | \hat{\Psi}_+(\eta,\bold{x}) \hat{\Psi}_+(\eta,\bold{y})^\dag| 0 \rangle &=& - \frac{\eta^3}{2} \mathbb{I}_2 \, \delta(\bold{x}-\bold{y})  ~, \\
\langle 0 | \hat{\Psi}_-(\eta,\bold{x}) \hat{\Psi}_-(\eta,\bold{y})^\dag| 0 \rangle &=& - \frac{\eta^3}{2} \mathbb{I}_2 \, \delta(\bold{x}-\bold{y}) ~. 
\end{eqnarray}
As expected from a parity invariant theory, the non-trivial two-point functions at finite separation involve mixed parity basis correlators.

\bibliographystyle{JHEP}
\bibliography{bib}

\end{document}